\documentclass[aps, 10pt, prb, twocolumn, amssymb, amsmath, amsfonts, showpacs, superscriptaddress, preprintnumbers, a4paper]{revtex4-2}  
\usepackage{amsmath, amssymb, amsthm}

\DeclareMathOperator{\Sgn}{sgn}
\usepackage{physics}
\usepackage{times}
\usepackage{graphicx, xcolor}
\usepackage[latin1]{inputenc}
\usepackage{comment}
\usepackage{ulem}
\usepackage[
  breaklinks,
  colorlinks=true,
  bookmarks=true,
  bookmarksopenlevel=3,
  bookmarksnumbered=true,
  menucolor=blue,
  linkcolor=blue,
  citecolor=blue,
  urlcolor=blue,
  pdfcreator={GhostScript},
  pdfkeywords={},
  pdfstartview=FitH]{hyperref}
\newtheorem{lemma}{Lemma}
\begin{document}
    
\def\lptms{Universit\'e Paris-Saclay, CNRS, LPTMS, 91405, Orsay, France.}
\def\lps{Universit\'e Paris-Saclay, CNRS, Laboratoire de Physiques des Solides, 91405, Orsay, France.}

\title{Exact many-body scars based on pairs or multimers in a chain of spinless fermions }

\author{Lorenzo Gotta}\email{lorenzo.gotta@universite-paris-saclay.fr}\affiliation{\lptms}
\author{Leonardo Mazza}\affiliation{\lptms}
\author{Pascal Simon}\affiliation{\lps}
\author{Guillaume Roux}\affiliation{\lptms}

\date{\today}

\begin{abstract}
We construct a 1D model Hamiltonian of spinless fermions for which the spinless analogue of $\eta$-pairing states are quantum many-body scars of the model.
These states are excited states and display subvolume entanglement entropy scaling; they form a tower of states that are equally spaced in energy (resulting in periodic oscillations in the Loschmidt echo and in the time evolution of observables for initial states prepared in a superposition of them) and are atypical in the sense that they weakly break the eigenstate thermalization hypothesis, while the other excited states are thermal.
We extend the approach by presenting models with a tower of scar states generated by multimers located at the edge of the Brillouin zone.
\end{abstract}

\maketitle

\section{Introduction} 

The question of thermalization in isolated many-body quantum systems typically relies on the eigenstate-thermalization-hypothesis (ETH)~\cite{Deutsch_1991,Srednicki_1994,Rigol_2008,Deutsch_2018}, which assumes that highly-excited eigenstates of nonintegrable quantum many-body systems act as effective baths for their small subsystems, thus allowing a conventional statistical ensemble description of local observables. An active research area stemming from the introduction of ETH has been the search for systems that violate such a conjecture and hence display anomalous thermalization behavior. 

The strategies towards the realization of systems lacking standard thermalization under unitary time evolution have relied on the presence of an extensive number of integrals of motion, thereby enforcing the time-evolved state to retain memory of its initial configuration. The two representatives of the aforesaid routes towards ETH-breaking are fine-tuned integrable systems, featuring an extensive set of global conserved quantities~\cite{Vidmar_2016,Essler_2016,Rigol_2007,Calabrese_2011,Pozsgay_2013,Fagotti_2013,Wright_2014,Ilievski_2015}, and strongly disordered systems, where the phenomenon of many-body-localization has been explained by introducing the idea of local integrals of motion~\cite{Basko_2006,Bardarson_2012,Serbyn_2013,Serbyn_2013b,Huse_2014,Luitz_2015,Chandran_2015,Ros_2015,Nandkishore_2015}. Moreover, signatures of breakdown of thermalization have been reported in Floquet quantum matter~\cite{Haldar_2018,Haldar_2021,Haldar_2022}, where the presence of an external driving can induce emergent conservation laws in clean interacting quantum many-body systems. 
  
More recently, an experiment on quantum quenches in a system of cold Rydberg atoms revealed persistent coherent oscillations in local observables when monitoring the time evolution starting from a specific initial state~\cite{Bernien_2017}. Such a feature was shown to rely on the presence of a measure-zero set of exceptional ETH-violating eigenstates, called quantum many-body scars~\cite{Turner_2018,Moudgalya_2021,Papic_2021,Chandran_2022}, embedded in an otherwise thermalizing spectrum~\cite{Turner_2018b,Iadecola_2019,Choi_2019,Lin_2019,Lin_2020}.  
A recent observation of many-body scarring in a quantum simulator has also been reported~\cite{Su_2022, Desaules_2022, Desaules_2022bis}. 
The latter have been later constructed in several nonintegrable models of spin-$1$~\cite{Moudgalya_2018,Moudgalya_2018b,Moudgalya_2020b,Lin_2019,Chattopadhyay_2020,Mark_2020b,Shiraishi_2017} and spin-$\frac{1}{2}$ chains~\cite{Shiraishi_2017,Iadecola_2020,Mark_2020b,Langlett_2021} or, more generically, spin-$S$ chains~\cite{Shibata_2020}, as well as in spinful fermionic systems in one and higher dimensions~\cite{Moudgalya_2020,Mark_2020,Desaules_2021} and in quantum Hall models~\cite{Moudgalya_2020c}. 
They all have the form of energetically equally-spaced towers of exact eigenstates of the Hamiltonian, and this peculiar spectral feature lies at the heart of the observed periodic revivals in the time evolution of suitably chosen local observables. Further examples of quantum many-body scars have been predicted in the framework of Floquet-engineered  systems~\cite{Mukherjee_2020,Zhao_2020,Mizuta_2020,Sugiura_2021}, lattice gauge theories~\cite{Banerjee_2021,Halimeh_2022,Aramthottil_2022}, flat-band models~\cite{McClarty_2020,Kuno_2020} and magnetically-frustrated systems~\cite{Lee_2020, McClarty_2020, Lee_2021}.   

A crucial step towards the understanding of quantum many-body scars was played by $\eta$-pairing states, firstly discovered by Yang as exact excited eigenstates of the Hubbard model with off-diagonal long-range order~\cite{Yang_1989}. Despite the $\eta$-pairing states not representing genuine many-body scars of the Hubbard model due to the presence of a hidden $\eta$-pairing $SU(2)$ symmetry, they display the prototypical algebraic properties of towers of scarred eigenstates, as they are generated by the repeated application of a ladder-like operator to a  weakly-entangled state, and a subvolume entanglement entropy scaling law~\cite{Vafek_2017}. Thus, the $\eta$-pairing states have inspired several works aiming at unveiling a universal scarring mechanism allowing for the microscopic emergence of many-body scars, as well as the search for Hubbard-like Hamiltonians with $\eta$-pairing-symmetry-breaking terms that preserve analytically tractable towers of $\eta$-pairing states as genuine many-body scars~\cite{Moudgalya_2020}.

In this work, we unveil the existence of an exact tower of scarred eigenstates of a spinless fermion Hamiltonian by generalizing the mechanism of $\eta$-pairing to the case of spinless fermions. 
Our analysis provides a further illustration of the characteristic properties of towers of scarred eigenstates studied in the literature and with one of the simplest particles: spinless fermions. The scarred eigenstates are characterized by ETH-violating properties such as the logarithmic scaling of the entanglement entropy in the size of the selected subsystem~\cite{Turner_2018b,Choi_2019,Chattopadhyay_2020,Schecter_2019,Iadecola_2020,Moudgalya_2020,Vafek_2017} and the off-diagonal long-range order~\cite{Schecter_2019,Iadecola_2020,Pakrouski_2020,Nakagawa_2022,Yoshida_2022} in the pair correlation function. Moreover, we show how the choice of a superposition of scarred eigenstates as the initial state of the time evolution leads to periodic revivals in the expectation values of local observables~\cite{Alhambra_2020,Schecter_2019,Iadecola_2020}. We highlight the peculiar feature of symmetry enhancement in the scarred subspace by drawing connections to the concept of quasi-symmetry~\cite{Ren_2021,ODea_2020} and its relation to many-body scar dynamics.

Remarkably, all these results can be extended to scar states characterised by multimers where $M>2$ particles are bound together; we discuss explicitly the case $M=3$, where trimers are fermionic and display some qualitative different features with respect to the bosonic case.

Despite the aforementioned phenomenology being known, we wish to highlight the aspects of our work that have not been significantly underlined in the preceding literature. Firstly, we extend pioneering results on many-body scars in spinful fermionic systems, where $\eta$-pairs represent the infinitely long-lived quasiparticles that underlie the corresponding tower of scarred eigenstates~\cite{Moudgalya_2020}. More precisely, we reveal how an analogous structure is realized in a system of spinless fermions. The extended spatial structure of the quasiparticles, namely pairs or multimers of spinless fermions, reflects itself into nonlocal expressions for the lowering operator in the scarred subspace (also discussed in Ref.~\cite{Iadecola_2020}) and for a nontrivial conserved quantity of the many-body scar dynamics.
Moreover, we highlight the fact that our paired scarred states can be interpreted within the framework of macroscopic quantum coherence in the grand-canonical ensemble by constructing a close analogue of bosonic coherent states~\cite{Shibata_2020}.
Our result differs from the one expected in the case of a purely bosonic mode as a result of the hard-core nature of the pairs.
We find that the fermionic case is even more peculiar: although we are able to write the scarred states analytically and to identify them in numerical simulations, we were not able to find an expression for the raising and lowering operators that is local. To the best of our knowledge, so far at least one of the two has always been found.

This article is organised as follows.
The first part of the article is devoted to scars based on pairs and it consists 
in Section~\ref{S1}. After introducing the model Hamiltonian and discussing the structure of the interaction term, we define the tower of scarred eigenstates and characterize their spectral and entanglement properties, thereby underlining their consequences on the dynamics. 
Our analysis is corroborated by numerical data.
The second part of the article consists in
Section~\ref{Sec:Multimers}, where we extend our results to the multimer case: we present a simple Hamiltonian that supports exact multimer scars and we argue with analytical and numerical results that the model is not integrable and that the scars are not generic.
The conclusions are presented in Sec.~\ref{S4}.

\section{Scars based on pairs in a spinless-fermion model}\label{S1}

We start our discussion focusing on a model that features exact many-body scar states that generalise $\eta$-pairing to a spinless fermion chain:
\begin{align}\label{Eq:hamiltonian}
\hat H =&  -t \sum_{j} \left[\hat c^{\dag}_j \hat c_{j+1}  +\text{H.c.} \right]-J \sum_j\left[ \hat c^{\dag}_j \hat n_{j+1} \hat c_{j+2}+\text{H.c.}\right] \nonumber \\
&-\mu \sum_{j} \hat n_j+J\sum_{j}
\left[\hat n_{j+1}(\hat n_j+\hat n_{j+2} )-2\hat n_j \hat n_{j+1} \hat n_{j+2} \right]; 
\end{align}
the fermionic creation and annihilation operators satisfy the canonical anticommutation relations $\{ c_i, c_j \} = 0$ and $\{ \hat c_i, \hat c^{\dag}_j\} = \delta_{i,j}$. 
The hopping amplitude is $t$, $\mu$ is the chemical potential and $J$ is the pair-hopping amplitude, associated to the motion of two neighboring particles; $J$ is also the parameter of different forms of density-density interactions. We take $J>0$ and $t>0$. 
This kind of correlated pair-hopping has been recently studied in a variety of works and is responsible for several phenomena related to pairing~\cite{Bariev_1991,Chhajlany_2016,Ruhman_2017, Gotta_2021, Gotta_2021bis, Gotta_2022}. 
The lattice size is denoted by $L$ and the Hamiltonian conserves the total number of particle operator $N$. 
Through this article, for simplicity, we always assume $L$ and $N$ to be even and we take open boundary conditions if not explicitly mentioned.
To ease readability, we write explicitly the bounds of summation only when they are non-trivial.

\subsection{Exact results on scar states and towers of states}\label{S2}

We now discuss a set of scarred eigenstates for $\hat H$. In order to do so, we define the tower of states for $k \in \mathbb \{0, 1, \ldots L/2 \}$:
\begin{align}\label{Eq:eta_pairs}
\ket{\psi_{k,\pi}}=\frac{1}{\sqrt{\binom{L-k}{k}}}\frac{(\hat\eta^{\dag}_{\pi})^{k}}{k!}\ket{\emptyset}, \, \text{ with } \,
\hat\eta^{\dag}_{\pi} = \sum_{j}e^{i\pi j} \hat c^{\dag}_j \hat c^{\dag}_{j+1}.
\end{align}
This set of states is the closest analogue of $\eta$-pairing in a spinless fermionic setup as they represent a condensate of pairs that have $\pi$ momentum, ie. they belong to the
edge of the first Brillouin zone. Such a state have a fixed number of pairs $k$, and thus a number of fermions $N= 2k$. 

The normalization factor follows from combinatorial considerations. Indeed, given $N$ fermions on a lattice of size $L$, one can map each fermionic configuration where particles form even-sized clusters to a spin configuration with $ k$ spin-up states on a spin chain of size $L-k$ via the rules $\ket{\bullet\bullet}\rightarrow\ket{\uparrow},\,\ket{\circ}\rightarrow\ket{\downarrow}$. Then, the number of fully-paired fermionic configurations on the original lattice equals the number of ways of distributing 
$k$ spins-up on a chain of length $L-k$, which is $\binom{L- k}{k}$.

The states introduced in Eq.~\eqref{Eq:eta_pairs} form a tower of energetically equally-spaced eigenstates of $\hat H$, satisfying the eigenvalue equation:
\begin{align}
\hat H \ket{\psi_{k,\pi}}=-2\mu k \ket{\psi_{k,\pi}}.\label{Eq:eigv_eq_pi}
\end{align}
This result is explicitly derived in Appendix~\ref{App:EigenvalueEquation}; very briefly, it follows from the destructive interference of the single fermions when single-particle hopping breaks a pair into two fermions (similar mechanisms have been also highlighted in other models, e.g.~spin-1 models~\cite{Iadecola_2019}).

Moreover, the $\ket{\psi_{k,\pi}}$ satisfy the standard restricted spectrum-generating algebra (RSGA) typical of the tower of states:
\begin{equation}\label{Eq:RSGA}
 [\hat H, \eta^\dagger_\pi ] \ket{\psi_{k,\pi}} = -2 \mu \eta^\dagger_\pi \ket{\psi_{k,\pi}},
\end{equation}
and as such they fit exactly in the standard theory of exact many-body scars with linearly-separated energies.
Additionally, the states $\ket{\psi_{k,\pi}}$ are the exact frustration-free ground states of the Hamiltonian 
$\hat H_J =+ (J/2) \sum_j L_j^\dagger L_j$ for $J>0$, where
\begin{equation}
\hat L_{j}=\hat n_j \hat n_{j+1}-\hat n_{j+1}\hat n_{j+2}+\hat c^{\dag}_{j+2}\hat n_{j+1}\hat c_j-\hat c^{\dag}_j\hat n_{j+1}\hat c_{j+2},
\end{equation}
which corresponds to the part proportional to $J$ of the model in Eq.~\eqref{Eq:hamiltonian}. Thus,
we can interpret them as scars obtained by deforming a frustration-free non-integrable model, the deformation being obtained by adding the single-particle hopping.

The fact that pairs located at momentum $\pi$ are eigenstates of the Hamiltonian means that they can be thought of as quasiparticles with infinite lifetime;
the equal energy spacing is instead associated to the fact that they are not interacting (see Appendix~\ref{App:CBA} for a coordinate Bethe Ansatz argument supporting the latter observation). It is enough to assume $t \gg |\mu|$  to place them in the middle of the spectrum of $\hat H$, of which they become exact eigenstates that lie at an extensive energy above the ground-state energy.

It is not difficult to observe that the $\ket{\psi_{k,\pi}}$  feature off-diagonal long-range order; let us introduce:
\begin{align}
P_{k}(r)=\bra{\psi_{k,\pi}} \hat c_j^\dagger \hat c_{j+1}^\dagger \hat c_{j+r} \hat c_{j+r+1} \ket{\psi_{k,\pi}},
\end{align}
to denote the pair correlation function evaluated on the state $\ket{\psi_{k,\pi}}$. Then, taking periodic boundary conditions and the thermodynamic limit at fixed density $n = 2 k/L$, we obtain
\begin{equation}\label{Eq:ODLRO}
 \lim_{L \to \infty}  P_{k}(r) =  (-1)^{r+1} \frac{n}{2-n}(1-n)^2,
\end{equation}
with $r>3$.
The explicit formula at finite size is given in Appendix~\ref{App:pair_corr}.
This observation alone is sufficient to motivate the fact that they are exceptional in the spectrum of the Hamiltonian and break ETH, as they violate the Mermin-Wagner theorem about thermal states in one dimension. 
As a further proof of ETH breaking, we will later also discuss the fact that the entanglement entropy of these states grows logarithmically with the subsystem length, instead of linearly,  as it is typically expected for thermal states.

\subsection{Algebraic properties}

As mentioned, the operators $\hat \eta^\dagger_{\pi}$ realise a RSGA in the subspace $\mathcal S$ spanned by the $\ket{ \psi_{k,\pi}}$. Indeed, the commutator $[\hat H , \hat\eta^{\dag}_{\pi}]$ reads:
\begin{align}
[\hat H,\hat \eta^{\dag}_{\pi}]=-2\mu\hat\eta^{\dag}_{\pi}+\hat O,
\end{align}
where the explicit expression for $\hat O$ is given in Appendix~\ref{App:O_operator}. There, we also show that the states $\ket{\psi_{k,\pi}}$ belong to the kernel of $\hat O$. Therefore, Eq.~\eqref{Eq:RSGA} is satisfied and, in turn, the eigenvalue equation~\eqref{Eq:eigv_eq_pi} is proven in a way that is different from that presented in Appendix~\ref{App:EigenvalueEquation}.

More interestingly, we observe peculiar consequences of the spatial structure of the pairs when considering a lowering operator $\hat\eta'_{\pi}$ satisfying $\hat\eta'_{\pi}\ket{\psi_{k,\pi}}\propto \ket{\psi_{k-1,\pi}}$ for $k \geq 1$ and $\hat\eta'_{\pi}\ket{\psi_{k=0,\pi}}=0$. The naive guess $\hat\eta'_{\pi}=\hat\eta_{\pi}$ fails, as one can notice by studying the explicit case $k=2,L=4$:
\begin{equation}\label{Eq:example}
\hat\eta_{\pi}\ket{\psi_{2,\pi}}=
\hat\eta_{\pi}\ket{\bullet \bullet \bullet \bullet}=
\ket{\bullet\circ\circ\bullet}-\ket{\circ\circ\bullet\bullet}-\ket{\bullet\bullet\circ\circ},
\end{equation}
where the filled dots indicate occupied sites, while empty dots denote empty sites. As one can infer from Eq.~\eqref{Eq:example}, the action of $\hat\eta_{\pi}$ on $\ket{\psi_{k,\pi}}$ generates configurations with unpaired fermions as soon as $k>1$, thus failing to reproduce the properties of a lowering operator inside the subspace $\mathcal{S}$.

In general, the explicit form of $\hat \eta'_\pi$ is rather complicated, we discuss here below for simplicity an expression that works if applied on states $\ket{\psi_{k,\pi}}$ for $k<L/3$, and that is non-local:
\begin{equation}\label{Eq:nonlocal_op}
\hat\eta'_{\pi} = \sum_{\ell=1}^{L-1}\frac{1}{\ell}\sum_{m=0}^{L-1}\frac{e^{2 \pi i \frac{m(\ell-\hat C)}L}}{L}\sum_{j=1}^{L-1}e^{-i\pi j} \hat P^{(j-1)} \hat c_{j+1}\hat c_j  .
\end{equation}
In Eq.~\eqref{Eq:nonlocal_op}, the operator $\hat C=\sum_{j=1}^{L-1}(1-\hat n_j)(1-\hat n_{j+1})$ counts the number of consecutive sites that are empty and
the sum over $m$ represents a Kronecker delta that selects the value of $\ell$ that is equal to the eigenvalue of the operator $\hat C$. 
The two operators $\hat c_{j+1}\hat c_j $ annihilate a pair at sites $j$ and $j+1$ and the projectors $\hat P^{(j-1)}$ check that the site $j$ is preceded by an even number of occupied sites and otherwise they annihilate the Fock state. Comparing with the sketch in Eq.~\eqref{Eq:example}, this term has the goal of avoiding that the unpaired configuration $\ket{\bullet\circ\circ\bullet}$ is generated from the initial state $\ket{\bullet \bullet \bullet \bullet}$.
It is possibly interesting to observe that there is a recursion relation obeyed by the projectors:
\begin{align}
&\hat P^{(0)}=1,\\
&\hat P^{(s)}=1-\hat n_s \hat P^{(s-1)}, \,\,\, 1\leq s \leq L .\nonumber
\end{align}

We claim that $\hat \eta'_{\pi}\ket{\psi_{k}}\propto \ket{\psi_{k-1,\pi}}$. Indeed, if one can show that each of the configurations contributing to the state $\ket{\psi_{k-1,\pi}}$ appears in the expression of the state $\hat \eta'_{\pi}\ket{\psi_{k}}$ with a unit coefficient (apart from overall normalization factors), then the proof is concluded. Consider any fully-paired Fock state $\ket{c}$ contributing to the state $\ket{\psi_{k-1,\pi}}$. The latter is generated whenever $\hat \eta'_{\pi}$ acts on a configuration contributing to $\ket{\psi_{k,\pi}}$ that can be obtained by adding a pair to the target configuration in $\ket{\psi_{k-1,\pi}}$. The number of such configurations with $k$ pairs equals the expectation value of $\hat C$ over the target configuration $\ket{c}$. Therefore, by dividing each contribution that results in $\ket{c}$ by the number of configurations in $\ket{\psi_{k,\pi}}$ that $\ket{c}$ can be reached by, one gets the desired result. This last operation is implemented by the operator expression that precedes the summation over the lattice sites in Eq.~\eqref{Eq:nonlocal_op}.

\subsection{Quasi-symmetries}

In this Section we draw connections with the quasi-symmetry picture of many-body scar subspaces~\cite{Ren_2021}. Since the spinless $\eta$-pairing states $\ket{\psi_{k,\pi}}$ that generate $\mathcal{S}$ are characterized by infinitely long-lived pair quasiparticles, we infer that the total number of pairs $\hat N_p$ is a conserved quantity under time evolution within $\mathcal{S}$. Its explicit form is once more nonlocal, and reads:
\begin{equation}
\hat N_p =\sum_{j=1}^{L-1} \hat{P}^{(j-1)}\hat n_j \hat n_{j+1}.
\end{equation} 
The quasi-symmetry property amounts then to the statement that:
\begin{equation}
\hat U_{\theta} \hat H \hat U_{\theta}^{\dag}|_{\mathcal{S}}= \hat H |_{\mathcal{S}},
\end{equation}
where $\hat U_{\theta}= e^{i\theta\hat N_p}$ is a unitary representation of $U(1)$. We conclude that the subspace $\mathcal{S}$ enjoys a nontrivial $U(1)$ quasi-symmetry linked to the infinite lifetime of the $\eta$-pairs. Similar considerations can be carried out in the case of the operator $\hat N_{stag}=\sum_{j=1}^{L} (-1)^j \hat n_j$, which gives rise to an additional $U(1)$ quasisymmetry of the subspace $S$ via the unitary representation $\hat U_{\varphi}= e^{i\varphi \hat N_{stag}}$, which enjoys the properties of being a tensor product representation over the Hilbert spaces attached to the lattice sites.

We underline that, according to the definition of quasi-symmetry of a degenerate subspace~\cite{Ren_2021}, the unitary representations are required to be tensor product representations over the Hilbert spaces attached to the lattice sites, in order to avoid including complicated transformations without a transparent physical meaning in the definition. While the latter condition is met by the unitary $\hat U_{\varphi}$, it is not satisfied by $\hat U_{\theta}$. However, given the clear physical meaning of the generator $\hat N_p$, we choose to include it in the discussion. 

\subsection{Dynamics and quantum coherence}

The consequences of Eq.~\eqref{Eq:eigv_eq_pi} on the dynamics of a generic superposition of the states $\ket{\psi_{k,\pi}}$ are easily computed; the existence of infinite revivals in the coherent dynamics that will be discussed below is a further consequence of ETH breaking. For an initial state of the form:
\begin{align}\label{Eq:superposition}
\ket{\psi(0)}=\sum_{k=0}^{L/2} m_{k}\ket{\psi_{k,\pi}}, \quad \sum_k |m_k|^2 =1,
\end{align}
the Loschmidt echo takes the form:
\begin{equation} \label{Eq:fidelity}
\mathcal L(t)=|\bra{\psi (0)}\ket{\psi (t)}|^2=\bigg |\sum_{k=0}^{L/2}|m_{k}|^2 e^{i\frac{2\mu k}{\hbar} t}\bigg |^2 ,
\end{equation}
and is periodic with period $T= \left|\pi\hbar/\mu \right|$.
Coherently with the interpretation of the states $\ket{\psi_{k,\pi}}$ as condensates of pairs, we find here that their time evolution is dictated by the chemical potential $\mu$~\cite{Stringari_2003}.

The time evolution of the expectation value of an operator is also easily computed. 
We take $\hat c_j \hat c_{j+1}$ as an example of operator
that has matrix elements between states whose number of pairs differs by one; assuming periodic boundary conditions, even $L$ and odd $L/2$ for simplicity, the dynamics of its expectation value reads:
\begin{align}\label{Eq:oscillating_corr}
&\langle\psi(t) |\hat c_j \hat c_{j+1}|\psi(t)\rangle =\\
&\quad (-1)^{j+1} e^{i\frac{2\mu t}{\hbar}}\sum_{k=0}^{L/2-2}  
\frac{ m^{*}_{k} m_{k+1}  \binom{L-2-k}{k}}{\sqrt{\frac{L}{L-k}\binom{L-k}{k} \frac{L}{L-k-1}\binom{L-k-1}{k+1}}},\nonumber
\end{align}
and exhibits a periodic oscillating behaviour. 
The complicated coefficients in terms of binomial appearing inside the summation take an easier expression when we consider the thermodynamic limit at fixed density $n = 2k/L$.
Noticeably, in agreement with Eq.~\eqref{Eq:ODLRO}, it is easy to show that:
\begin{equation}
\lim_{L \to \infty}\bra{\psi_{k,\pi}} \hat c_j \hat c_{j+1} \ket{\psi_{k+1,\pi}}=(-1)^{j+1}\sqrt{\frac{n}{2-n}}(1-n),
\end{equation}
which allows to conclude that the thermodynamic limit of the pair correlation function, $\lim_{L \to \infty} P_k (r)$, equals
\begin{equation}
\lim_{L \to \infty} \bra{\psi_{k+1,\pi}} \hat c_j^\dagger \hat c_{j+1}^\dagger\ket{\psi_{k,\pi}}\bra{\psi_{k,\pi}} \hat c_{j+r} \hat c_{j+r+1} \ket{\psi_{k+1,\pi}}
\end{equation}
when $r>3$. Once more, this expression certifies that the states $\ket{\psi_{k,\pi}}$ feature off-diagonal long-range order.

If the states $\ket{\psi_{k,\pi}}$ can be considered as many-body states with macroscopic coherence and fixed number of particles, states of the form~\eqref{Eq:superposition} can be used to discuss the macroscopic quantum coherence in the more usual grand-canonical ensemble.
We can introduce the $\alpha$ states
\begin{align}
\ket{\alpha} = \mathcal{N}_\alpha \, e^{\alpha \hat \eta_{\pi}^{\dag}}\ket{0}, \qquad \alpha \in \mathbb C
\end{align}
where $\mathcal{N}_\alpha$ is a normalization constant, and by applying the time-evolution operator on $\ket{\alpha}$, one can easily verify that they remain of the same form, and that the parameter $\alpha$ obeys the time-evolution relation:
\begin{equation}
 \alpha(t) = e^{i \frac{2 \mu}{\hbar} t} \, \alpha(0).
\end{equation}
It is tempting to interpret the $\alpha$ states as the coherent states of a quantum harmonic oscillator, but we stress that even if we assume infinite size, the $\hat \eta_\pi$ and $\hat \eta_{\pi}^\dagger$ do not satisfy the canonical commutation relation, and for instance  $\hat \eta_\pi \ket{\alpha} \neq \alpha \ket{\alpha}$.
This follows from the considerations presented above on the algebraic properties of the $\hat \eta_\pi^{(\dagger)}$.

The oscillatory behavior of the coherence parameter $\alpha$ demonstrates transparently that the state $\ket{\alpha}$ returns to itself after a period $T$ and naturally translates into periodic oscillations in the time evolution of suitably chosen local observables, as demonstrated more generally in Eq.~\eqref{Eq:oscillating_corr}. 

We probe the macroscopic coherence of the state $\ket{\alpha}$ by evaluating the expression in Eq.~\eqref{Eq:oscillating_corr} for the choice $\ket{\psi(t)}=\ket{\alpha(t)}$. The result takes the following form in the limit $L\rightarrow +\infty$ (see Appendix~\ref{App:Coherence}):
\begin{equation}\label{Eq:coherence}
\bra{\alpha(t)}e^{i\pi j}\hat c_{j+1} \hat c_{j}\ket{\alpha(t)}= \frac{2\, \alpha(t)}{\sqrt{1+4|\alpha|^2}\left(1+\sqrt{1+4|\alpha|^2}\right)}.
\end{equation}
While for small values of $\alpha$ the result reproduces the value obtained for the coherent state obtained from a single bosonic mode, the term in the denominator of Eq.~\eqref{Eq:coherence} corrects the result for larger values of $\alpha$ and arises from the hard-core nature of the pairs that populate the system.

\subsection{Entanglement}

In this subsection we compute the scaling of the entanglement entropy of the $\ket{\psi_{k,\pi}}$ for a bipartition of the system into two halves of length $L/2$.
We show that they are entanglement outliers, as they display a logarithmic scaling of the half-chain entanglement entropy; as we have already mentioned, the scaling is not the typical scaling of a thermal state and thus is a consequence of ETH breaking. 
To this end, we consider the density matrix $\rho_{k,\pi}=\ket{\psi_{k,\pi}}\hspace{-0.1cm} \bra{\psi_{k,\pi}}$ on a system with $L$ sites and we aim to compute the reduced density matrix for the first $\frac{L}{2}$ sites, i.e., $\rho^{\left(\frac{L}{2}\right)}_{k,\pi}=\text{Tr}_{\left[\frac{L}{2}+1,\dots,L \right]}(\rho_{k,\pi})$. 
We choose for simplicity the second Renyi entropy, which is defined as follows:
\begin{equation}\label{Eq:S_vN}
S_{\frac{L}{2}, k, \pi}= -\log\left\{\text{Tr}\left[ \left(\rho^{\left (\frac{L}{2} \right) }_{k,\pi} \right)^2\right]\right\}.
\end{equation}
An analytical generic formula for any $k$ can be obtained in terms of binomial coefficients, and reads:
\begin{widetext}
\begin{equation} \label{Eq:Renyi_entropy}
 S_{\frac L2,k , \pi} = -\log\left\{
 \sum_{l=\max\{0,\lceil\frac{N-L/2}{2}\rceil \}}^{\min\{\frac{N}{2},\lfloor\frac{L/2}{2}\rfloor \}}
 \left( \frac{\binom{\frac{L}{2}-\frac{N}{2}+l}{\frac{N}{2}-l}\binom{\frac{L}{2}-l}{l}}{\binom{L-\frac{N}{2}}{\frac{N}{2}}} \right)^2 + 
 \sum_{l=\max\{0,\lceil\frac{N-L/2-1}{2}\rceil \}}^{\min\{\frac{N}{2}-1,\lfloor\frac{L/2-1}{2}\rfloor \}} \left(
 \frac{\binom{\frac{L}{2}-\frac{N}{2}+l}{\frac{N}{2}-l-1}\binom{\frac{L}{2}-1-l}{l}}{\binom{L-\frac{N}{2}}{\frac{N}{2}}} \right)^2\right\}.
\end{equation}
\end{widetext}
A more readable analytical expression can be found taking the thermodynamic limit $L \to \infty$ and $k \to \infty$ and fixing the ratio $2k/L = n = 2/3$. In this limit, in Appendix~\ref{App:Entanglement} we show that the formula is well approximated by:
\begin{equation} \label{Eq:asymptotics_entropy}
 S_{\frac{L}{2}, k=\frac{L}{3}, \pi} \xrightarrow{L \to \infty} \frac{1}{2}\log\left(\frac{6\pi L}{25}\right). 
\end{equation}

\begin{figure*}

\includegraphics[width=\textwidth]{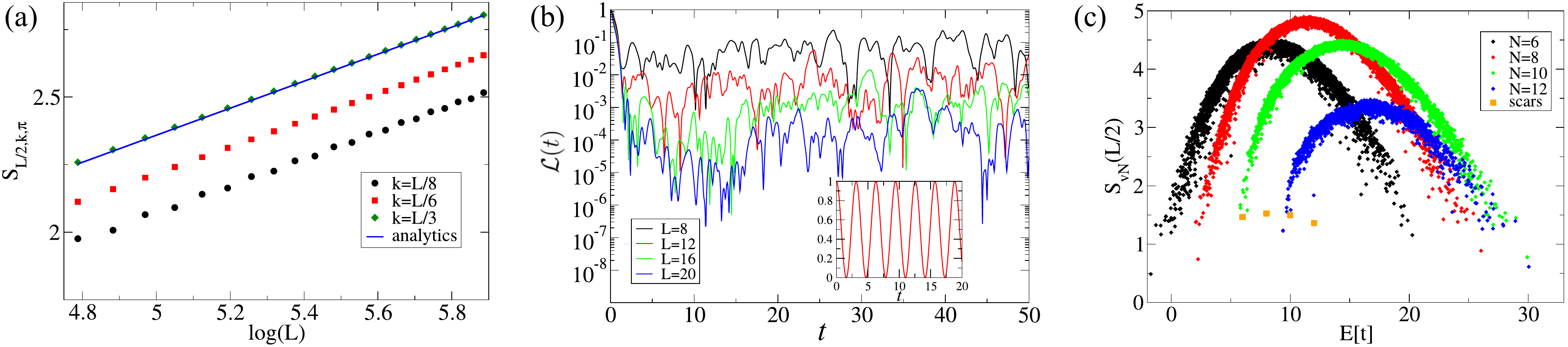}
\caption{(a) Scaling of the half-chain entanglement entropy of the states $\ket{\psi_{k,\pi}}$ for fillings $n=2k/L=1/4$, $1/3$, $2/3$ according to Eq.~\eqref{Eq:Renyi_entropy}. The blue line is the asymptotic expression~\eqref{Eq:asymptotics_entropy}.
(b) Loschmidt echo $\mathcal{L}(t)$ as a function of time $t$ for the Hamiltonian parameters $t=1$, $J=2$, $\mu=-1$, starting from a product state $\ket{\psi_2}$ (see text). Inset: same for a superposition of scarred eigenstates $\ket{\psi_1}$ (see text).
(c) Half-chain entanglement entropy of the eigenstates of Hamiltonian~\eqref{Eq:hamiltonian} with $t=J=1$, $\mu=-1$ for $L=16$ and  $N=6,8,10,12$ particles.
}
\label{Fig:entanglement_scaling}
\label{Fig:fidelity}
\label{Fig:Sent}
\end{figure*}

For more clarity, we have evaluated the resulting entanglement entropy scaling law in Fig.~\ref{Fig:entanglement_scaling}(a), where the analytical prediction in Eq.~\eqref{Eq:Renyi_entropy} is plotted as a function of $\log L$ for several choices of the system filling, i.e., of the number of pairs. The figure confirms the agreement with the scaling for $n=2/3$ in Eq.~\eqref{Eq:asymptotics_entropy} and demonstrates a scaling as $\log L$ for other filling choices. 
As already mentioned, a logarithmic scaling of the entanglement entropy signals a non-ETH state, and shows the exceptional character of the $\ket{\psi_{k,\pi}}$.

\subsection{Numerical analysis}\label{S3}

We proceed by providing numerical benchmarks of the scarred eigenstates discussed in the previous sections by performing exact diagonalization simulations with the QuSpin package~\cite{Quspin_1,Quspin_2}. We start by presenting the behaviour of the Loschmidt echo $\mathcal L(t)$ when the system is initialized either in the superposition of scarred eigenstates $\ket{\psi_1}=\frac{1}{\sqrt{2}}\left(\ket{\psi_{1,\pi}}+\ket{\psi_{2,\pi}}\right)$ or in the generic product states $\ket{\psi_2}=\prod_{j=1}^{L/4}\hat c^{\dag}_{2j}\ket{0}$. The data presented in Fig.~\ref{Fig:fidelity}(Center) show that, while the superposition of scarred eigenstates shows exact revivals, as predicted exactly via Eq.~\eqref{Eq:fidelity}, the coherent dynamics of a generic product state displays a phenomenology that is consistent with the loss of memory of the initial state, as generically expected for a thermalizing isolated many-body quantum system. The data confirm therefore that the revivals associated to the existence of an exact tower of states embedded in the spectrum is atypical and not observed when the dynamics of a generic intial state is monitored.

A further check is provided by plotting the half-chain entanglement entropy of a system described by Hamiltonian~\eqref{Eq:hamiltonian} on a lattice of size $L=16$ for $N=6,8,10,12$. The points highlighted in orange in Fig.~\ref{Fig:Sent}(c), which refer to the half-chain entanglement entropy of the scarred eigenstates $\{\ket{\psi_{k,\pi}}\}_{k=3}^{6}$, point towards the anomalously low entanglement of scarred eigenstates in comparison to other generic excited states. This finding is consistent with the non-thermal nature of the unveiled quantum many-body scars and confirms their nature of exceptional states embedded in an otherwise ETH-satisfying spectrum.

\begin{figure*}[t]
 \includegraphics[width=\textwidth]{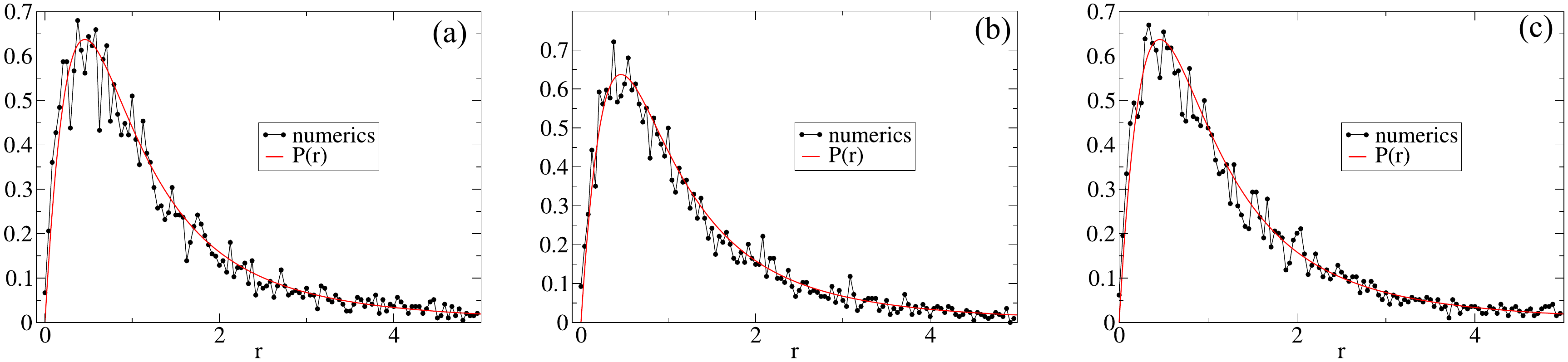}
 \caption{(a) Level statistics probe $P(r)$ in the bulk of the spectrum of~\eqref{Eq:hamiltonian} with OBC with $t=J=1$, $\mu=-1$ for size $L=18$ and $N=6$, in the inversion symmetry sector $I=-1$ (black dots) compared to the GOE prediction (red solid line).
 (b) Same plot for the Hamiltonian $\hat H_3$ in Eq.~\eqref{Eq:Hamiltonian:H3} with the same parameters.
 (c) Same plot for $\hat H_3$ but setting $J=0$ and keeping other
 parameters identical.}
\label{Fig:LSS}
 \label{Fig:LSS_trimers}
 \label{Fig:LSS_J_0_trimers}
 \end{figure*}

We conclude this section by demonstrating the nonintegrability of Hamiltonian~\eqref{Eq:hamiltonian} by means of the study of level-spacing statistics. More specifically, we compute the probability density function of the ratio of consecutive level spacings $r_n=s_n/s_{n-1}$~\cite{Oganesyan_2007, Atas_2013}, where $s_n=E_{n+1}-E_n$ is the difference between two consecutive energy levels $E_n$ and $E_{n+1}$ in the spectrum. The comparison between the numerical data and the Wigner-Dyson probability distribution for the GOE ensemble provided in Fig.~\ref{Fig:LSS}(a) shows a neat quantitative agreement. Moreover, the average of the level-spacing ratio $\tilde{r}_n=\min(s_n,s_{n-1})/\max(s_n,s_{n-1})$ obtained from the numerical data equals $0.52822\dots$, is perfectly compatible with the theoretical value $\langle \tilde{r}\rangle=0.53590\dots$. We are thus able to conclude that the model is not integrable.

\subsection{Relation with previous work}

Recently, an article has presented results on a class of spin models that are connected to ours~\cite{Shibata_2020}. One of the models considered (in the notation of the article, it corresponds to $n=2$) is 
\begin{equation}
 H = 
 \sum_{j=1}^{L} \left(S_j^+ S_{j+1}^- + H.c. \right) + 
 h \sum_{j=1}^L S_j^z + H_{pert,2},
\end{equation}
where $S^\alpha_j$ are spin-1/2 operators. 
The Hamiltonian $H_{pert,2}$ is parametrised by three sets of coefficients $c^{(i)}_j$, with $i=1,2,3$ and $j = 1,2 \ldots L$; if we take $c_j^{(1)} = c_j^{(3)}=0$ and $c_j^{(2)}=2J$, we obtain:
\begin{equation}
 H_{pert,2} = J \sum_{j=1}^L \left( \ket{011} + \ket{110} \right) \left( \bra{011} + \bra{110} \right).
\end{equation}
If we apply the Jordan-Wigner transformation to this model, we obtain the model written in Eq.~\eqref{Eq:hamiltonian} upon setting $t=1$ and $h = -\mu$.
Hence, Hamiltonian in Eq.~\eqref{Eq:hamiltonian} maps to a specific instance of the model discussed in Ref.~\cite{Shibata_2020}.
Our work, presented directly in the fermionic form, puts more emphasis on pairing nature of the model and of the scars, and lends itself to multimer generalisations, as it is discussed in the next section. Our point of view and complementary results shed new lights on possible route for identifying scar models.

\section{Scars based on multimers in a spinless-fermion model} 
\label{Sec:Multimers}
In the following, we generalize our findings on many-body scars with $\eta$-pairing to the case of multimers of arbitrary size $M \geq 2$ by constructing a family of Hamiltonians $\hat H_M$, such that states composed of many $\eta$-multimers of size $M$ are exact many-body scars of the Hamiltonian $\hat H_M$. 

The Hamiltonian is the sum of three terms: $\hat H_M = \hat H_t + \hat H_\mu + \hat H_J$ and their explicit forms in open boundary conditions read:
\begin{subequations}
\begin{align}\label{Eq:scars_M}
\hat H_t =& -t  \sum_{j=1}^{L-1} \left[\hat c^{\dag}_j \hat c_{j+1}+H.c. \right]-t \hspace{-0.25cm} \sum_{j=1}^{L-M+1}\left[\hat M^{\prime \dag}_{j+1} \hat M_{j}^\prime+H.c.\right]; \\
\hat H_\mu =& -\mu\sum_{j=1}^L \hat n_j; \\
\hat H_J =&J\sum_{j=1}^{L-M}\left(\prod_{l=0}^{M-1}\hat n_{j+l}+\prod_{l=1}^{M}\hat n_{j+l}-2\prod_{l=0}^{M}\hat n_{j+l} \right. \nonumber \\ & \qquad \qquad \left.+ \left[\hat M^{\dag}_j \hat M_{j+1} +H.c.\right]\right).
\end{align}
\end{subequations}
In the Hamiltonian $\hat H_t$ we recognize a single-particle and a $M-1$-particle hopping term, with 
\begin{equation}
\hat M^{\prime\dag}_j=\prod_{l=0}^{M-2}\hat c^{\dag}_{j+l}=\hat c^{\dag}_j\dots \hat c^{\dag}_{j+M-2},
\end{equation}
whereas $\hat H_\mu$ is just a chemical potential term.
The Hamiltonian $\hat H_J$ is a frustration-free positive Hamiltonian of the form $J\sum_j \hat L^{\dag (M)}_j\hat L_j^{(M)}$, comprising both density-density interactions and a $M$-particle hopping term:
\begin{equation}
\hat M^{\dag}_j=\prod_{l=0}^{M-1}\hat c^{\dag}_{j+l}=\hat c^{\dag}_j\dots \hat c^{\dag}_{j+M-1}.
\end{equation}
The explicit expression of $\hat L^{(M)}_j$, that is inessential for this discussion, is given in Appendix~\ref{App:Multimer}.
For completeness, we mention that multimer Hamiltonians and variants thereof have been discussed in Refs.~\cite{Mazza_2018, Gotta_2022}.

The Hamiltonian $\hat H_M$ is the generalisation of the Hamiltonian $\hat H$ introduced in Eq.~\eqref{Eq:hamiltonian} and it reduces exactly to it for $M=2$.
A particularly interesting property is the fact that for $M \geq 3$ the simpler Hamiltonian $\hat H_t+ \hat H_\mu$ is already not integrable; for $M=2$ this would not be true as it would be a free-fermion model.
We verify this first statement with a numerical analysis for the case $M=3$, and for reading convenience we write here the explicit Hamiltonian:
\begin{widetext}
\begin{align}
\hat H_3 =& -t\sum_{j=1}^{L-1}\left(\hat c^{\dag}_j \hat c_{j+1}+H.c.\right)-\mu\sum_{j=1}^L \hat n_j+t\sum_{j=1}^{L-2}\left(\hat c^{\dag}_{j+2} \hat n_{j+1}\hat c_j +H.c. \right) \nonumber\\
&+J\sum_{j=1}^{L-3}\left[\hat n_j \hat n_{j+1} \hat n_{j+2} +\hat n_{j+1}\hat n_{j+2}\hat n_{j+3}-2\hat n_j \hat n_{j+1}\hat n_{j+2}\hat n_{j+3}+\left(\hat c^{\dag}_j \hat n_{j+1}\hat n_{j+2}\hat c_{j+3} +H.c. \right) \right].
\label{Eq:Hamiltonian:H3}
\end{align}
\end{widetext}
First of all, we  study its level-spacing statistics, as in the case of $\eta$-pairs. The results shown in Fig.~\ref{Fig:LSS_trimers}(b) and (c) showcase the agreement between the Wigner-Dyson distribution and the numerical data. For $J \neq 0$, the value of $\langle \tilde{r}\rangle$ is computed from the numerics to be $0.5296\ldots$, which is compatible with its theoretical prediction.
For $J=0$, the average of $\langle \tilde{r}\rangle$ computed numerically takes the value $0.5326\ldots$. We conclude that the deformation of the Hamiltonian proportional to $J$ is not necessary to make the model nonintegrable when working with $M=3$ and we expect this to be true in general for $M>2$.

\subsection{Exact results on scar states}\label{Subsec:Definition:Scar} 

We assume open boundary conditions and introduce the  towers of states $\ket*{\psi_{k,\pi}^{(M)}}$ that generalise those presented in Eq.~\eqref{Eq:eta_pairs} to multimers. 
The goal of this section is to define and characterise the properties of these states; in Sec.~\ref{Sec:Numerics:Trimer} we will present numerical evidence that they are scar states, and an analytical proof of that will follow in Sec.~\ref{Subsection:proof_multimers}.

For $M$ even, the states are defined as condensates of multimers at $k=\pi$:
\begin{equation}
\ket*{\psi_{k,\pi}^{(M)}}=\frac{\left(\hat\eta^{ (M) \dag}_{\pi}\right)^k}{k!\sqrt{\binom{L-(M-1)k}{k}}}\ket{\emptyset}, \quad \hat\eta^{ (M) \dag}_{\pi}=\sum_{j}e^{i\pi j}\hat M_j^\dagger.
\end{equation}
The physical properties that the states $\ket*{\psi_{k,\pi}^{(M)}}$ display are an extension of the ones unveiled in the case of pairs. More precisely, these states display long-range order signaled by a nonvanishing value of the large distance behavior of the multimer correlator $\bra*{\psi_{k,\pi}^{(M)}}\hat M^{\dag}_j \hat M_{j+r}\ket*{\psi_{k,\pi}^{(M)}}$. Such a feature is a consequence of the statistics of the underlying quasiparticles: indeed, since the multimer creation operators $\hat M^{\dag}_j$ commute for even $M$ when they have nonintersecting support, the even-sized multimers in such states behave as a condensate of hard-core bosons.

On the other hand, for $M$ odd, the multimers are fermionic, and the notion of condensation cannot be applied. In this case the  states $\ket*{\psi_{k,\pi}^{(M)}}$ are the equal-amplitude superposition of all multimer states weighted by a phase factor. To make this precise, we need to define what is a multimer state, namely a real-space Fock state of the spinless fermions where fermions bunch in groups whose length is a multiple of $M$. For $M=3$, for instance, 
$\ket{ \circ \bullet \bullet \bullet \circ \circ \circ \circ \circ}$ and $\ket{ \circ \bullet \bullet \bullet \bullet \bullet \bullet \circ \circ}$ are multimer states, whereas 
$\ket{ \circ \bullet \bullet \bullet \bullet \circ \circ \circ \circ}$ is not. To each multimer state we can associate the phase factor $e^{i \pi \sum_j m_j}$, where $m_j$ is the site where the first fermion of the $j$-th multimer is located. For the two multimer states given above, in the first case $m_1=2$, and in the second case $m_1=2$ and $m_2=5$. The linear superposition of all multimer states multiplied by the given phase factors defines the exact scars; these states generalise the properties of the bosonic states defined above.

\begin{figure}
\centering
\includegraphics[width =\columnwidth]{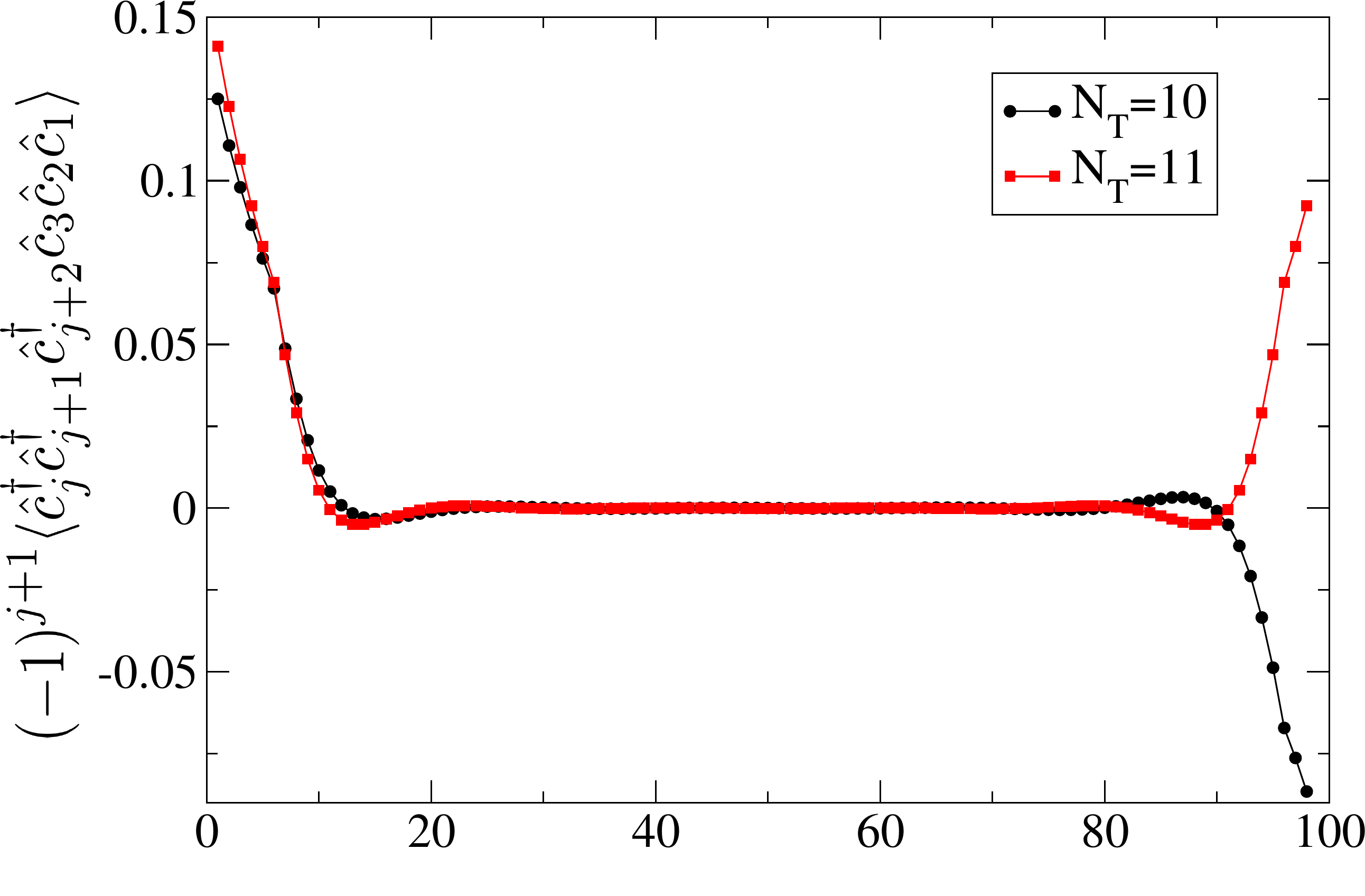}
\caption{Trimer correlation function $(-1)^{j+1}\langle\psi_{k,\pi}^{(3)}|\hat c^{\dag}_j \hat c^{\dag}_{j+1}\hat c^{\dag}_{j+2}\hat c_3 \hat c_2 \hat c_1|\psi_{k,\pi}^{(3)}\rangle$ on a lattice of size $L=100$ for $k=N_T=10,11$ trimers.}
\label{Fig:trimer_corr}
\end{figure}

It is interesting to observe that in the fermionic case the tower of states are not generated by a local raising operator, as it happens in the hard-core bosonic case. This observation can be traced back to the fermionic anticommutation relation $\{\hat M^{\dag}_j , \hat M^{\dag}_l\}=0$ when $|j-l|\geq M$, which in turn implies that the naive guess $\hat\eta^{ (M) \dag}_{\pi}=\sum_{j}e^{i\pi j}\hat M_j^\dagger$ fails because the latter operator squares to zero. The correct form of the raising operator has the following structure, generalizing the Jordan-Wigner string:
\begin{equation}
\hat\eta^{ (M) \dag}_{\pi}=\sum_{j}e^{i\pi j}e^{i\pi\hat N_{j}^{(M)}}\hat M_j^\dagger,
\end{equation}
where $\hat N_{j}^{(M)}$ is a nonlocal operator that outputs the number of multimers of size $M$ that are present before site $j$. A construction of such an operator is presented in Subsection~\ref{Subsection:quasisymmetries}.

Moreover, the correlation properties of states $\ket*{\psi_{k,\pi}^{(M)}}$ with odd $M$ display edge properties that depend on the total parity associated to the number of  multimers. For instance, the edge-to-edge correlation function reads:
\begin{align}
\bra*{\psi_{k,\pi}^{(M)}}\hat c^{\dag}_{L-2}\hat c^{\dag}_{L-1}\hat c^{\dag}_L \hat c_3 \hat c_2 \hat c_1\ket*{\psi_{k,\pi}^{(M)}}
= (-1)^{L+k} \frac{\binom{L-4-2k}{k-1}}{\binom{L-2k}{k}},
\end{align}
provided $k\leq 1+\lfloor(L-6)/3\rfloor$, and vanishes otherwise. More generally, we compute the trimer correlation function on a system with OBC (the explicit expression is provided in Appendix~\ref{App:trimer_corr}) and present it in Fig.~\ref{Fig:trimer_corr}: the result shows a revival of correlations at the right edge of the system, whose sign depends on the parity of the total number of trimers in the system. Such a feature is reminiscent of the behavior of the single-particle correlator $\langle \hat c^{\dag}_j \hat c_1\rangle$ for the ground states of the Kitaev chain;
it hints at the fact that the equal-weight superposition of the states $\ket*{\psi_{k,\pi}^{(3)}}$ with fixed parity of the number of trimers, i.e., with $k=2n$ or $k=2n+1$, could possess the nontrivial topological properties of the ground state of the Kitaev chain in its topological phase. These considerations are a consequence of the fermionic nature of the quasiparticles associated to the tower of scarred eigenstates with an odd value of $M$.

\subsection{Numerical signatures of the trimer scars}\label{Sec:Numerics:Trimer}

In order to verify the existence of multimer scars, we focus on the specific case $M=3$ and we employ numerical tools: we will show the existence of atypical eigenstates with energies $E_k =-M\mu k$, $k=0,\dots,\lfloor L/M\rfloor$. 

\begin{figure}
\centering
\includegraphics[width =\columnwidth]{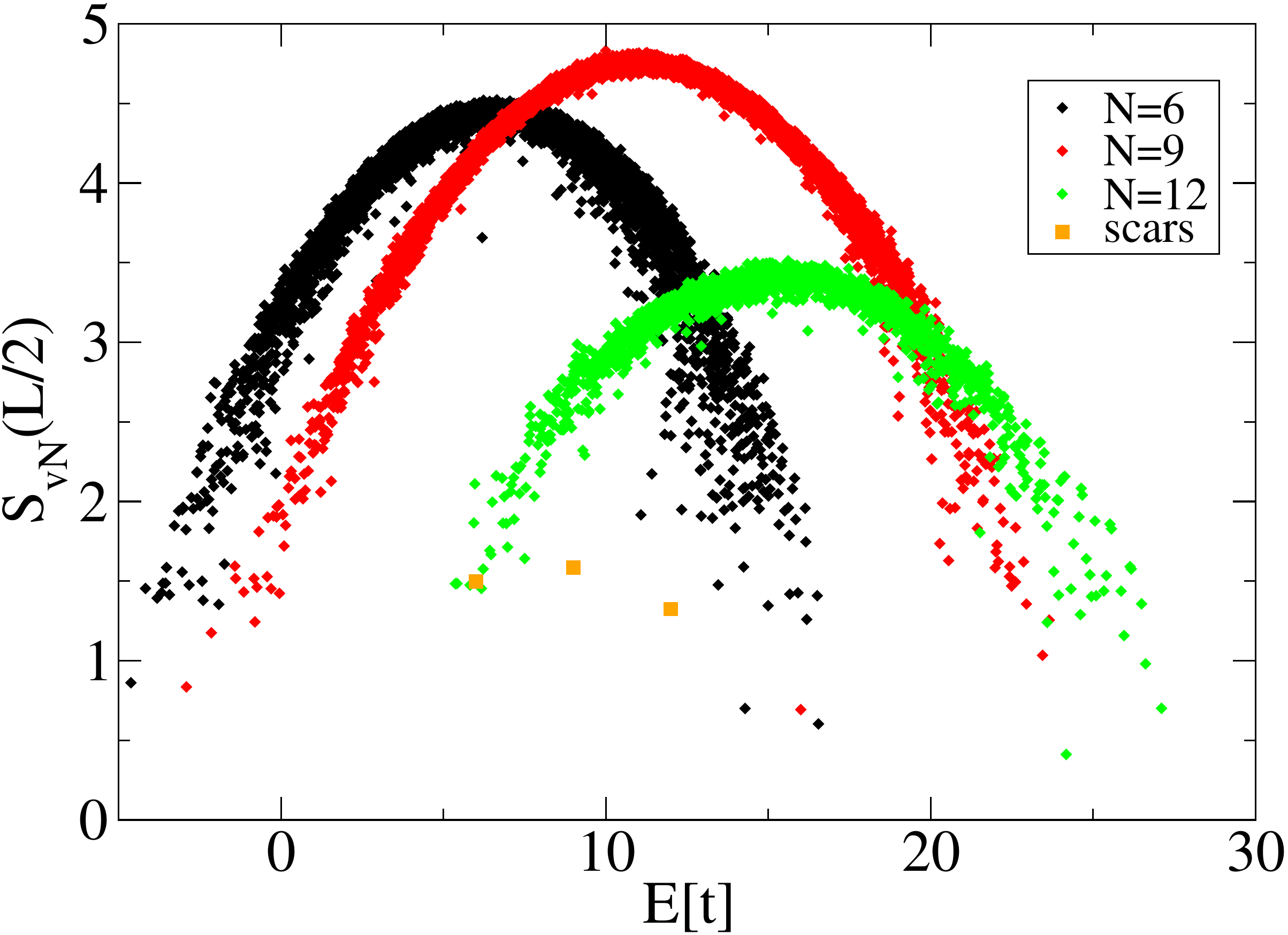}
\caption{Half-chain von Neumann entanglement entropy of the eigenstates of the Hamiltonian $\hat H_3$ with $t = J = 1$, $\mu = -1$ on a lattice of size $L = 16$  in the sectors with $N = 6, 9, 12$ particles.}
\label{Fig:entropy_trimers}
\end{figure}

\begin{figure}
\centering
\includegraphics[width =\columnwidth]{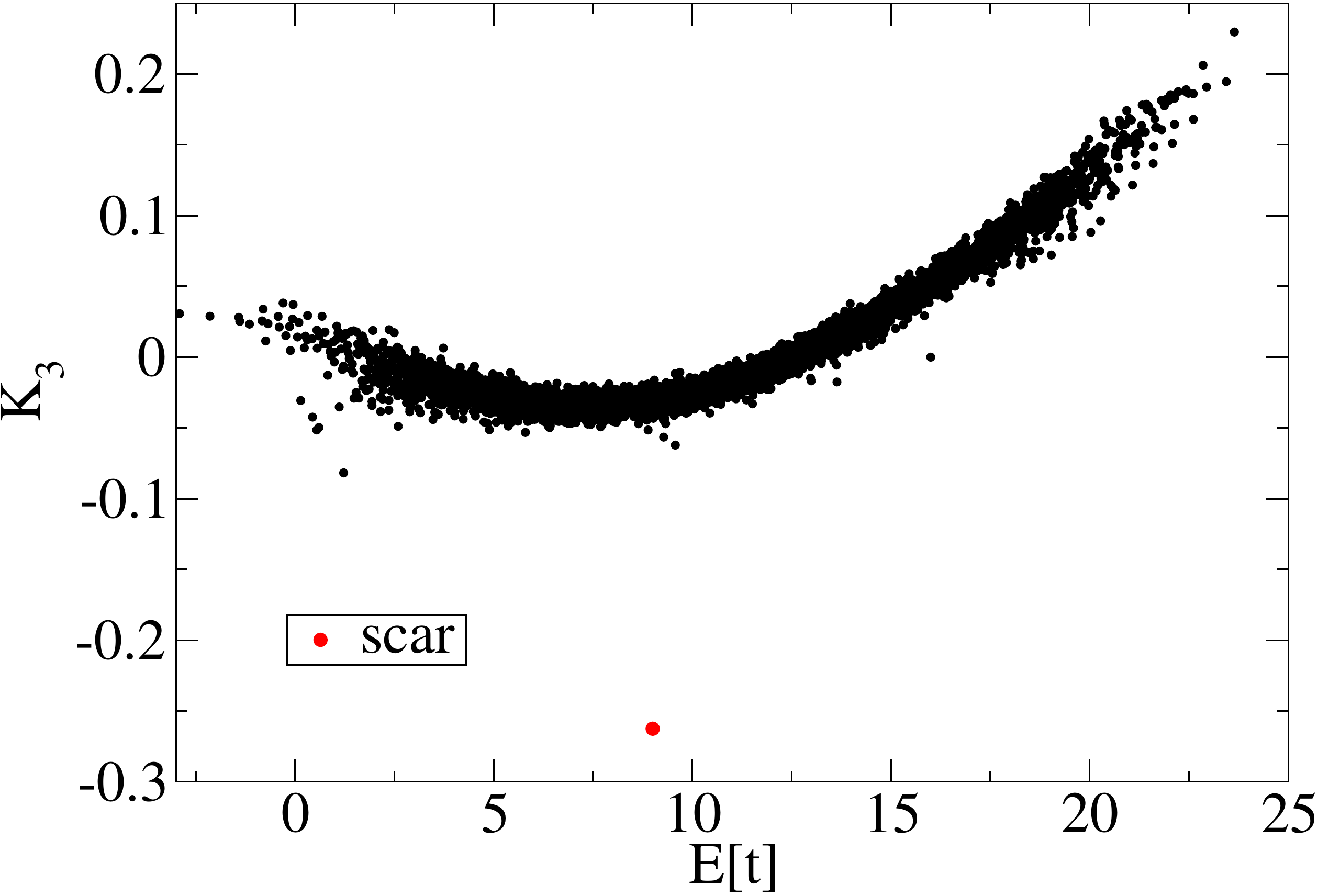}
\caption{Expectation value $K_3$ over the eigenstates of Hamiltonian $\hat H_3$ for $t=J=1$, $\mu=-1$ for $L=16$ and $N=9$.}
\label{Fig:density}
\end{figure}

We test the half-chain von Neumann entanglement entropy in Fig. \ref{Fig:entropy_trimers}, where the atypical eigenstates belonging to the tower of states defined in Eq.~\eqref{Eq:scars_M} with $M=3$ emerge as entanglement outliers with respect to the typical behavior observed for generic highly-excited eigenstates.
We also show in Fig.~\ref{Fig:density} the expectation value of a generic local observable related to the relevant quasiparticles, in this case the total trimer-hopping energy density 
\begin{equation}
K_3=\frac{1}{L}\sum_{j=1}^{L-3}\langle \hat c^{\dag}_{j}\hat n_{j+1} \hat n_{j+2}\hat c_{j+3}+\text{H.c.}\rangle.
\end{equation}
The plot shows that $K_3$
takes an anomalous value when evaluated over the scarred eigenstate $\ket*{\psi_{3,\pi}^{(3)}}$, whereas it behaves as smooth function of energy for the other eigenstates, thereby supporting  the validity of the weak ETH-breaking associated to the scars that we are presenting.

\subsection{Proof of the exact multimer scar states}\label{Subsection:proof_multimers}

In this section we show that the $\ket*{\psi_{k,\pi}^{(M)}}$ introduced in Sec.~\ref{Subsec:Definition:Scar} are scar states with energies $E_k$.
We recall that they are defined as the normalized equal-amplitude superposition of all Fock configurations obtained by distributing $k$ trimers over the system, each multiplied by the phase factor $e^{i\pi\sum_{l=1}^k j_l}$, where $j_l$ is the position of the first fermion in the $l^{th}$ multimer. The Hamiltonian $\hat H_t$ annihilates the states $\ket*{\psi_{k,\pi}^{(M)}}$ by means of the destructive interference among single-particle hopping and multimer hopping. Similarly, it is easy to see that $\hat H_{\mu}\ket*{\psi_{k,\pi}^{(M)}}=-M\mu k\ket*{\psi_{k,\pi}^{(M)}}$. Hence, we focus on the problem of showing that the states $\ket*{\psi_{k,\pi}^{(M)}}$ are annihilated by $\hat H_J$. 

By performing the transformation $\hat c_j\rightarrow e^{-i\frac{\pi}{M}j}\hat c_j$ via the unitary operator $\hat U_M =\prod_{j=1}^{L}e^{i\frac{\pi}{M} j \hat n_j}$ to remove the $\eta$-pairing phase factor from each term in the superposition that defines $\ket*{\psi_{k,\pi}^{(M)}}$, we need to prove that the states $\ket*{\psi_{k,0}^{(M)}}$ are eigenstates of $\hat U_M\hat H_J\hat U^{\dag}_M=J\sum_{j=1}^{L-M}\hat H_j$, where:
\begin{align}
\hat H_j = \prod_{l=0}^{M-1}\hat n_j +\prod_{l=1}^M \hat n_{j+l} -2\prod_{l=0}^M \hat n_{j+l}-(\hat M^{\dag}_j \hat M_{j+1}+h.c.).
\end{align}   

\begin{widetext}
We show now that $\hat H_j \ket*{\psi_{k,0}^{(M)}}=0$ for $j=1,\dots, L-M$. Let us denote by $\mathcal{C}_M^{(k)}$ the set of all Fock space configurations obtained by distributing $k$ multimers of size $M$ over the lattice, each multiplied by the normalization factor of the state $\ket*{\psi_{k,0}^{(M)}}$. In other words, $\mathcal{C}_M^{(k)}$ is the collection of all Fock position basis states that contribute to $\ket*{\psi_{k,0}^{(M)}}$. Let us further introduce the following subsets of $\mathcal{C}_M^{(k)}$: 
\begin{align}\label{Eq:annihilation}
&\mathcal{S}_{j_1,j_2}=\{ \ket{\psi}\in\mathcal{C}_M^{(k)}: \bra{\psi}\hat n_l\ket{\psi}=1,\, j_1\leq l \leq j_2\},\\
&\mathcal{S}_{\alpha; j_1,j_2}=\{ \ket{\psi}\in\mathcal{C}_M^{(k)}:\bra{\psi}\hat n_{j_1-1}\ket{\psi}=\alpha,\, \bra{\psi}\hat n_l\ket{\psi}=1,\, j_1\leq l \leq j_2\},\\
&\mathcal{S}_{j_1,j_2;\alpha}=\{ \ket{\psi}\in\mathcal{C}_M^{(k)}: \bra{\psi}\hat n_{j_2+1}\ket{\psi}=\alpha,\, \bra{\psi}\hat n_l\ket{\psi}=1,\, j_1\leq l \leq j_2\},
\end{align}
with $\alpha=0$. 
Then, one can write:
\begin{align}\label{Eq:annihilation}
\hat H_j \ket*{\psi_{k,0}^{(M)}}=\sum_{\ket{\psi}\in\mathcal{S}_{j,j+M-1}} \ket{\psi}+\sum_{\ket{\psi}\in\mathcal{S}_{j+1,j+M}} \ket{\psi}-2\!\!\!\sum_{\ket{\psi}\in\mathcal{S}_{j,j+M}} \ket{\psi}-\!\!\!\sum_{\ket{\psi}\in\mathcal{S}_{0;j+1,j+M}} \ket*{\psi^{(L)}}-\!\!\!\sum_{\ket{\psi}\in\mathcal{S}_{j,j+M-1;0}} \ket*{\psi^{(R)}},
\end{align}
where $\ket*{\psi^{(L)}}$ is obtained from the corresponding state $\ket{\psi}\in\mathcal{S}_{0;j+1,j+M}$ by moving the multimer placed on sites $j+1,\dots,j+M$ to the sites $j,\dots,j+M-1$ and leaving the site $j+M$ empty. As a result, $\ket*{\psi^{(L)}}\in \mathcal{S}_{j,j+M-1;0}$. Similarly, the state $\ket*{\psi^{(R)}}$ is obtained from the corresponding state $\ket{\psi}\in\mathcal{S}_{j,j+M-1;0}$ by moving the multimer placed on sites $j,\dots,j+M-1$ to the sites $j+1,\dots,j+M$ and leaving the site $j$ empty. Thus, $\ket*{\psi^{(R)}}\in \mathcal{S}_{0;j+1,j+M}$. Noticing that:
\begin{equation}
\sum_{\ket{\psi}\in\mathcal{S}_{0;j+1,j+M}} \ket*{\psi^{(L)}}=\sum_{\ket{\psi}\in\mathcal{S}_{j,j+M-1;0}} \ket{\psi},\qquad \qquad
\sum_{\ket{\psi}\in\mathcal{S}_{j,j+M-1;0}} \ket*{\psi^{(R)}}=\sum_{\ket{\psi}\in\mathcal{S}_{0;j+1,j+M}} \ket{\psi},
\end{equation}
and rewriting the first two terms in Eq.~\eqref{Eq:annihilation} by making explicit reference to the occupation of site $j+M$ and site $j$, respectively, one obtains:
\begin{align}
\hat H_j \ket*{\psi_{k,0}^{(M)}}=&\sum_{\ket{\psi}\in\mathcal{S}_{j,j+M-1,0}} \ket{\psi}+\sum_{\ket{\psi}\in\mathcal{S}_{j,j+M}} \ket{\psi}+\sum_{\ket{\psi}\in\mathcal{S}_{0,j+1,j+M}} \ket{\psi}\nonumber\\
&\quad+\sum_{\ket{\psi}\in\mathcal{S}_{j,j+M}} \ket{\psi}-2\sum_{\ket{\psi}\in\mathcal{S}_{j,j+M}} \ket{\psi}-\sum_{\ket{\psi}\in\mathcal{S}_{j,j+M-1;0}}\ket{\psi}-\sum_{\ket{\psi}\in\mathcal{S}_{0;j+1,j+M}} \ket{\psi}=0.
\end{align}
The result is therefore proven.
\end{widetext}

\subsection{Quasi-symmetries}\label{Subsection:quasisymmetries}

We extend to the general multimer case the observation of symmetry enhancement in the scarred subspace spanned by the states $\ket*{\psi_{k,\pi}^{(M)}}$ by constructing examples of observables conserved by the dynamics within the scarred subspace in analogy with the analysis carried out in the case of pairs. Firstly, the number of multimers of size $M$, $\hat N_M$, commutes with the Hamiltonian when their action is restricted to the scarred subspace. The latter is formally defined via the following nonlocal operator:
\begin{align}
\hat N_M=\sum_{j=1}^{L-M+1} \hat P_{M}^{(j-1)}\prod_{l=0}^{M-1}\hat n_{j+l},
\end{align}
where $\hat P_{T}^{(j-1)}$ is a projection operator that does not vanish if and only if the site $j$ is preceded by a number of occupied sites that is a multiple of $M$, i.e., if it is preceded by multimers of size $M$ (including the case of no particle on site $j-1$). It is defined recursively by setting:
\begin{align}
& \hat P_{M}^{(s)} = 1,\qquad 0\leq s\leq M-1,\\
&  \hat P_{M}^{(s)} = 1-\hat n_s + \prod_{l=0}^{M-1}\hat n_{s-M+1+l} \hat P_{M}^{(s-M)}, \;\; s \geq M.
\end{align}
The enhanced symmetry properties of the scarred subspace are thus expressed mathematically as:
\begin{equation}
\hat U_{\hat N_M,\phi} \hat H \hat U_{\hat N_M}^{\dag}|_{\mathcal{S}_M}=\hat H,
\end{equation}
where $\mathcal{S}_M$ is the subspace spanned by the scar states based on $\eta$-multimers of size $M$ and $\hat U_{\hat N_M,\phi}=e^{i\phi \hat N_M}$ is the associated unitary representation of $U(1)$.

Similarly, the quantity:
\begin{equation}
\hat N_{stag,M}=\sum_j e^{i\frac{2\pi}{M}j}\hat n_j
\end{equation}
vanishes when evaluated on the scar states. Thus, the unitary representation $\hat U_{stag,M,\phi}=e^{i\phi \hat N_{stag,M}}$ of $U(1)$ leaves the scarred eigenspace invariant and further enjoys the property of being a tensor product representation over all lattice sites.

\section{Conclusions}\label{S4}

In this article, we have studied several models of spinless fermions with exact many-body scars that are based on pairs or multimer bound states.
In the first part, we have focused on the case of condensates of pairs.
We have characterized exactly the spectral and entanglement properties of the tower of eigenstates responsible for their emergence, thereby proving that the latter display subvolume entanglement entropy scaling and that they are energetically equally-spaced. 
The aforementioned results are corroborated by the numerical analysis of the Loschmidt echo and of the half-chain entanglement entropy, which give clear evidence of the exceptional character of the dynamical properties exhibited by the scarred eigenstates. The latter are indeed atypical, as generic excited eigenstates of the model Hamiltonian are expected to possess standard thermalization properties described within the framework of ETH. Finally, the level-spacing statistics reveals that the system is not integrable and hence that the ETH-breaking involves only a measure-zero set of eigenstates of the Hamiltonian.
The results are then generalised in the second part of the article, where we focus on multimers. The Hamiltonian that we propose is simpler than in the pair case, whereas the scars generalise in many aspects the properties of the pair condensates. We have highlighted the fact that for fermionic multimers there cannot be condensation and we could not find local raising or lowering operators generating the tower of states.

The unveiled results open the route towards further investigations. On one side, it will be interesting to investigate whether such a construction is amenable to generalizations to  higher-dimensional setups. The spatial structure of the pair and multimer complicates the algebraic relations and makes such an extension non-trivial.
On the other side, a stimulating challenge for the future would consist in devising quantum-state engineering protocols to prepare the system in the discovered scar states or in a state that is sufficiently close to it to make the revivals visible in an experiment with a quantum simulator.

\acknowledgements 
We acknowledge enlightening discussions with H.~Katsura.
We acknowledge funding by LabEx PALM (ANR-10-LABX-0039-PALM). This work has been supported by Region Ile-de-France in the framework of the DIM Sirteq.

\appendix

\onecolumngrid
\section{Proof of Eq.~\eqref{Eq:eigv_eq_pi}}\label{App:EigenvalueEquation}
\subsection{Proof by direct verification}
In order to assert the validity of Eq.~\eqref{Eq:eigv_eq_pi}, we consider the Hamiltonian $\hat H_0=\hat U_0 \hat H \hat U_0^{\dag}$ and prove the analogous relation:
\begin{equation}\label{Eq:eigv_eq_0}
\hat H_0 \ket{\psi_{k,0}}=-2\mu k \ket{\psi_{k,0}}.
\end{equation}
Combining Eq.~\eqref{Eq:eigv_eq_0} and Eq.~\eqref{Eq:psi_k_vs_phi_k} with the definition of $\hat H_0$, we obtain the desired result presented in Eq.~\eqref{Eq:eigv_eq_pi}.

To this end, let us consider more explicitly the expression of $\hat H_0$:
\begin{align}\label{Eq:H_0}
\hat H_0 =& it\sum_{j=1}^{L-1}\left[\hat c^{\dag}_j \hat c_{j+1}-c^{\dag}_{j+1}\hat c_j \right]-\mu\sum_{j=1}^L \hat n_j 
+2J\sum_{j=1}^{L-2}[\hat n_j\hat n_{j+1}+\hat n_{j+1} \hat n_{j+2}
-2\hat n_j \hat n_{j+1} \hat n_{j+2}
+(\hat c^{\dag}_j \hat n_{j+1} \hat c_{j+2}+\text{H.c.})].
\end{align}

The action of the single-particle hopping term in Eq.~\eqref{Eq:H_0} on the states $\ket{\psi_{k,0}}$ can be evaluated as follows:
\begin{align}
&it\sum_{j=1}^{L-1}\left[\hat c^{\dag}_j \hat c_{j+1}-c^{\dag}_{j+1}\hat c_j \right]\ket{\psi_{k,0}}
=\quad it\left[\sum_{j=1}^{L-2}\hat c^{\dag}_j \hat c_{j+1}-\sum_{j=2}^{L-1}\hat c^{\dag}_{j+1} \hat c_j \right]\ket{\psi_{k,0}}
= it\sum_{j=1}^{L-2}\left[(\hat c^{\dag}_j-\hat c^{\dag}_{j+2})\hat c_{j+1} \right]\ket{\psi_{k,0}},
\end{align}
where we have discarded the terms $\hat c^{\dag}_2 \hat c_1$ and $\hat c^{\dag}_{L-1} \hat c_L$ when going from the first to the second row since their action vanishes on fully paired Fock basis configurations. After the above manipulations, and denoting as $\mathcal{C}$ the set of all pair configurations contributing to the equal-weight superposition defining the state $\ket{\psi_{k,0}}$, let us evaluate:
\begin{align}
\left (\hat c^{\dag}_j-\hat c^{\dag}_{j+2}\right )\hat c_{j+1}\ket{\psi_{k,0}}=
\frac{1}{k!\sqrt{\binom{L-k}{k}}}\bigg[\sum_{\substack{c\in \mathcal{C}:\\ \langle \hat n_j\rangle=0,\\
\langle\hat n_{j+1}\rangle=1,\\
\langle\hat n_{j+2}\rangle =1}}\ket{\dots \bullet\circ\bullet\dots}-\sum_{\substack{c\in \mathcal{C}:\\ \langle \hat n_j\rangle=1,\\
\langle\hat n_{j+1}\rangle=1,\\
\langle\hat n_{j+2}\rangle =0}}\ket{\dots \bullet\circ\bullet\dots}\bigg]=0
\end{align}
As the above relation holds for $j=1,\dots,L-2$, one obtains:
\begin{equation}
 it\sum_{j=1}^{L-1}\left[\hat c^{\dag}_j \hat c_{j+1}-c^{\dag}_{j+1}\hat c_j \right]\ket{\psi_{k,0}}=0.
\end{equation}

The action of the chemical potential on the states $\ket{\psi_{k,\pi}}$ is trivial, as it amounts to counting the number of particles in the given state, and reads:
\begin{equation}
-\mu\sum_{j=1}^L \hat n_j \ket{\psi_{k,0}}=-2\mu k \ket{\psi_{k,0}}.
\end{equation}

Finally, in order to evaluate the action of the interacting term on $\ket{\psi_{k,0}}$, it is convenient to rewrite it as:
\begin{align}
2J\sum_{j=1}^{L-2}[&\hat n_j\hat n_{j+1}+\hat n_{j+1} \hat n_{j+2}
-2\hat n_j \hat n_{j+1} \hat n_{j+2}
+(\hat c^{\dag}_j \hat n_{j+1} \hat c_{j+2}+H.c.)]= 2J\sum_{j=1}^{L-2}\hat U_0 \hat L^{\dag}_j \hat U_0^{\dag}\hat U_0\hat L_j \hat U_0^{\dag},
\end{align}
where:
\begin{align}
\hat U_0\hat L_j \hat U_0^{\dag}=\hat n_j \hat n_{j+1}-\hat n_{j+1}\hat n_{j+2}-\hat c^{\dag}_{j+2}\hat n_{j+1}\hat c_j+\hat c^{\dag}_j\hat n_{j+1}\hat c_{j+2}.
\end{align}
Hence, it suffices to prove that $\hat U_0\hat L_j \hat U_0^{\dag}\ket{\psi_{k,0}}=0$. By denoting as $\mathcal{C}$ the set of all pair configurations contributing to the equal-weight superposition defining the state $\ket{\psi_{k,0}}$, we obtain:
\begin{align}
\hat U_0\hat L_j \hat U_0^{\dag}\ket{\psi_{k,0}}=\frac{1}{k!\sqrt{\binom{L-k}{k}}}\bigg[\sum_{\substack{c\in \mathcal{C}:\\ \langle \hat n_j\rangle=1,\\
\langle\hat n_{j+1}\rangle=1}}\ket{c} -\sum_{\substack{c\in \mathcal{C}:\\ \langle \hat n_{j+1}\rangle=1,\\
\langle\hat n_{j+2}\rangle=1}}\ket{c}
+\sum_{\substack{c\in \mathcal{C}:\\ \langle \hat n_j\rangle=0,\\
\langle\hat n_{j+1}\rangle=1,\\ \langle \hat n_{j+2} \rangle =1}}\ket{c}-\sum_{\substack{c\in \mathcal{C}:\\ \langle \hat n_j\rangle=1,\\
\langle\hat n_{j+1}\rangle=1,\\ \langle \hat n_{j+2} \rangle =0}}\ket{c}\bigg]=0,
\end{align} 
where the final result is obtained by combining the first and last summation and the second and third summation, respectively. The result presented in Eq.~\eqref{Eq:eigv_eq_pi} is thus proved.

\subsection{Algebraic proof}

We follow yet another, more rigorous route to prove Eq.~\eqref{Eq:eigv_eq_pi}. We aim at showing that, for $k=0,\dots,\lfloor \frac{L}{2} \rfloor$, the following is true:
\begin{align}\label{Eq:alg_proof}
&\left(-t \sum_{j=1}^{L-1} [\hat c^{\dag}_j \hat c_{j+1}  +H.c.]+J\sum_{j=1}^{L-2}\hat L^{\dag}_{j} \hat L_{j}\right)\ket{\psi_{k,\pi}}=0. 
 \end{align}
We start by considering the single-particle hopping term. In this case, it is straightforward to show that:
\begin{align}
&\left[-t \sum_{j=1}^{L-1} \left(\hat c^{\dag}_j \hat c_{j+1}  +H.c.\right),\hat \eta^{\dag}_{\pi}\right]=-\hat c^{\dag}_1\hat c^{\dag}_3
+\sum_{j=2}^{L-2}(-1)^j\left( \hat c^{\dag}_{j-1}\hat c^{\dag}_{j+1} +\hat c^{\dag}_j \hat c^{\dag}_{j+2}\right)+(-1)^{L-1}\hat c^{\dag}_{L-2}\hat c^{\dag}_L=0,
\end{align}
which trivially implies:
\begin{equation}
-t \sum_{j=1}^{L-1} \left(\hat c^{\dag}_j \hat c_{j+1}  +\text{H.c.}\right)\ket{\psi_{k,\pi}}=0,\qquad k=0,\dots,\lfloor \frac{L}{2} \rfloor
\end{equation}

On the other hand, if we denote $\hat H_{int}=J\sum_{j=1}^{L-2}\hat L^{\dag}_{j} \hat L_{j}$, in order to prove $\hat H_{int}\ket{\psi_{k,\pi}}=0$, we need a preparatory lemma.
\begin{lemma}
Let us consider a Hamiltonian $\hat H$ and a set of nonzero states $\{ \left(\hat A^{\dag}\right)^k \ket{\emptyset}\}_{k=0}^N$ obtained by repeated application of the operator $\hat A^{\dag}$ to the vacuum. Let us further define:
\begin{align}
&\hat H_0 = \hat H,\\
&\hat H_1 = [\hat H, \hat A^{\dag}],\\
& \hat H_k = [ \hat H_{k-1}, \hat A^{\dag}],\,\, 2\leq k \leq N.
\end{align}

Then, if $\hat H_k \ket{\emptyset}=0$ for $k=0,\dots,N$, one has:
\begin{equation}
\hat H \left (\hat A^{\dag} \right)^k\ket{\emptyset}=0,\,\,\, k=0,\dots,N
\end{equation}
Proof.
\end{lemma}
We prove the lemma by showing by induction that:
\begin{align}
\hat H  (\hat A^{\dag})^n=\sum_{k=0}^n\binom{n}{k}(\hat A^{\dag})^k\hat H_{n-k},\,\,\, k=0,\dots,N.
\end{align}
The base case $k=0$ is trivially verified. Thus, let us assume the result is proven for $0\leq k \leq n$ and show that it holds as a result for $k=n+1$ as well. We perform the following manipulations:
\begin{align}
&\hat H (\hat A^{\dag})^{n+1}=\left[\sum_{k=0}^n\binom{n}{k}(\hat A^{\dag})^k\hat H_{n-k} \right]\hat A^{\dag}=\sum_{k=0}^n \binom{n}{k}(\hat A^{\dag})^k\left[\hat H_{n+1-k}+\hat A^{\dag}\hat H_{n-k} \right]\\
&\qquad\qquad= \hat H_{n+1}+\sum_{k=1}^n\left[\binom{n}{k}+\binom{n}{k-1}\right](\hat A^{\dag})^k \hat H_{n+1-k}+(\hat A^{\dag})^{n+1}\hat H_0 = \sum_{k=0}^{n+1}\binom{n+1}{k}(\hat A^{\dag})^k\hat H_{n+1-k},\nonumber
\end{align}
thus proving the result. The lemma follows trivially from the assumption that $\hat H_k \ket{\emptyset}=0$ for $k=0,\dots,N$.

We wish to apply the above lemma to the Hamiltonian $\hat H_{int}$ and the set of states $\{(\hat\eta^{\dag}_{\pi})^k\ket{\emptyset}\}_{k=0}^{\lfloor \frac{L}{2}\rfloor}$. To this end, let us consider the parameter-dependent state:
\begin{equation}
\ket{\psi (\alpha)} = e^{\alpha \hat \eta^{\dag}_{\pi}}\hat H_{int} e^{-\alpha \hat \eta^{\dag}_{\pi}}\ket{\emptyset}.
\end{equation}
On one hand, it can be shown that $\ket{\psi(\alpha)}=0$ by rewriting $\ket{\psi(\alpha)}$ as:
\begin{align}
&\ket{\psi(\alpha)}= e^{\alpha \hat \eta^{\dag}_{\pi}}\left(\sum_{j=1}^{L-2}\hat L^{\dag}_j \hat L_j\right)  e^{-\alpha \hat \eta^{\dag}_{\pi}}\ket{\emptyset}=\prod_{j=1}^{3}e^{\alpha (-1)^j \hat c^{\dag}_j \hat c^{\dag}_{j+1}}\hat L^{\dag}_1 \hat L_1\prod_{j=1}^{3}e^{-\alpha (-1)^j \hat c^{\dag}_j \hat c^{\dag}_{j+1}}\ket{\emptyset} \\
&\qquad\quad+\sum_{j=2}^{L-3}\prod_{k=j-1}^{j+2}e^{\alpha (-1)^k \hat c^{\dag}_k \hat c^{\dag}_{k+1}}\hat L^{\dag}_j \hat L_j \prod_{k=j-1}^{j+2}e^{-\alpha (-1)^k \hat c^{\dag}_k \hat c^{\dag}_{k+1}}\ket{\emptyset}+\prod_{j=L-3}^{L-1}e^{\alpha (-1)^j \hat c^{\dag}_j \hat c^{\dag}_{j+1}}\hat L^{\dag}_{L-2} \hat L_{L-2}\prod_{j=L-3}^{L-1}e^{-\alpha (-1)^j \hat c^{\dag}_j \hat c^{\dag}_{j+1}}\ket{\emptyset}\nonumber
\end{align}
and showing that each term in the above summation vanishes.

On the other hand, the Baker-Campbell-Hausdorff formula allows to express the latter as:
\begin{equation}
\ket{\psi (\alpha)} =\hat H_{int} \ket{\emptyset}+\alpha[\hat \eta^{\dag}_{\pi}, \hat H_{int}]\ket{\emptyset}+\frac{\alpha^2}{2!}[\hat \eta^{\dag}_{\pi},[\hat \eta^{\dag}_{\pi},\hat H_{int}]]\ket{\emptyset}+\dots,
\end{equation}
up to order $\lfloor \frac{L}{2} \rfloor$. As $\ket{\psi (\alpha)}$ vanishes, each of the states multiplying the corresponding power of $\alpha$ must vanish. Hence, the conditions of the lemma are satisfied and Eq.~\eqref{Eq:alg_proof} is proved, which in turn implies the validity of Eq.~\eqref{Eq:eigv_eq_pi}.
\medskip

\section{Coordinate Bethe Ansatz in the fully paired subspace for the interacting term in Hamiltonian~\eqref{Eq:hamiltonian}}\label{App:CBA}

We start from the spinless fermion Hamiltonian~\eqref{Eq:hamiltonian} with $t=\mu=0$ and rewrite it in PBC with a change in the sign of the pair hopping for the terms across the bond among sites $L$ and $1$:

\begin{align}
\hat H&=J\sum_{j=1}^{L-2} \left[\hat n_j \hat n_{j+1} +\hat n_{j+1}\hat n_{j+2}-2\hat n_j \hat n_{j+1} \hat n_{j+2}-(\hat c_j^{\dag} \hat n_{j+1}\hat c_{j+2}+H.c.)\right]\\
&+J\left[\hat n_{L-1} \hat n_{L} +\hat n_{L}\hat n_{1}-2\hat n_{L-1} \hat n_{L} \hat n_{1} +(\hat c_{L-1}^{\dag} \hat n_{L}\hat c_{1}+H.c.)\right]\\
&+J\left[\hat n_L\hat n_{1} +\hat n_{1}\hat n_{2}-2\hat n_L \hat n_{1} \hat n_{2} + (\hat c_L^{\dag} \hat n_{1}\hat c_{2}+H.c.)\right].
\end{align}
When rewritten in spin-$1/2$ language and in the sector of even parity (that the fully paired subspace belongs to), it takes the form:
\begin{equation}\label{Eq:interacting_spin}
\hat H=J\sum_{j=1}^L \left[\hat n_j \hat n_{j+1} +\hat n_{j+1}\hat n_{j+2}-2\hat n_j \hat n_{j+1} \hat n_{j+2} + (\hat \sigma_j^+ \hat n_{j+1}\hat\sigma_{j+2}^{-}+H.c.)\right],
\end{equation}
where $\hat n_j =\frac{1+\hat\sigma_j^z}{2}$. In the following, we apply the coordinate Bethe Ansatz technique to the subspace spanned by fully paired configurations and write down Bethe equations for the momenta of the pairs first in the $1$-pair problem and then in the $2$-pair one, showing the lack of interactions among $\eta$-pairs.
\subsection{$1$-pair problem}

We search for a generic eigenstate in the $1$-pair subspace by writing it in the form:
\begin{equation}
\ket{\psi}=\sum_{j=1}^L a(j) \hat\sigma_j^+\hat\sigma_{j+1}^+ \ket{\downarrow},
\end{equation}
where $\ket{\downarrow}$ is the spin down ferromagnetic state.
The Schr\"odinger equation $\hat H\ket{\psi}=E\ket{\psi}$ and the periodic boundary conditions read:
\begin{align}
&2J a(j) +Ja(j-1) +Ja(j+1)=Ea(j), \\
&a(L+j)=a(j).
\end{align}
Looking for a solution of the form $a(j)=Ae^{ikj}$, one obtains the conditions:
\begin{align}
E=2J+2J\cos(k),\quad\text{with}\qquad
k=\frac{2\pi}{L}n,\,\,\, n=0,\dots,L-1.
\end{align}
When $L$ is even, the $1$-pair eta-pairing state $\ket{\psi}\propto \left(\sum_{j=1}^L e^{i\pi j}\hat\sigma_j^+\hat\sigma_{j+1}^+ \right)\ket{\downarrow}$ is recovered.

\subsection{$2$-pair problem}

We search for a generic eigenstate in the $2$-pair subspace by writing it in the form:
\begin{equation}
\ket{\psi}=\sum_{1\leq j_1<j_2\leq L} a(j_1,j_2) \hat\sigma_{j_1}^+\hat\sigma_{j_1+1}^+ \hat\sigma_{j_2}^+\hat\sigma_{j_2+1}^+ \ket{\downarrow},
\end{equation}
where one should notice that the term multiplying $a(j_1,j_1+1)$ vanishes. 
The Schroedinger equation $\hat H\ket{\psi}=E\ket{\psi}$ reads now:
\begin{align}
&4J a(j_1,j_2)+Ja(j_1-1,j_2)+Ja(j_1+1,j_2) +Ja(j_1,j_2-1)+Ja(j_1,j_2+1)=Ea(j_1,j_2),\,\,\, j_2> j_1+2 \label{Eq:2p:1}\\
&2J a(j_1,j_1+2) +J a(j_1-1,j_2) +J a(j_1,j_1+3) = E a(j_1,j_1+2), \,\,\,j_2=j_1+2, \label{Eq:2p:2}
\end{align}
while PBC are enforced through the equation:
\begin{equation} \label{PBC_2p}
a(j_1,j_2)=a(j_2,j_1+L).
\end{equation}
The Ansatz for the coefficients $a(j_1,j_2)$ takes the form:
\begin{equation}
a(j_1,j_2)=A_{12} e^{i(k_1j_1+k_2j_2)}+A_{21}e^{i(k_2j_1+k_1j_2)}.
\end{equation}
The above Ansatz solves Eq.~\eqref{Eq:2p:1} with energy $E=4J +2J\cos(k_1)+2J \cos(k_2)$, while Eq.~\eqref{Eq:2p:2} is solved by adding to it the terms such that it takes the same form as Eq.~\eqref{Eq:2p:1} and setting them to zero. The result of this procedure leads to the condition:
\begin{equation} \label{Eq:2p:3}
2J a(j_1,j_1+2)+Ja(j_1+1,j_1+2)+Ja(j_1,j_1+1)=0,
\end{equation}
that amounts to requiring that:
\begin{equation}
\frac{A_{21}}{A_{12}}=-\frac{2e^{i2k_2}+e^{i(k_1+2k_2)} +e^{ik_2}}{2e^{i2k_1}+e^{i(k_2+2k_1)} +e^{ik_1}}:=S(k_1,k_2),
\end{equation}
where we have defined the scattering matrix $S(k_1,k_2)$ via the expression to its left.

Finally, imposing PBC, one obtains the relations:
\begin{align}
&A_{12}=A_{21}e^{ik_1L},\\
&A_{21}=A_{12} e^{ik_2L},
\end{align}
which, owing to the property $S(k,k)=-1$, can be rewritten in the compact form:
\begin{equation}\label{Eq:2p:4}
\prod_{l=1}^2 S(k_j,k_l) = -e^{-ik_j L}; \,\,\, j=1,2.
\end{equation}
The result can be shown to generalize to the nontrivial three-pair case, thus proving the Bethe-Ansatz solvability of model~\eqref{Eq:interacting_spin} in the subspace spanned by fully-paired configurations.

It should be noticed that the two-pair $\eta$-pairing state, obtained for $k_1=k_2=\pi$, satisfies Eq.~\eqref{Eq:2p:3} for all values of $A_{12},A_{21}$ and the PBC in Eq.~\eqref{PBC_2p} for an even value of $L$, which impose $A_{12}=A_{21}$. The scattering matrix is ill-defined in this case, as the coefficient $a(j_1,j_2)$ reduces to:
\begin{equation}
a(j_1,j_2)\propto e^{i\pi j_1}e^{i\pi j_2},
\end{equation}
i.e., it factorizes into independent plane waves with quasimomentum $\pi$.
\medskip

\section{Finite-size formula for $P_k(r)$}\label{App:pair_corr}
We present here an explicit finite-size formula for $P_k(r)$ with $r>3$. Straightforward combinatorial considerations give the result:
\begin{align}
P_k(r)=(-1)^{r+1}\frac{\sum_{l=\max\left(0,k-1-\lfloor \frac{L-r-2}{2}\rfloor\right)}^{\min(k-1,\lfloor \frac{r-2}{2} \rfloor)}\binom{r-2-l}{l}\binom{L-1-r-k+l}{k-l-1}}{\frac{L}{L-k}\binom{L-k}{k}}.
\end{align}
\medskip

\section{The operator $\hat O$ annihilates the tower of states}\label{App:O_operator}

The operator $\hat O$ takes the form $\hat O=J\sum_{j=2}^{L-2}\hat O_j$, where:
\begin{align}
\hat O_j=e^{i\pi(j-1)}(1-2\hat n_{j-1})\hat n_j \hat c^{\dag}_{j+1}\hat c^{\dag}_{j+2}
+e^{i\pi(j-1)}\hat c^{\dag}_{j-1}\hat c^{\dag}_j \hat n_{j+1}(1-2\hat n_{j+2})
+e^{i\pi j}\hat c^{\dag}_{j-1}(\hat n_j +\hat n_{j+1})\hat c^{\dag}_{j+2}.
\end{align}
Hence, it suffices to show that $\hat O_j\ket{\psi_{k,\pi}}=0$ for $2\leq j \leq L-2$. We adopt the notation $\sum_{c:\ket{x_1,\dots,x_4}} e^{i\pi \Sgn(c)}\ket{y_1,\dots,y_4}$ to denote the sum over all Fock states $\ket{c}$ contributing to the state $\ket{\psi_{k,\pi}}$ that have the form $\ket{x_1,\dots,x_4}$ on sites $j-1,j,j+1,j+2$ prior to the application of $\hat O_j$ and the form $\ket{y_1,\dots,y_4}$ after the application of $\hat O_j$, and where $\Sgn(c)=\sum_{m=1}^k j_m$, $j_m$ being the position of the first fermion of the $m^{th}$ pair in the state $\ket{\psi_{k,\pi}}$.
Then, it is easy to obtain:
\begin{align}
\hat O_j \ket{\psi_{k,\pi}}=e^{i\pi j}\left(\sum_{c:\ket{\bullet\bullet\circ\circ}}e^{i\pi \Sgn(c)}\ket{\bullet\bullet\bullet\bullet}+\sum_{c:\ket{\circ\circ\bullet\bullet}}e^{i\pi \Sgn(c)}\ket{\bullet\bullet\bullet\bullet}+2\sum_{c:\ket{\circ\bullet\bullet\circ}}e^{i\pi \Sgn(c)}\ket{\bullet\bullet\bullet\bullet}\right).
\end{align}
The number of configurations contributing to the state $\ket{\psi_{k,\pi}}$ that locally, on sites $j-1,j,j+1,j+2$, have the form $\ket{\bullet\bullet\circ\circ}$ is equal to the number of those with the forms $\ket{\circ\circ\bullet\bullet}$ and $\ket{\circ\bullet\bullet\circ}$, and they can be put in a one-to-one correspondence with each other by mapping each configuration in of the aforesaid three collections to the one that is identical to it up to the different occupation of the highlighted sites $j-1,j,j+1,j+2$. Since the local configuration resulting from the application of $\hat O_j$ is the same in all three cases, the three summations are carried over the same set of configurations. On the other hand, the sign of each of the configurations in the last summation is the opposite of the sign of the corresponding ones in the first two summations, leading to the desired result $\hat O_j\ket{\psi_{k,\pi}}=0$.
\medskip

\section{Macroscopic coherence of the state $\ket{\alpha}$}\label{App:Coherence}

We derive here explicitly the result shown in Eq.~\eqref{Eq:coherence}. The explicit expression for the state $\ket{\alpha}$ reads:
\begin{align}\label{Eq:alpha_state}
\ket{\alpha}= \mathcal{N}\sum_{k=0}^{ L/2-1}\sqrt{\frac{L}{L-k}\binom{L-k}{k}}\alpha^k\ket{\psi_{k,\pi}},
\end{align}
where the normalization constant $\mathcal{N}$ satisfies:
\begin{align}
|\mathcal{N}|^2=\frac{1}{\sum_{k=0}^{L/2-1}\frac{L}{L-k}\binom{L-k}{k}|\alpha|^{2k}}
\end{align}
which results from imposing the normalization condition $\bra{\alpha}\ket{\alpha}=1$. 

Plugging the expression of the expansion coefficients in Eq.~\eqref{Eq:alpha_state} into Eq.~\eqref{Eq:oscillating_corr}, one obtains:
\begin{align}\label{Eq:exact_expr}
\bra{\alpha(t)}e^{i\pi j} \hat c_{j+1}\hat c_j \ket{\alpha(t)} = \alpha e^{i\frac{2\mu t}{\hbar}} \frac{\sum_{k=0}^{L/2-2}\binom{L-k-2}{k}|\alpha|^{2k}}{\sum_{k=0}^{L/2-1}\frac{L}{L-k}\binom{L-k}{k}|\alpha|^{2k}}.
\end{align}
If we assume self-consistently that the sums in the numerator and denominator of Eq.~\eqref{Eq:exact_expr} will be dominated by terms with $k=O(L)$ and apply Stirling's approximation $n!\approx \sqrt{2\pi n} n^n e^{-n}$, one obtains, as a function of the rescaled variable $x=\frac{3k}{L}$:
\begin{align}
&\frac{L}{L-k}\binom{L-k}{k}|\alpha|^{2k}\approx g(x)e^{\frac{L}{3}f(x)},\\
&\binom{L-k-2}{k}|\alpha|^{2k}\approx h(x)e^{\frac{L}{3}f(x)}
\end{align}
in the limit of large $L$, where we have introduced the function:
\begin{align}
&f(x)=(3-x)\log(3-x)-x\log x-(3-2x)\log(3-2x)+(2\log |\alpha|)x,\\
& g(x) = 3\sqrt{\frac{3}{2\pi L}}\sqrt{\frac{1}{x(3-x)(3-2x)}},\\
& h(x) =\sqrt{\frac{3}{2\pi L}}\sqrt{\frac{(3-2x)^3}{x(3-x)^3}}.
\end{align} 
We proceed by converting the summations over $k$ in Eq.~\eqref{Eq:exact_expr} into continuous integrals over $x$ and applying the saddle-point integration technique, in order to get to the final result:
\begin{align}\label{Eq:coherence_result}
\bra{\alpha(t)}e^{i\pi j} \hat c_{j+1}\hat c_j \ket{\alpha(t)} = \alpha e^{i\frac{2\mu t}{\hbar}}\frac{\int_0^{\frac{3}{2}}h(x) e^{\frac{L}{3}f(x)}\dd x }{\int_0^{\frac{3}{2}}g(x) e^{\frac{L}{3}f(x)}\dd x }\approx \frac{h(x^*)}{g(x^*)}=\frac{1}{3} \frac{(3-2x^*)^2}{3-x^*},
\end{align}
where:
\begin{equation}
x^*=\frac{3}{2}\left(1-\frac{1}{\sqrt{1+4|\alpha|^2}} \right)
\end{equation} 
satisfies $f'(x^*)=0$. Plugging the expression of $x^*$ into Eq.~\eqref{Eq:coherence_result}, one recovers Eq.~\eqref{Eq:coherence}.
\medskip

\section{Entanglement entropy of the states $\ket{\psi_{k,\pi}}$}\label{App:Entanglement}

For the sake of convenience, we introduce the states:
\begin{align}
\ket{\psi_{k,q}}=\frac{1}{\sqrt{\binom{L-k}{k}}}\frac{(\hat\eta^{\dag}_{q})^{k}}{k!}\ket{\emptyset}, \, \text{ with } \,
\hat\eta^{\dag}_{q} = \sum_{j=1}^{L-1}e^{iqj} \hat c^{\dag}_j \hat c^{\dag}_{j+1}.
\end{align}
Furthermore, we describe a general relation between the states $\ket{\psi_{k,\pi}}$ and the states $\ket{\psi_{k,q}}$. By introducing the unitary operator $\hat U_{q}=\prod_{j=1}^{L}e^{i\frac{\pi +q}{2}j\hat n_j}$, one can show that the following relation holds:
\begin{equation}
\hat U_q \hat \eta^{\dag}_{\pi}\hat U^{\dag}_q = e^{i\frac{\pi+q}{2}}\hat \eta^{\dag}_q,
 \end{equation}
 which in turn implies that:
 \begin{equation}\label{Eq:psi_k_vs_phi_k}
 \ket{\psi_{k,q}}=e^{-i\frac{\pi +q}{2}k}\hat U_q \ket{\psi_{k,\pi}}
 \end{equation}
Since the state $\ket{\psi_{k,\pi}}$ is related to the state $\ket{\psi_{k,q}}$ by a unitary transformation, we underline that the states $\ket{\psi_{k,\pi}}$ are eigenstates of $\hat H$ if and only if the states $\ket{\psi_{k,q}}$ are eigenstates of $\hat U_q \hat H \hat U^{\dag}_q$.

We start by evaluating the half-chain entanglement entropy for the states $\ket{\psi_{k,0}}$. We can distinguish among the configurations in which no pair is placed on the sites $\frac{L}{2}$ and $\frac{L}{2}+1$ and the configurations in which this is instead the case:
\begin{align}
\ket{\psi_{k,0}}=&\frac{1}{\sqrt{\binom{L-k}{k}}}\Bigg[\sum_{n=\max\{0,\lceil\frac{N-L/2}{2}\rceil \}}^{\min\{\frac{N}{2},\lfloor\frac{L/2}{2}\rfloor \}}\sum_{\substack{\{\vec{j}^{(n)}_{P,[1,\dots,\frac{L}{2}]} \}\\ \{\vec{j'}^{\left(\frac{N}{2}-n\right)}_{P,[\frac{L}{2}+1,\dots,L]}\}}} \ket{\vec{j}^{(n)}_{P,[1,\dots,\frac{L}{2}]}}\ket{\vec{j'}^{\left(\frac{N}{2}-n\right)}_{P,[\frac{L}{2}+1,\dots,L]}}+\\
+&\sum_{l=\max\{0,\lceil\frac{N-L/2-1}{2}\rceil \}}^{\min\{\frac{N}{2}-1,\lfloor\frac{L/2-1}{2}\rfloor \}}\sum_{\substack{\{\vec{j}^{(n)}_{P,[1,\dots,\frac{L}{2}-1]} \}\\ \{\vec{j'}^{\left(\frac{N}{2}-1-n\right)}_{P,[\frac{L}{2}+2,\dots,L]}\}}} \ket{\vec{j}^{(n)}_{P,[1,\dots,\frac{L}{2}-1]};\bullet}\ket{\bullet;\vec{j'^{\left(\frac{N}{2}-1-n\right)}}^{\left(\frac{N}{2}-n\right)}_{P,[\frac{L}{2}+2,\dots,L]}}\Bigg],\nonumber
\end{align}
where the symbol $\bullet$ indicates an occupied site and the notation of the form $\ket{\vec{j}^{(n)}_{P,[1,\dots,\frac{L}{2}}}$ denote the equal-weight (with unit weight, hence unnormalized) superposition of all possible distributions of $n$ pairs over the sites $i=1,\dots,\frac{L}{2}$. By denoting as $\mathcal{F}_R$ the set of Fock configurations on the right half of the chain, the expression of the partial density matrix $\rho_{k,0}^{\left(\frac{L}{2}\right)}=\text{Tr}_{[\frac{L}{2}+1,\dots,L]}(\ket{\psi_{k,0}}\hspace{-0.1cm}\bra{\psi_{k,0}})$ takes then the form:
\begin{align}
&\rho_{k,0}^{\left(\frac{L}{2}\right)}=\frac{1}{\binom{L-k}{k}}\sum_{k\in \mathcal{F}_{R}}\Bigg[\sum_{l=\max\{0,\lceil\frac{N-L/2}{2}\rceil \}}^{\min\{\frac{N}{2},\lfloor\frac{L/2}{2}\rfloor \}}\sum_{\substack{\{\vec{j}^{(n)}_{P,[1,\dots,\frac{L}{2}]} \}\\ \{\vec{l}^{(n)}_{P,[1,\dots,\frac{L}{2}]}\}}}\sum_{\substack{\{\vec{j'}^{\left(\frac{N}{2}-n\right)}_{P,[\frac{L}{2}+1,\dots,L]} \}\\ \{\vec{l'}^{\left(\frac{N}{2}-n\right)}_{P,[\frac{L}{2}+1,\dots,L]}\}}}\ket{\vec{j}^{(n)}_{P,[1,\dots,\frac{L}{2}]}}\bra{\vec{l}^{(n)}_{P,[1,\dots,\frac{L}{2}]}}\nonumber\\
&\qquad\times\bra{k}\ket{\vec{j'^{\left(\frac{N}{2}-n\right)}}_{P,[\frac{L}{2}+1,\dots,L]}}\bra{\vec{l'^{\left(\frac{N}{2}-n\right)}}_{P,[\frac{L}{2}+1,\dots,L]}}\ket{k}\nonumber\\
&\qquad +\sum_{l=\max\{0,\lceil\frac{N-L/2-1}{2}\rceil \}}^{\min\{\frac{N}{2}-1,\lfloor\frac{L/2-1}{2}\rfloor \}}\sum_{\substack{\{\vec{j}^{(n)}_{P,[1,\dots,\frac{L}{2}-1]} \}\\ \{\vec{l}^{(n)}_{P,[1,\dots,\frac{L}{2}-1]}\}}}\sum_{\substack{\{\vec{j'}^{\left(\frac{N}{2}-n\right)}_{P,[\frac{L}{2}+2,\dots,L]} \}\\ \{\vec{l'}^{\left(\frac{N}{2}-n\right)}_{P,[\frac{L}{2}+2,\dots,L]}\}}}\ket{\vec{j}^{(n)}_{P,[1,\dots,\frac{L}{2}-1]};\bullet}\bra{\vec{l}^{(n)}_{P,[1,\dots,\frac{L}{2}-1]};\bullet}\nonumber\\
&\qquad\times\bra{k}\ket{\vec{j'}^{\left(\frac{N}{2}-1-n\right)}_{P,[\frac{L}{2}+2,\dots,L]};\bullet}\bra{\bullet;\vec{l'}^{\left(\frac{N}{2}-1-n\right)}_{P,[\frac{L}{2}+2,\dots,L]}}\ket{k}\Bigg].\nonumber
\end{align}
For every configuration $k\in\mathcal{F}_R$ that results in a nonvanishing scalar product in the above expression, one obtains the unnormalized equal-weight superposition with unit weight of all configurations in the left half of the system that are compatible with it. After normalizing the latter and counting all configurations $k\in\mathcal{F}_R$ that give a nonzero contribution, one obtains:
\begin{align} 
\rho_{k,0}^{\left(\frac{L}{2}\right)}
=\label{Eq:density_matrix}&\sum_{l=\max\{0,\lceil\frac{N-L/2}{2}\rceil \}}^{\min\{\frac{N}{2},\lfloor\frac{L/2}{2}\rfloor \}}\ket{\varphi_l^{[1,\dots,\frac{L}{2}]}}\bra{\varphi_l^{[1,\dots,\frac{L}{2}]}}\frac{\binom{\frac{L}{2}-\frac{N}{2}+l}{\frac{N}{2}-l}\binom{\frac{L}{2}-l}{l}}{\binom{L-\frac{N}{2}}{\frac{N}{2}}}\\
&+\sum_{l=\max\{0,\lceil\frac{N-L/2-1}{2}\rceil \}}^{\min\{\frac{N}{2}-1,\lfloor\frac{L/2-1}{2}\rfloor \}}\ket{\varphi_l^{[1,\dots,\frac{L}{2}-1]}}\ket{\bullet}\bra{\bullet}\bra{\varphi_l^{[1,\dots,\frac{L}{2}-1]}}\frac{\binom{\frac{L}{2}-\frac{N}{2}+l}{\frac{N}{2}-l-1}\binom{\frac{L}{2}-1-l}{l}}{\binom{L-\frac{N}{2}}{\frac{N}{2}}},\nonumber
\end{align}
where $\ket{\varphi_l^{[1,\dots,s]}}$ denotes the normalized equal-weight superposition of all Fock states with $l$ pairs distributed over $s$ lattice sites and $\ket{\bullet}$ indicates the occupation of site $\frac{L}{2}$. 
The eigenvalues $\lambda_s$ of the reduced density matrix $\rho_{k,0}$ can be read off directly Eq.~\eqref{Eq:density_matrix} as the coefficients of each term of the summations in Eq.~\eqref{Eq:density_matrix}, and the second Renyi entropy can be computed accordingly as $S_{\frac{L}{2}, k, 0}=-\log\left(\sum_s\lambda_s^2\right)$. 

We now proceed to show that the half-chain entanglement entropy of the states $\ket{\psi_{k,\pi}}$ defined in Eq.~\eqref{Eq:eta_pairs} equals the one of the states $\ket{\psi_{k,0}}$ introduced via Eq.~\eqref{Eq:psi_k_vs_phi_k}, thus showing that the subvolume entanglement scaling law holds for both towers of states. To this end, we recall the previously defined unitary operator $\hat U_{0} = \prod_{j=1}^L e^{i\frac{\pi}{2}j \hat n_j}$. 
After using Eq.~\eqref{Eq:psi_k_vs_phi_k} and noticing that $\hat U_0$ factorizes as $\hat U_0=\hat U_{0,\left[1,\frac{L}{2}\right]}\hat U_{0,\left[\frac{L}{2}+1, L\right]}$, where $\hat U_{0,\left[1,\frac{L}{2}\right]}=\prod_{j=1}^{\frac{L}{2}} e^{i\frac{\pi}{2}j \hat n_j}$ and $\hat U_{0,\left[\frac{L}{2}+1,L\right]}=\prod_{j=\frac{L}{2}+1}^{L} e^{i\frac{\pi}{2}j \hat n_j}$, the half-chain reduced density matrix for a generic state $\ket{\psi_{k,\pi}}$ can then be expressed as:
\begin{align}
\rho_{k,\pi}^{(\frac{L}{2})}=\frac{1}{(k!)^2\binom{L-k}{k}}\hat U_{0,\left[1,\frac{L}{2}\right]} \text{Tr}_{[\frac{L}{2}+1,\dots,L]}\left[ \hat U_{0,\left[\frac{L}{2}+1,L\right]}(\hat\eta ^{\dag}_0)^k\ket{\emptyset}\bra{\emptyset}(\hat \eta_0)^k \hat U^{\dag}_{0,\left[\frac{L}{2}+1,L\right]}\right]\hat U^{\dag}_{0,\left[1,\frac{L}{2}\right]}
\end{align}
By making use of the cyclic invariance property of the trace, the reduced density matrix for the states $\ket{\psi_{k,\pi}}$ is manifestly shown to be related to the corresponding quantity for the states $\ket{\psi_{k,0}}$ via a similarity transformation implemented by a unitary operator, i.e.:
 \begin{equation}
\rho_{k,\pi}^{(\frac{L}{2})}=\hat U_{0,\left[1,\frac{L}{2}\right]}Tr_{[\frac{L}{2}+1,\dots,L]}\left[\ket{\psi_{k,0}}\bra{\psi_{k,0}}\right] \hat U^{\dag}_{0,\left[1,\frac{L}{2}\right]},
 \end{equation}
which leaves the entanglement entropy unaffected.

We are now in a position to estimate the large-$L$ scaling of the second Renyi entropy of the states $\ket{\psi_{k,\pi}}$ analytically in the case $k=L/3$, designed to ensure that the binomial coefficients in the numerator of the combinatorial coefficient in the first row of Eq.~\eqref{Eq:density_matrix} are peaked around the same value of $l=L/6=O(L)$. After introducing the rescaled variable $x=6l/L$, one obtains by means of the Stirling approximation the asymptotic behaviors:
\begin{align}
&\frac{\binom{\frac{L}{2}-\frac{N}{2}+l}{\frac{N}{2}-l}\binom{\frac{L}{2}-l}{l}}{\binom{L-\frac{N}{2}}{\frac{N}{2}}}\approx g(x)e^{\frac{L}{6}f(x)},\\
&\frac{\binom{\frac{L}{2}-\frac{N}{2}+l}{\frac{N}{2}-l-1}\binom{\frac{L}{2}-1-l}{l}}{\binom{L-\frac{N}{2}}{\frac{N}{2}}}\approx h(x) e^{\frac{L}{6}f(x)},
\end{align}
where the large $L$ limit has been taken and we have introduced the functions:
\begin{align}
&f(x)= (1+x)\log(1+x) +(3-x)\log(3-x)-(2-x)\log(2-x)\\
&\quad\qquad -(2x-1)\log(2x-1)-x\log x-(3-2x)\log (3-2x)-4\log 2,\nonumber\\ 
& g(x) =\sqrt{\frac{3}{\pi L}\frac{(1+x)(3-x)}{x(2-x)(2x-1)(3-2x)}},\\
& h(x)= \frac{(2-x)(3-2x)}{(3-x)(2x-1)}g(x).
\end{align}
Armed with these expressions, we proceed to evaluate the argument of Eq.~\eqref{Eq:Renyi_entropy} by converting the discrete sums over $l$ into continuous integrals over $x$ and applying the saddle-point integration technique:
\begin{align}\label{Eq:entropy_saddle_point}
 & \sum_{l=\max\{0,\lceil\frac{N-L/2}{2}\rceil \}}^{\min\{\frac{N}{2},\lfloor\frac{L/2}{2}\rfloor \}}
 \left( \frac{\binom{\frac{L}{2}-\frac{N}{2}+l}{\frac{N}{2}-l}\binom{\frac{L}{2}-l}{l}}{\binom{L-\frac{N}{2}}{\frac{N}{2}}} \right)^2 + 
 \sum_{l=\max\{0,\lceil\frac{N-L/2-1}{2}\rceil \}}^{\min\{\frac{N}{2}-1,\lfloor\frac{L/2-1}{2}\rfloor \}} \left(
 \frac{\binom{\frac{L}{2}-\frac{N}{2}+l}{\frac{N}{2}-l-1}\binom{\frac{L}{2}-1-l}{l}}{\binom{L-\frac{N}{2}}{\frac{N}{2}}} \right)^2\\
&\qquad\approx \frac{L}{6}\left(\int_{\frac{1}{2}}^{\frac{3}{2}} dx\, g^{2}(x) e^{\frac{L}{3}f(x)}+\int_{\frac{1}{2}}^{\frac{3}{2}}dx\, h^{2}(x) e^{\frac{L}{3}f(x)}\right)\approx \frac{L}{6}e^{\frac{L}{3}f(x^*)}\sqrt{\frac{6\pi}{L|f''(x^*)|}}\left(g^2(x^*)+h^2(x^*) \right),\nonumber
\end{align}
where $x^{*}=1$ satisfies $f'(x^*)=0$ and $f''(x^*)<0$. After a straightforward substitution of the numerical value of $x^*$ in Eq.~\eqref{Eq:entropy_saddle_point}, one gets the result $\frac{5}{\sqrt{6\pi L}}$, which in turns gives the logarithmic scaling in Eq.~\eqref{Eq:asymptotics_entropy} once plugged into Eq.~\eqref{Eq:Renyi_entropy}.

As a final consistency check, we verify that the large $L$ asymptotic behavior of the eigenvalues of the half-chain reduced density matrix $\rho_{k,\pi}^{\left(\frac{L}{2}\right)}$ preserves the normalization condition that they are subject to. Specifically, we evaluate $\text{Tr}\left[\rho_{k,\pi}^{\left(\frac{L}{2}\right)}\right]$, namely:
\begin{align}
& \sum_{l=\max\{0,\lceil\frac{N-L/2}{2}\rceil \}}^{\min\{\frac{N}{2},\lfloor\frac{L/2}{2}\rfloor \}}
  \frac{\binom{\frac{L}{2}-\frac{N}{2}+l}{\frac{N}{2}-l}\binom{\frac{L}{2}-l}{l}}{\binom{L-\frac{N}{2}}{\frac{N}{2}}}  + 
 \sum_{l=\max\{0,\lceil\frac{N-L/2-1}{2}\rceil \}}^{\min\{\frac{N}{2}-1,\lfloor\frac{L/2-1}{2}\rfloor \}} 
 \frac{\binom{\frac{L}{2}-\frac{N}{2}+l}{\frac{N}{2}-l-1}\binom{\frac{L}{2}-1-l}{l}}{\binom{L-\frac{N}{2}}{\frac{N}{2}}} \\
 & \qquad \approx \frac{L}{6}\left(\int_{\frac{1}{2}}^{\frac{3}{2}} dx\, g(x) e^{\frac{L}{6}f(x)}+\int_{\frac{1}{2}}^{\frac{3}{2}}dx\, h(x) e^{\frac{L}{6}f(x)}\right)\approx \frac{L}{6}e^{\frac{L}{6}f(x^*)}\sqrt{\frac{12\pi}{L|f''(x^*)|}}\left(g(x^*)+h(x^*) \right)=1,\nonumber
\end{align}
consistently with the expected result.

\section{Some additional results on multimer scars}\label{App:Multimer}

In Sec.~\ref{Sec:Multimers}
we have introduced the Hamiltonian $\hat H_J$ as a frustration-free positive Hamiltonian of the form $J\sum_j \hat L^{\dag (M)}_j\hat L_j^{(M)}$. 
The explicit expression of $\hat L^{(M)}_j$ reads
\begin{align}
\hat L_j^{(M)}=&\prod_{l=0}^{M-1} \hat n_{j+l}- \prod_{l=0}^{M-1}\hat n_{j+l+1} -(-1)^M \hat c^{\dag}_j \left(\prod_{l=1}^{M-1}\hat n_{j+l} \right)\hat c_{j+M}+(-1)^M \hat c^{\dag}_{j+M} \left(\prod_{l=1}^{M-1}\hat n_{j+l} \right)\hat c_{j} = \nonumber \\
=&
\prod_{l=0}^{M-1} \hat n_{j+l}- \prod_{l=0}^{M-1}\hat n_{j+l+1}+ \hat M_j^\dagger \hat M_{j+1}- \hat M_{j+1}^\dagger \hat M_j
\end{align}

\section{Explicit expression of the trimer correlations}\label{App:trimer_corr}

The trimer correlation function for $7\leq j\leq L-5$ reads:
\begin{align}
\bra{\psi_{k,\pi}^{(3)}}\hat c^{\dag}_j \hat c^{\dag}_{j+1} \hat c^{\dag}_{j+2} \hat c_3 \hat c_2 \hat c_1 \ket{\psi_{k,\pi}^{(3)}}=&\frac{(-1)^{j+1}}{\binom{L-2k}{k}}\biggl[\sum_{l=\max\{ 0,\lceil(k-1-\lfloor (L-j-2)/3\rfloor )/2\rceil\}}^{\min\{\lfloor (k-1)/2\rfloor ,\lfloor (j-4)/6\rfloor \}}\binom{j-4-4l}{2l}\binom{L-j-2-2(k-1-2l)}{k-1-2l}+\nonumber\\
&-\sum_{k=\max\{1,\lceil (k-\lfloor (L-j-2)/3\rfloor)/2 \rceil \}}^{\min\{\lfloor k/2\rfloor,\lfloor (1+\lfloor (j-4)/3 \rfloor)/2 \rfloor \}} \binom{j-4-2(2l-1)}{2l-1}\binom{L-j-2-2(k-1-2l+1)}{k-1-2l+1} \biggr],
\end{align}
while, for $1\leq j \leq 6$, it takes the form:
\begin{align}
\bra{\psi_{k,\pi}^{(3)}}\hat c^{\dag}_j \hat c^{\dag}_{j+1} \hat c^{\dag}_{j+2} \hat c_3 \hat c_2 \hat c_1 \ket{\psi_{k,\pi}^{(3)}}=(-1)^{j+1}\frac{\binom{L-j-2-2(k-1)}{k-1}}{\binom{L-2k}{k}},
\end{align}
and, for $L-4\leq j \leq L-2$, it is given by:
\begin{equation}
\bra{\psi_{k,\pi}^{(3)}}\hat c^{\dag}_j \hat c^{\dag}_{j+1} \hat c^{\dag}_{j+2} \hat c_3 \hat c_2 \hat c_1 \ket{\psi_{k,\pi}^{(3)}}=(-1)^{j+1}(-1)^{k-1}\frac{\binom{j-4-2(k-1)}{k-1}}{\binom{L-2k}{k}}.
\end{equation}

\twocolumngrid

\bibliographystyle{apsrev4-1}

\begin{thebibliography}{80}%
\makeatletter
\providecommand \@ifxundefined [1]{%
 \@ifx{#1\undefined}
}%
\providecommand \@ifnum [1]{%
 \ifnum #1\expandafter \@firstoftwo
 \else \expandafter \@secondoftwo
 \fi
}%
\providecommand \@ifx [1]{%
 \ifx #1\expandafter \@firstoftwo
 \else \expandafter \@secondoftwo
 \fi
}%
\providecommand \natexlab [1]{#1}%
\providecommand \enquote  [1]{``#1''}%
\providecommand \bibnamefont  [1]{#1}%
\providecommand \bibfnamefont [1]{#1}%
\providecommand \citenamefont [1]{#1}%
\providecommand \href@noop [0]{\@secondoftwo}%
\providecommand \href [0]{\begingroup \@sanitize@url \@href}%
\providecommand \@href[1]{\@@startlink{#1}\@@href}%
\providecommand \@@href[1]{\endgroup#1\@@endlink}%
\providecommand \@sanitize@url [0]{\catcode `\\12\catcode `\$12\catcode
  `\&12\catcode `\#12\catcode `\^12\catcode `\_12\catcode `\%12\relax}%
\providecommand \@@startlink[1]{}%
\providecommand \@@endlink[0]{}%
\providecommand \url  [0]{\begingroup\@sanitize@url \@url }%
\providecommand \@url [1]{\endgroup\@href {#1}{\urlprefix }}%
\providecommand \urlprefix  [0]{URL }%
\providecommand \Eprint [0]{\href }%
\providecommand \doibase [0]{http://dx.doi.org/}%
\providecommand \selectlanguage [0]{\@gobble}%
\providecommand \bibinfo  [0]{\@secondoftwo}%
\providecommand \bibfield  [0]{\@secondoftwo}%
\providecommand \translation [1]{[#1]}%
\providecommand \BibitemOpen [0]{}%
\providecommand \bibitemStop [0]{}%
\providecommand \bibitemNoStop [0]{.\EOS\space}%
\providecommand \EOS [0]{\spacefactor3000\relax}%
\providecommand \BibitemShut  [1]{\csname bibitem#1\endcsname}%
\let\auto@bib@innerbib\@empty
\bibitem [{\citenamefont {Deutsch}(1991)}]{Deutsch_1991}%
  \BibitemOpen
  \bibfield  {author} {\bibinfo {author} {\bibfnamefont {J.~M.}\ \bibnamefont
  {Deutsch}},\ }\href {\doibase 10.1103/PhysRevA.43.2046} {\bibfield  {journal}
  {\bibinfo  {journal} {Phys. Rev. A}\ }\textbf {\bibinfo {volume} {43}},\
  \bibinfo {pages} {2046} (\bibinfo {year} {1991})}\BibitemShut {NoStop}%
\bibitem [{\citenamefont {Srednicki}(1994)}]{Srednicki_1994}%
  \BibitemOpen
  \bibfield  {author} {\bibinfo {author} {\bibfnamefont {M.}~\bibnamefont
  {Srednicki}},\ }\href {\doibase 10.1103/PhysRevE.50.888} {\bibfield
  {journal} {\bibinfo  {journal} {Phys. Rev. E}\ }\textbf {\bibinfo {volume}
  {50}},\ \bibinfo {pages} {888} (\bibinfo {year} {1994})}\BibitemShut
  {NoStop}%
\bibitem [{\citenamefont {Rigol}\ \emph {et~al.}(2008)\citenamefont {Rigol},
  \citenamefont {Dunjko},\ and\ \citenamefont {Olshanii}}]{Rigol_2008}%
  \BibitemOpen
  \bibfield  {author} {\bibinfo {author} {\bibfnamefont {M.}~\bibnamefont
  {Rigol}}, \bibinfo {author} {\bibfnamefont {V.}~\bibnamefont {Dunjko}}, \
  and\ \bibinfo {author} {\bibfnamefont {M.}~\bibnamefont {Olshanii}},\
  }\href@noop {} {\bibfield  {journal} {\bibinfo  {journal} {Nature}\ }\textbf
  {\bibinfo {volume} {452}},\ \bibinfo {pages} {854} (\bibinfo {year}
  {2008})}\BibitemShut {NoStop}%
\bibitem [{\citenamefont {Deutsch}(2018)}]{Deutsch_2018}%
  \BibitemOpen
  \bibfield  {author} {\bibinfo {author} {\bibfnamefont {J.~M.}\ \bibnamefont
  {Deutsch}},\ }\href@noop {} {\bibfield  {journal} {\bibinfo  {journal}
  {Reports on Progress in Physics}\ }\textbf {\bibinfo {volume} {81}},\
  \bibinfo {pages} {082001} (\bibinfo {year} {2018})}\BibitemShut {NoStop}%
\bibitem [{\citenamefont {Vidmar}\ and\ \citenamefont
  {Rigol}(2016)}]{Vidmar_2016}%
  \BibitemOpen
  \bibfield  {author} {\bibinfo {author} {\bibfnamefont {L.}~\bibnamefont
  {Vidmar}}\ and\ \bibinfo {author} {\bibfnamefont {M.}~\bibnamefont {Rigol}},\
  }\href {\doibase 10.1088/1742-5468/2016/06/064007} {\bibfield  {journal}
  {\bibinfo  {journal} {Journal of Statistical Mechanics: Theory and
  Experiment}\ }\textbf {\bibinfo {volume} {2016}},\ \bibinfo {pages} {064007}
  (\bibinfo {year} {2016})}\BibitemShut {NoStop}%
\bibitem [{\citenamefont {Essler}\ and\ \citenamefont
  {Fagotti}(2016)}]{Essler_2016}%
  \BibitemOpen
  \bibfield  {author} {\bibinfo {author} {\bibfnamefont {F.~H.~L.}\
  \bibnamefont {Essler}}\ and\ \bibinfo {author} {\bibfnamefont
  {M.}~\bibnamefont {Fagotti}},\ }\href {\doibase
  10.1088/1742-5468/2016/06/064002} {\bibfield  {journal} {\bibinfo  {journal}
  {Journal of Statistical Mechanics: Theory and Experiment}\ }\textbf {\bibinfo
  {volume} {2016}},\ \bibinfo {pages} {064002} (\bibinfo {year}
  {2016})}\BibitemShut {NoStop}%
\bibitem [{\citenamefont {Rigol}\ \emph {et~al.}(2007)\citenamefont {Rigol},
  \citenamefont {Dunjko}, \citenamefont {Yurovsky},\ and\ \citenamefont
  {Olshanii}}]{Rigol_2007}%
  \BibitemOpen
  \bibfield  {author} {\bibinfo {author} {\bibfnamefont {M.}~\bibnamefont
  {Rigol}}, \bibinfo {author} {\bibfnamefont {V.}~\bibnamefont {Dunjko}},
  \bibinfo {author} {\bibfnamefont {V.}~\bibnamefont {Yurovsky}}, \ and\
  \bibinfo {author} {\bibfnamefont {M.}~\bibnamefont {Olshanii}},\ }\href
  {\doibase 10.1103/PhysRevLett.98.050405} {\bibfield  {journal} {\bibinfo
  {journal} {Phys. Rev. Lett.}\ }\textbf {\bibinfo {volume} {98}},\ \bibinfo
  {pages} {050405} (\bibinfo {year} {2007})}\BibitemShut {NoStop}%
\bibitem [{\citenamefont {Calabrese}\ \emph {et~al.}(2011)\citenamefont
  {Calabrese}, \citenamefont {Essler},\ and\ \citenamefont
  {Fagotti}}]{Calabrese_2011}%
  \BibitemOpen
  \bibfield  {author} {\bibinfo {author} {\bibfnamefont {P.}~\bibnamefont
  {Calabrese}}, \bibinfo {author} {\bibfnamefont {F.~H.~L.}\ \bibnamefont
  {Essler}}, \ and\ \bibinfo {author} {\bibfnamefont {M.}~\bibnamefont
  {Fagotti}},\ }\href {\doibase 10.1103/PhysRevLett.106.227203} {\bibfield
  {journal} {\bibinfo  {journal} {Phys. Rev. Lett.}\ }\textbf {\bibinfo
  {volume} {106}},\ \bibinfo {pages} {227203} (\bibinfo {year}
  {2011})}\BibitemShut {NoStop}%
\bibitem [{\citenamefont {Pozsgay}(2013)}]{Pozsgay_2013}%
  \BibitemOpen
  \bibfield  {author} {\bibinfo {author} {\bibfnamefont {B.}~\bibnamefont
  {Pozsgay}},\ }\href {\doibase 10.1088/1742-5468/2013/07/p07003} {\bibfield
  {journal} {\bibinfo  {journal} {Journal of Statistical Mechanics: Theory and
  Experiment}\ }\textbf {\bibinfo {volume} {2013}},\ \bibinfo {pages} {P07003}
  (\bibinfo {year} {2013})}\BibitemShut {NoStop}%
\bibitem [{\citenamefont {Fagotti}\ and\ \citenamefont
  {Essler}(2013)}]{Fagotti_2013}%
  \BibitemOpen
  \bibfield  {author} {\bibinfo {author} {\bibfnamefont {M.}~\bibnamefont
  {Fagotti}}\ and\ \bibinfo {author} {\bibfnamefont {F.~H.~L.}\ \bibnamefont
  {Essler}},\ }\href {\doibase 10.1088/1742-5468/2013/07/p07012} {\bibfield
  {journal} {\bibinfo  {journal} {Journal of Statistical Mechanics: Theory and
  Experiment}\ }\textbf {\bibinfo {volume} {2013}},\ \bibinfo {pages} {P07012}
  (\bibinfo {year} {2013})}\BibitemShut {NoStop}%
\bibitem [{\citenamefont {Wright}\ \emph {et~al.}(2014)\citenamefont {Wright},
  \citenamefont {Rigol}, \citenamefont {Davis},\ and\ \citenamefont
  {Kheruntsyan}}]{Wright_2014}%
  \BibitemOpen
  \bibfield  {author} {\bibinfo {author} {\bibfnamefont {T.~M.}\ \bibnamefont
  {Wright}}, \bibinfo {author} {\bibfnamefont {M.}~\bibnamefont {Rigol}},
  \bibinfo {author} {\bibfnamefont {M.~J.}\ \bibnamefont {Davis}}, \ and\
  \bibinfo {author} {\bibfnamefont {K.~V.}\ \bibnamefont {Kheruntsyan}},\
  }\href {\doibase 10.1103/PhysRevLett.113.050601} {\bibfield  {journal}
  {\bibinfo  {journal} {Phys. Rev. Lett.}\ }\textbf {\bibinfo {volume} {113}},\
  \bibinfo {pages} {050601} (\bibinfo {year} {2014})}\BibitemShut {NoStop}%
\bibitem [{\citenamefont {Ilievski}\ \emph {et~al.}(2015)\citenamefont
  {Ilievski}, \citenamefont {De~Nardis}, \citenamefont {Wouters}, \citenamefont
  {Caux}, \citenamefont {Essler},\ and\ \citenamefont
  {Prosen}}]{Ilievski_2015}%
  \BibitemOpen
  \bibfield  {author} {\bibinfo {author} {\bibfnamefont {E.}~\bibnamefont
  {Ilievski}}, \bibinfo {author} {\bibfnamefont {J.}~\bibnamefont {De~Nardis}},
  \bibinfo {author} {\bibfnamefont {B.}~\bibnamefont {Wouters}}, \bibinfo
  {author} {\bibfnamefont {J.-S.}\ \bibnamefont {Caux}}, \bibinfo {author}
  {\bibfnamefont {F.~H.~L.}\ \bibnamefont {Essler}}, \ and\ \bibinfo {author}
  {\bibfnamefont {T.}~\bibnamefont {Prosen}},\ }\href {\doibase
  10.1103/PhysRevLett.115.157201} {\bibfield  {journal} {\bibinfo  {journal}
  {Phys. Rev. Lett.}\ }\textbf {\bibinfo {volume} {115}},\ \bibinfo {pages}
  {157201} (\bibinfo {year} {2015})}\BibitemShut {NoStop}%
\bibitem [{\citenamefont {Basko}\ \emph {et~al.}(2006)\citenamefont {Basko},
  \citenamefont {Aleiner},\ and\ \citenamefont {Altshuler}}]{Basko_2006}%
  \BibitemOpen
  \bibfield  {author} {\bibinfo {author} {\bibfnamefont {D.}~\bibnamefont
  {Basko}}, \bibinfo {author} {\bibfnamefont {I.}~\bibnamefont {Aleiner}}, \
  and\ \bibinfo {author} {\bibfnamefont {B.}~\bibnamefont {Altshuler}},\ }\href
  {\doibase https://doi.org/10.1016/j.aop.2005.11.014} {\bibfield  {journal}
  {\bibinfo  {journal} {Annals of Physics}\ }\textbf {\bibinfo {volume}
  {321}},\ \bibinfo {pages} {1126} (\bibinfo {year} {2006})}\BibitemShut
  {NoStop}%
\bibitem [{\citenamefont {Bardarson}\ \emph {et~al.}(2012)\citenamefont
  {Bardarson}, \citenamefont {Pollmann},\ and\ \citenamefont
  {Moore}}]{Bardarson_2012}%
  \BibitemOpen
  \bibfield  {author} {\bibinfo {author} {\bibfnamefont {J.~H.}\ \bibnamefont
  {Bardarson}}, \bibinfo {author} {\bibfnamefont {F.}~\bibnamefont {Pollmann}},
  \ and\ \bibinfo {author} {\bibfnamefont {J.~E.}\ \bibnamefont {Moore}},\
  }\href {\doibase 10.1103/PhysRevLett.109.017202} {\bibfield  {journal}
  {\bibinfo  {journal} {Phys. Rev. Lett.}\ }\textbf {\bibinfo {volume} {109}},\
  \bibinfo {pages} {017202} (\bibinfo {year} {2012})}\BibitemShut {NoStop}%
\bibitem [{\citenamefont {Serbyn}\ \emph
  {et~al.}(2013{\natexlab{a}})\citenamefont {Serbyn}, \citenamefont
  {Papi\ifmmode~\acute{c}\else \'{c}\fi{}},\ and\ \citenamefont
  {Abanin}}]{Serbyn_2013}%
  \BibitemOpen
  \bibfield  {author} {\bibinfo {author} {\bibfnamefont {M.}~\bibnamefont
  {Serbyn}}, \bibinfo {author} {\bibfnamefont {Z.}~\bibnamefont
  {Papi\ifmmode~\acute{c}\else \'{c}\fi{}}}, \ and\ \bibinfo {author}
  {\bibfnamefont {D.~A.}\ \bibnamefont {Abanin}},\ }\href {\doibase
  10.1103/PhysRevLett.110.260601} {\bibfield  {journal} {\bibinfo  {journal}
  {Phys. Rev. Lett.}\ }\textbf {\bibinfo {volume} {110}},\ \bibinfo {pages}
  {260601} (\bibinfo {year} {2013}{\natexlab{a}})}\BibitemShut {NoStop}%
\bibitem [{\citenamefont {Serbyn}\ \emph
  {et~al.}(2013{\natexlab{b}})\citenamefont {Serbyn}, \citenamefont
  {Papi\ifmmode~\acute{c}\else \'{c}\fi{}},\ and\ \citenamefont
  {Abanin}}]{Serbyn_2013b}%
  \BibitemOpen
  \bibfield  {author} {\bibinfo {author} {\bibfnamefont {M.}~\bibnamefont
  {Serbyn}}, \bibinfo {author} {\bibfnamefont {Z.}~\bibnamefont
  {Papi\ifmmode~\acute{c}\else \'{c}\fi{}}}, \ and\ \bibinfo {author}
  {\bibfnamefont {D.~A.}\ \bibnamefont {Abanin}},\ }\href {\doibase
  10.1103/PhysRevLett.111.127201} {\bibfield  {journal} {\bibinfo  {journal}
  {Phys. Rev. Lett.}\ }\textbf {\bibinfo {volume} {111}},\ \bibinfo {pages}
  {127201} (\bibinfo {year} {2013}{\natexlab{b}})}\BibitemShut {NoStop}%
\bibitem [{\citenamefont {Huse}\ \emph {et~al.}(2014)\citenamefont {Huse},
  \citenamefont {Nandkishore},\ and\ \citenamefont {Oganesyan}}]{Huse_2014}%
  \BibitemOpen
  \bibfield  {author} {\bibinfo {author} {\bibfnamefont {D.~A.}\ \bibnamefont
  {Huse}}, \bibinfo {author} {\bibfnamefont {R.}~\bibnamefont {Nandkishore}}, \
  and\ \bibinfo {author} {\bibfnamefont {V.}~\bibnamefont {Oganesyan}},\ }\href
  {\doibase 10.1103/PhysRevB.90.174202} {\bibfield  {journal} {\bibinfo
  {journal} {Phys. Rev. B}\ }\textbf {\bibinfo {volume} {90}},\ \bibinfo
  {pages} {174202} (\bibinfo {year} {2014})}\BibitemShut {NoStop}%
\bibitem [{\citenamefont {Luitz}\ \emph {et~al.}(2015)\citenamefont {Luitz},
  \citenamefont {Laflorencie},\ and\ \citenamefont {Alet}}]{Luitz_2015}%
  \BibitemOpen
  \bibfield  {author} {\bibinfo {author} {\bibfnamefont {D.~J.}\ \bibnamefont
  {Luitz}}, \bibinfo {author} {\bibfnamefont {N.}~\bibnamefont {Laflorencie}},
  \ and\ \bibinfo {author} {\bibfnamefont {F.}~\bibnamefont {Alet}},\ }\href
  {\doibase 10.1103/PhysRevB.91.081103} {\bibfield  {journal} {\bibinfo
  {journal} {Phys. Rev. B}\ }\textbf {\bibinfo {volume} {91}},\ \bibinfo
  {pages} {081103} (\bibinfo {year} {2015})}\BibitemShut {NoStop}%
\bibitem [{\citenamefont {Chandran}\ \emph {et~al.}(2015)\citenamefont
  {Chandran}, \citenamefont {Kim}, \citenamefont {Vidal},\ and\ \citenamefont
  {Abanin}}]{Chandran_2015}%
  \BibitemOpen
  \bibfield  {author} {\bibinfo {author} {\bibfnamefont {A.}~\bibnamefont
  {Chandran}}, \bibinfo {author} {\bibfnamefont {I.~H.}\ \bibnamefont {Kim}},
  \bibinfo {author} {\bibfnamefont {G.}~\bibnamefont {Vidal}}, \ and\ \bibinfo
  {author} {\bibfnamefont {D.~A.}\ \bibnamefont {Abanin}},\ }\href {\doibase
  10.1103/PhysRevB.91.085425} {\bibfield  {journal} {\bibinfo  {journal} {Phys.
  Rev. B}\ }\textbf {\bibinfo {volume} {91}},\ \bibinfo {pages} {085425}
  (\bibinfo {year} {2015})}\BibitemShut {NoStop}%
\bibitem [{\citenamefont {Ros}\ \emph {et~al.}(2015)\citenamefont {Ros},
  \citenamefont {M\"{u}ller},\ and\ \citenamefont {Scardicchio}}]{Ros_2015}%
  \BibitemOpen
  \bibfield  {author} {\bibinfo {author} {\bibfnamefont {V.}~\bibnamefont
  {Ros}}, \bibinfo {author} {\bibfnamefont {M.}~\bibnamefont {M\"{u}ller}}, \
  and\ \bibinfo {author} {\bibfnamefont {A.}~\bibnamefont {Scardicchio}},\
  }\href {\doibase https://doi.org/10.1016/j.nuclphysb.2014.12.014} {\bibfield
  {journal} {\bibinfo  {journal} {Nuclear Physics B}\ }\textbf {\bibinfo
  {volume} {891}},\ \bibinfo {pages} {420} (\bibinfo {year}
  {2015})}\BibitemShut {NoStop}%
\bibitem [{\citenamefont {Nandkishore}\ and\ \citenamefont
  {Huse}(2015)}]{Nandkishore_2015}%
  \BibitemOpen
  \bibfield  {author} {\bibinfo {author} {\bibfnamefont {R.}~\bibnamefont
  {Nandkishore}}\ and\ \bibinfo {author} {\bibfnamefont {D.~A.}\ \bibnamefont
  {Huse}},\ }\href {\doibase 10.1146/annurev-conmatphys-031214-014726}
  {\bibfield  {journal} {\bibinfo  {journal} {Annual Review of Condensed Matter
  Physics}\ }\textbf {\bibinfo {volume} {6}},\ \bibinfo {pages} {15} (\bibinfo
  {year} {2015})}\BibitemShut {NoStop}%
  \bibitem [{\citenamefont {Haldar}\ \emph {et~al.}(2018)\citenamefont {Haldar},
  \citenamefont {Moessner},\ and\ \citenamefont {Das}}]{Haldar_2018}%
  \BibitemOpen
  \bibfield  {author} {\bibinfo {author} {\bibfnamefont {A.}~\bibnamefont
  {Haldar}}, \bibinfo {author} {\bibfnamefont {R.}~\bibnamefont {Moessner}},\
  and\ \bibinfo {author} {\bibfnamefont {A.}~\bibnamefont {Das}},\ }\href
  {https://doi.org/10.1103/PhysRevB.97.245122} {\bibfield  {journal} {\bibinfo
  {journal} {Phys. Rev. B}\ }\textbf {\bibinfo {volume} {97}},\ \bibinfo
  {pages} {245122} (\bibinfo {year} {2018})}\BibitemShut {NoStop}%
\bibitem [{\citenamefont {Haldar}\ \emph {et~al.}(2021)\citenamefont {Haldar},
  \citenamefont {Sen}, \citenamefont {Moessner},\ and\ \citenamefont
  {Das}}]{Haldar_2021}%
  \BibitemOpen
  \bibfield  {author} {\bibinfo {author} {\bibfnamefont {A.}~\bibnamefont
  {Haldar}}, \bibinfo {author} {\bibfnamefont {D.}~\bibnamefont {Sen}},
  \bibinfo {author} {\bibfnamefont {R.}~\bibnamefont {Moessner}},\ and\
  \bibinfo {author} {\bibfnamefont {A.}~\bibnamefont {Das}},\ }\href
  {https://doi.org/10.1103/PhysRevX.11.021008} {\bibfield  {journal} {\bibinfo
  {journal} {Phys. Rev. X}\ }\textbf {\bibinfo {volume} {11}},\ \bibinfo
  {pages} {021008} (\bibinfo {year} {2021})}\BibitemShut {NoStop}%
\bibitem [{\citenamefont {Haldar}\ and\ \citenamefont
  {Das}(2022)}]{Haldar_2022}%
  \BibitemOpen
  \bibfield  {author} {\bibinfo {author} {\bibfnamefont {A.}~\bibnamefont
  {Haldar}}\ and\ \bibinfo {author} {\bibfnamefont {A.}~\bibnamefont {Das}},\
  }\href {https://doi.org/10.1088/1361-648X/ac03d2} {\bibfield  {journal}
  {\bibinfo  {journal} {Journal of Physics: Condensed Matter}\ }\textbf
  {\bibinfo {volume} {34}},\ \bibinfo {pages} {234001} (\bibinfo {year}
  {2022})}\BibitemShut {NoStop}%
\bibitem [{\citenamefont {Bernien}\ \emph {et~al.}(2017)\citenamefont
  {Bernien}, \citenamefont {Schwartz}, \citenamefont {Keesling}, \citenamefont
  {Levine}, \citenamefont {Omran}, \citenamefont {Pichler}, \citenamefont
  {Choi}, \citenamefont {Zibrov}, \citenamefont {Endres}, \citenamefont
  {Greiner} \emph {et~al.}}]{Bernien_2017}%
  \BibitemOpen
  \bibfield  {author} {\bibinfo {author} {\bibfnamefont {H.}~\bibnamefont
  {Bernien}}, \bibinfo {author} {\bibfnamefont {S.}~\bibnamefont {Schwartz}},
  \bibinfo {author} {\bibfnamefont {A.}~\bibnamefont {Keesling}}, \bibinfo
  {author} {\bibfnamefont {H.}~\bibnamefont {Levine}}, \bibinfo {author}
  {\bibfnamefont {A.}~\bibnamefont {Omran}}, \bibinfo {author} {\bibfnamefont
  {H.}~\bibnamefont {Pichler}}, \bibinfo {author} {\bibfnamefont
  {S.}~\bibnamefont {Choi}}, \bibinfo {author} {\bibfnamefont {A.~S.}\
  \bibnamefont {Zibrov}}, \bibinfo {author} {\bibfnamefont {M.}~\bibnamefont
  {Endres}}, \bibinfo {author} {\bibfnamefont {M.}~\bibnamefont {Greiner}},
  \emph {et~al.},\ }\href@noop {} {\bibfield  {journal} {\bibinfo  {journal}
  {Nature}\ }\textbf {\bibinfo {volume} {551}},\ \bibinfo {pages} {579}
  (\bibinfo {year} {2017})}\BibitemShut {NoStop}%
\bibitem [{\citenamefont {Turner}\ \emph
  {et~al.}(2018{\natexlab{a}})\citenamefont {Turner}, \citenamefont
  {Michailidis}, \citenamefont {Abanin}, \citenamefont {Serbyn},\ and\
  \citenamefont {Papi\'c}}]{Turner_2018}%
  \BibitemOpen
  \bibfield  {author} {\bibinfo {author} {\bibfnamefont {C.~J.}\ \bibnamefont
  {Turner}}, \bibinfo {author} {\bibfnamefont {A.~A.}\ \bibnamefont
  {Michailidis}}, \bibinfo {author} {\bibfnamefont {D.~A.}\ \bibnamefont
  {Abanin}}, \bibinfo {author} {\bibfnamefont {M.}~\bibnamefont {Serbyn}}, \
  and\ \bibinfo {author} {\bibfnamefont {Z.}~\bibnamefont {Papi\'c}},\ }\href
  {\doibase 10.1038/s41567-018-0137-5} {\bibfield  {journal} {\bibinfo
  {journal} {Nature Physics}\ }\textbf {\bibinfo {volume} {14}},\ \bibinfo
  {pages} {745} (\bibinfo {year} {2018}{\natexlab{a}})}\BibitemShut {NoStop}%
\bibitem [{\citenamefont {Moudgalya}\ \emph {et~al.}(2022)\citenamefont
  {Moudgalya}, \citenamefont {Bernevig},\ and\ \citenamefont
  {Regnault}}]{Moudgalya_2021}%
  \BibitemOpen
  \bibfield  {author} {\bibinfo {author} {\bibfnamefont {S.}~\bibnamefont
  {Moudgalya}}, \bibinfo {author} {\bibfnamefont {B.~A.}\ \bibnamefont
  {Bernevig}}, \ and\ \bibinfo {author} {\bibfnamefont {N.}~\bibnamefont
  {Regnault}},\ }\href {\doibase 10.1088/1361-6633/ac73a0} {\bibfield
  {journal} {\bibinfo  {journal} {Reports on Progress in Physics}\ }\textbf
  {\bibinfo {volume} {85}},\ \bibinfo {pages} {086501} (\bibinfo {year}
  {2022})}\BibitemShut {NoStop}%
\bibitem [{\citenamefont {Papi{\'c}}(2021)}]{Papic_2021}%
  \BibitemOpen
  \bibfield  {author} {\bibinfo {author} {\bibfnamefont {Z.}~\bibnamefont
  {Papi{\'c}}},\ }\href {\doibase https://doi.org/10.48550/arXiv.2108.03460}
  {\bibfield  {journal} {\bibinfo  {journal} {arXiv preprint arXiv:2108.03460}\
  } (\bibinfo {year} {2021}),\
  https://doi.org/10.48550/arXiv.2108.03460}\BibitemShut {NoStop}%
\bibitem [{\citenamefont {Chandran}\ \emph {et~al.}(2022)\citenamefont
  {Chandran}, \citenamefont {Iadecola}, \citenamefont {Khemani},\ and\
  \citenamefont {Moessner}}]{Chandran_2022}%
  \BibitemOpen
  \bibfield  {author} {\bibinfo {author} {\bibfnamefont {A.}~\bibnamefont
  {Chandran}}, \bibinfo {author} {\bibfnamefont {T.}~\bibnamefont {Iadecola}},
  \bibinfo {author} {\bibfnamefont {V.}~\bibnamefont {Khemani}}, \ and\
  \bibinfo {author} {\bibfnamefont {R.}~\bibnamefont {Moessner}},\ }\href
  {\doibase https://doi.org/10.48550/ARXIV.2206.11528} {\bibfield  {journal}
  {\bibinfo  {journal} {arXiv}\ } (\bibinfo {year} {2022}),\
  https://doi.org/10.48550/ARXIV.2206.11528}\BibitemShut {NoStop}%
\bibitem [{\citenamefont {Turner}\ \emph
  {et~al.}(2018{\natexlab{b}})\citenamefont {Turner}, \citenamefont
  {Michailidis}, \citenamefont {Abanin}, \citenamefont {Serbyn},\ and\
  \citenamefont {Papi\ifmmode~\acute{c}\else \'{c}\fi{}}}]{Turner_2018b}%
  \BibitemOpen
  \bibfield  {author} {\bibinfo {author} {\bibfnamefont {C.~J.}\ \bibnamefont
  {Turner}}, \bibinfo {author} {\bibfnamefont {A.~A.}\ \bibnamefont
  {Michailidis}}, \bibinfo {author} {\bibfnamefont {D.~A.}\ \bibnamefont
  {Abanin}}, \bibinfo {author} {\bibfnamefont {M.}~\bibnamefont {Serbyn}}, \
  and\ \bibinfo {author} {\bibfnamefont {Z.}~\bibnamefont
  {Papi\ifmmode~\acute{c}\else \'{c}\fi{}}},\ }\href {\doibase
  10.1103/PhysRevB.98.155134} {\bibfield  {journal} {\bibinfo  {journal} {Phys.
  Rev. B}\ }\textbf {\bibinfo {volume} {98}},\ \bibinfo {pages} {155134}
  (\bibinfo {year} {2018}{\natexlab{b}})}\BibitemShut {NoStop}%
\bibitem [{\citenamefont {Iadecola}\ \emph {et~al.}(2019)\citenamefont
  {Iadecola}, \citenamefont {Schecter},\ and\ \citenamefont
  {Xu}}]{Iadecola_2019}%
  \BibitemOpen
  \bibfield  {author} {\bibinfo {author} {\bibfnamefont {T.}~\bibnamefont
  {Iadecola}}, \bibinfo {author} {\bibfnamefont {M.}~\bibnamefont {Schecter}},
  \ and\ \bibinfo {author} {\bibfnamefont {S.}~\bibnamefont {Xu}},\ }\href
  {\doibase 10.1103/PhysRevB.100.184312} {\bibfield  {journal} {\bibinfo
  {journal} {Phys. Rev. B}\ }\textbf {\bibinfo {volume} {100}},\ \bibinfo
  {pages} {184312} (\bibinfo {year} {2019})}\BibitemShut {NoStop}%
\bibitem [{\citenamefont {Choi}\ \emph {et~al.}(2019)\citenamefont {Choi},
  \citenamefont {Turner}, \citenamefont {Pichler}, \citenamefont {Ho},
  \citenamefont {Michailidis}, \citenamefont {Papi\ifmmode~\acute{c}\else
  \'{c}\fi{}}, \citenamefont {Serbyn}, \citenamefont {Lukin},\ and\
  \citenamefont {Abanin}}]{Choi_2019}%
  \BibitemOpen
  \bibfield  {author} {\bibinfo {author} {\bibfnamefont {S.}~\bibnamefont
  {Choi}}, \bibinfo {author} {\bibfnamefont {C.~J.}\ \bibnamefont {Turner}},
  \bibinfo {author} {\bibfnamefont {H.}~\bibnamefont {Pichler}}, \bibinfo
  {author} {\bibfnamefont {W.~W.}\ \bibnamefont {Ho}}, \bibinfo {author}
  {\bibfnamefont {A.~A.}\ \bibnamefont {Michailidis}}, \bibinfo {author}
  {\bibfnamefont {Z.}~\bibnamefont {Papi\ifmmode~\acute{c}\else \'{c}\fi{}}},
  \bibinfo {author} {\bibfnamefont {M.}~\bibnamefont {Serbyn}}, \bibinfo
  {author} {\bibfnamefont {M.~D.}\ \bibnamefont {Lukin}}, \ and\ \bibinfo
  {author} {\bibfnamefont {D.~A.}\ \bibnamefont {Abanin}},\ }\href {\doibase
  10.1103/PhysRevLett.122.220603} {\bibfield  {journal} {\bibinfo  {journal}
  {Phys. Rev. Lett.}\ }\textbf {\bibinfo {volume} {122}},\ \bibinfo {pages}
  {220603} (\bibinfo {year} {2019})}\BibitemShut {NoStop}%
\bibitem [{\citenamefont {Lin}\ and\ \citenamefont
  {Motrunich}(2019)}]{Lin_2019}%
  \BibitemOpen
  \bibfield  {author} {\bibinfo {author} {\bibfnamefont {C.-J.}\ \bibnamefont
  {Lin}}\ and\ \bibinfo {author} {\bibfnamefont {O.~I.}\ \bibnamefont
  {Motrunich}},\ }\href {\doibase 10.1103/PhysRevLett.122.173401} {\bibfield
  {journal} {\bibinfo  {journal} {Phys. Rev. Lett.}\ }\textbf {\bibinfo
  {volume} {122}},\ \bibinfo {pages} {173401} (\bibinfo {year}
  {2019})}\BibitemShut {NoStop}%
\bibitem [{\citenamefont {Lin}\ \emph {et~al.}(2020)\citenamefont {Lin},
  \citenamefont {Calvera},\ and\ \citenamefont {Hsieh}}]{Lin_2020}%
  \BibitemOpen
  \bibfield  {author} {\bibinfo {author} {\bibfnamefont {C.-J.}\ \bibnamefont
  {Lin}}, \bibinfo {author} {\bibfnamefont {V.}~\bibnamefont {Calvera}}, \ and\
  \bibinfo {author} {\bibfnamefont {T.~H.}\ \bibnamefont {Hsieh}},\ }\href
  {\doibase 10.1103/PhysRevB.101.220304} {\bibfield  {journal} {\bibinfo
  {journal} {Phys. Rev. B}\ }\textbf {\bibinfo {volume} {101}},\ \bibinfo
  {pages} {220304} (\bibinfo {year} {2020})}\BibitemShut {NoStop}%
\bibitem [{\citenamefont {Su}\ \emph {et~al.}(2022)\citenamefont {Su},
  \citenamefont {Sun}, \citenamefont {Hudomal}, \citenamefont {Desaules},
  \citenamefont {Zhou}, \citenamefont {Yang}, \citenamefont {Halimeh},
  \citenamefont {Yuan}, \citenamefont {Papi\ifmmode~\acute{c}\else
  \'{c}\fi{}},\ and\ \citenamefont {Pan}}]{Su_2022}%
  \BibitemOpen
  \bibfield  {author} {\bibinfo {author} {\bibfnamefont {G.-X.}\ \bibnamefont
  {Su}}, \bibinfo {author} {\bibfnamefont {H.}~\bibnamefont {Sun}}, \bibinfo
  {author} {\bibfnamefont {A.}~\bibnamefont {Hudomal}}, \bibinfo {author}
  {\bibfnamefont {J.-Y.}\ \bibnamefont {Desaules}}, \bibinfo {author}
  {\bibfnamefont {Z.-Y.}\ \bibnamefont {Zhou}}, \bibinfo {author}
  {\bibfnamefont {B.}~\bibnamefont {Yang}}, \bibinfo {author} {\bibfnamefont
  {J.~C.}\ \bibnamefont {Halimeh}}, \bibinfo {author} {\bibfnamefont {Z.-S.}\
  \bibnamefont {Yuan}}, \bibinfo {author} {\bibfnamefont {Z.}~\bibnamefont
  {Papi\ifmmode~\acute{c}\else \'{c}\fi{}}}, \ and\ \bibinfo {author}
  {\bibfnamefont {J.-W.}\ \bibnamefont {Pan}},\ }\href {\doibase
  https://doi.org/10.48550/arXiv.2201.00821} {\bibfield  {journal} {\bibinfo
  {journal} {arXiv}\ } (\bibinfo {year} {2022}),\
  https://doi.org/10.48550/arXiv.2201.00821}\BibitemShut {NoStop}%
\bibitem [{\citenamefont {Desaules}\ \emph
  {et~al.}(2022{\natexlab{a}})\citenamefont {Desaules}, \citenamefont
  {Banerjee}, \citenamefont {Hudomal}, \citenamefont
  {Papi\ifmmode~\acute{c}\else \'{c}\fi{}}, \citenamefont {Sen},\ and\
  \citenamefont {Halimeh}}]{Desaules_2022}%
  \BibitemOpen
  \bibfield  {author} {\bibinfo {author} {\bibfnamefont {J.-Y.}\ \bibnamefont
  {Desaules}}, \bibinfo {author} {\bibfnamefont {D.}~\bibnamefont {Banerjee}},
  \bibinfo {author} {\bibfnamefont {A.}~\bibnamefont {Hudomal}}, \bibinfo
  {author} {\bibfnamefont {Z.}~\bibnamefont {Papi\ifmmode~\acute{c}\else
  \'{c}\fi{}}}, \bibinfo {author} {\bibfnamefont {A.}~\bibnamefont {Sen}}, \
  and\ \bibinfo {author} {\bibfnamefont {J.~C.}\ \bibnamefont {Halimeh}},\
  }\href {\doibase https://doi.org/10.48550/arXiv.2203.08830} {\bibfield
  {journal} {\bibinfo  {journal} {arXiv}\ } (\bibinfo {year}
  {2022}{\natexlab{a}}),\
  https://doi.org/10.48550/arXiv.2203.08830}\BibitemShut {NoStop}%
\bibitem [{\citenamefont {Desaules}\ \emph
  {et~al.}(2022{\natexlab{b}})\citenamefont {Desaules}, \citenamefont
  {Hudomal}, \citenamefont {Banerjee}, \citenamefont {Sen}, \citenamefont
  {Papi\ifmmode~\acute{c}\else \'{c}\fi{}},\ and\ \citenamefont
  {Halimeh}}]{Desaules_2022bis}%
  \BibitemOpen
  \bibfield  {author} {\bibinfo {author} {\bibfnamefont {J.-Y.}\ \bibnamefont
  {Desaules}}, \bibinfo {author} {\bibfnamefont {A.}~\bibnamefont {Hudomal}},
  \bibinfo {author} {\bibfnamefont {D.}~\bibnamefont {Banerjee}}, \bibinfo
  {author} {\bibfnamefont {A.}~\bibnamefont {Sen}}, \bibinfo {author}
  {\bibfnamefont {Z.}~\bibnamefont {Papi\ifmmode~\acute{c}\else \'{c}\fi{}}}, \
  and\ \bibinfo {author} {\bibfnamefont {J.~C.}\ \bibnamefont {Halimeh}},\
  }\href {\doibase https://doi.org/10.48550/arXiv.2204.01745} {\bibfield
  {journal} {\bibinfo  {journal} {arXiv}\ } (\bibinfo {year}
  {2022}{\natexlab{b}}),\
  https://doi.org/10.48550/arXiv.2204.01745}\BibitemShut {NoStop}%
\bibitem [{\citenamefont {Moudgalya}\ \emph
  {et~al.}(2018{\natexlab{a}})\citenamefont {Moudgalya}, \citenamefont
  {Rachel}, \citenamefont {Bernevig},\ and\ \citenamefont
  {Regnault}}]{Moudgalya_2018}%
  \BibitemOpen
  \bibfield  {author} {\bibinfo {author} {\bibfnamefont {S.}~\bibnamefont
  {Moudgalya}}, \bibinfo {author} {\bibfnamefont {S.}~\bibnamefont {Rachel}},
  \bibinfo {author} {\bibfnamefont {B.~A.}\ \bibnamefont {Bernevig}}, \ and\
  \bibinfo {author} {\bibfnamefont {N.}~\bibnamefont {Regnault}},\ }\href
  {\doibase 10.1103/PhysRevB.98.235155} {\bibfield  {journal} {\bibinfo
  {journal} {Phys. Rev. B}\ }\textbf {\bibinfo {volume} {98}},\ \bibinfo
  {pages} {235155} (\bibinfo {year} {2018}{\natexlab{a}})}\BibitemShut
  {NoStop}%
\bibitem [{\citenamefont {Moudgalya}\ \emph
  {et~al.}(2018{\natexlab{b}})\citenamefont {Moudgalya}, \citenamefont
  {Regnault},\ and\ \citenamefont {Bernevig}}]{Moudgalya_2018b}%
  \BibitemOpen
  \bibfield  {author} {\bibinfo {author} {\bibfnamefont {S.}~\bibnamefont
  {Moudgalya}}, \bibinfo {author} {\bibfnamefont {N.}~\bibnamefont {Regnault}},
  \ and\ \bibinfo {author} {\bibfnamefont {B.~A.}\ \bibnamefont {Bernevig}},\
  }\href {\doibase 10.1103/PhysRevB.98.235156} {\bibfield  {journal} {\bibinfo
  {journal} {Phys. Rev. B}\ }\textbf {\bibinfo {volume} {98}},\ \bibinfo
  {pages} {235156} (\bibinfo {year} {2018}{\natexlab{b}})}\BibitemShut
  {NoStop}%
\bibitem [{\citenamefont {Moudgalya}\ \emph
  {et~al.}(2020{\natexlab{a}})\citenamefont {Moudgalya}, \citenamefont
  {O\'Brien}, \citenamefont {Bernevig}, \citenamefont {Fendley},\ and\
  \citenamefont {Regnault}}]{Moudgalya_2020b}%
  \BibitemOpen
  \bibfield  {author} {\bibinfo {author} {\bibfnamefont {S.}~\bibnamefont
  {Moudgalya}}, \bibinfo {author} {\bibfnamefont {E.}~\bibnamefont {O\'Brien}},
  \bibinfo {author} {\bibfnamefont {B.~A.}\ \bibnamefont {Bernevig}}, \bibinfo
  {author} {\bibfnamefont {P.}~\bibnamefont {Fendley}}, \ and\ \bibinfo
  {author} {\bibfnamefont {N.}~\bibnamefont {Regnault}},\ }\href {\doibase
  10.1103/PhysRevB.102.085120} {\bibfield  {journal} {\bibinfo  {journal}
  {Phys. Rev. B}\ }\textbf {\bibinfo {volume} {102}},\ \bibinfo {pages}
  {085120} (\bibinfo {year} {2020}{\natexlab{a}})}\BibitemShut {NoStop}%
\bibitem [{\citenamefont {Chattopadhyay}\ \emph {et~al.}(2020)\citenamefont
  {Chattopadhyay}, \citenamefont {Pichler}, \citenamefont {Lukin},\ and\
  \citenamefont {Ho}}]{Chattopadhyay_2020}%
  \BibitemOpen
  \bibfield  {author} {\bibinfo {author} {\bibfnamefont {S.}~\bibnamefont
  {Chattopadhyay}}, \bibinfo {author} {\bibfnamefont {H.}~\bibnamefont
  {Pichler}}, \bibinfo {author} {\bibfnamefont {M.~D.}\ \bibnamefont {Lukin}},
  \ and\ \bibinfo {author} {\bibfnamefont {W.~W.}\ \bibnamefont {Ho}},\ }\href
  {\doibase 10.1103/PhysRevB.101.174308} {\bibfield  {journal} {\bibinfo
  {journal} {Phys. Rev. B}\ }\textbf {\bibinfo {volume} {101}},\ \bibinfo
  {pages} {174308} (\bibinfo {year} {2020})}\BibitemShut {NoStop}%
\bibitem [{\citenamefont {Mark}\ \emph {et~al.}(2020)\citenamefont {Mark},
  \citenamefont {Lin},\ and\ \citenamefont {Motrunich}}]{Mark_2020b}%
  \BibitemOpen
  \bibfield  {author} {\bibinfo {author} {\bibfnamefont {D.~K.}\ \bibnamefont
  {Mark}}, \bibinfo {author} {\bibfnamefont {C.-J.}\ \bibnamefont {Lin}}, \
  and\ \bibinfo {author} {\bibfnamefont {O.~I.}\ \bibnamefont {Motrunich}},\
  }\href {\doibase 10.1103/PhysRevB.101.195131} {\bibfield  {journal} {\bibinfo
   {journal} {Phys. Rev. B}\ }\textbf {\bibinfo {volume} {101}},\ \bibinfo
  {pages} {195131} (\bibinfo {year} {2020})}\BibitemShut {NoStop}%
\bibitem [{\citenamefont {Shiraishi}\ and\ \citenamefont
  {Mori}(2017)}]{Shiraishi_2017}%
  \BibitemOpen
  \bibfield  {author} {\bibinfo {author} {\bibfnamefont {N.}~\bibnamefont
  {Shiraishi}}\ and\ \bibinfo {author} {\bibfnamefont {T.}~\bibnamefont
  {Mori}},\ }\href {\doibase 10.1103/PhysRevLett.119.030601} {\bibfield
  {journal} {\bibinfo  {journal} {Phys. Rev. Lett.}\ }\textbf {\bibinfo
  {volume} {119}},\ \bibinfo {pages} {030601} (\bibinfo {year}
  {2017})}\BibitemShut {NoStop}%
\bibitem [{\citenamefont {Iadecola}\ and\ \citenamefont
  {Schecter}(2020)}]{Iadecola_2020}%
  \BibitemOpen
  \bibfield  {author} {\bibinfo {author} {\bibfnamefont {T.}~\bibnamefont
  {Iadecola}}\ and\ \bibinfo {author} {\bibfnamefont {M.}~\bibnamefont
  {Schecter}},\ }\href {\doibase 10.1103/PhysRevB.101.024306} {\bibfield
  {journal} {\bibinfo  {journal} {Phys. Rev. B}\ }\textbf {\bibinfo {volume}
  {101}},\ \bibinfo {pages} {024306} (\bibinfo {year} {2020})}\BibitemShut
  {NoStop}%
\bibitem [{\citenamefont {Langlett}\ and\ \citenamefont
  {Xu}(2021)}]{Langlett_2021}%
  \BibitemOpen
  \bibfield  {author} {\bibinfo {author} {\bibfnamefont {C.~M.}\ \bibnamefont
  {Langlett}}\ and\ \bibinfo {author} {\bibfnamefont {S.}~\bibnamefont {Xu}},\
  }\href {\doibase 10.1103/PhysRevB.103.L220304} {\bibfield  {journal}
  {\bibinfo  {journal} {Phys. Rev. B}\ }\textbf {\bibinfo {volume} {103}},\
  \bibinfo {pages} {L220304} (\bibinfo {year} {2021})}\BibitemShut {NoStop}%
\bibitem [{\citenamefont {Shibata}\ \emph {et~al.}(2020)\citenamefont
  {Shibata}, \citenamefont {Yoshioka},\ and\ \citenamefont
  {Katsura}}]{Shibata_2020}%
  \BibitemOpen
  \bibfield  {author} {\bibinfo {author} {\bibfnamefont {N.}~\bibnamefont
  {Shibata}}, \bibinfo {author} {\bibfnamefont {N.}~\bibnamefont {Yoshioka}}, \
  and\ \bibinfo {author} {\bibfnamefont {H.}~\bibnamefont {Katsura}},\ }\href
  {\doibase 10.1103/PhysRevLett.124.180604} {\bibfield  {journal} {\bibinfo
  {journal} {Phys. Rev. Lett.}\ }\textbf {\bibinfo {volume} {124}},\ \bibinfo
  {pages} {180604} (\bibinfo {year} {2020})}\BibitemShut {NoStop}%
\bibitem [{\citenamefont {Moudgalya}\ \emph
  {et~al.}(2020{\natexlab{b}})\citenamefont {Moudgalya}, \citenamefont
  {Regnault},\ and\ \citenamefont {Bernevig}}]{Moudgalya_2020}%
  \BibitemOpen
  \bibfield  {author} {\bibinfo {author} {\bibfnamefont {S.}~\bibnamefont
  {Moudgalya}}, \bibinfo {author} {\bibfnamefont {N.}~\bibnamefont {Regnault}},
  \ and\ \bibinfo {author} {\bibfnamefont {B.~A.}\ \bibnamefont {Bernevig}},\
  }\href {\doibase 10.1103/PhysRevB.102.085140} {\bibfield  {journal} {\bibinfo
   {journal} {Phys. Rev. B}\ }\textbf {\bibinfo {volume} {102}},\ \bibinfo
  {pages} {085140} (\bibinfo {year} {2020}{\natexlab{b}})}\BibitemShut
  {NoStop}%
\bibitem [{\citenamefont {Mark}\ and\ \citenamefont
  {Motrunich}(2020)}]{Mark_2020}%
  \BibitemOpen
  \bibfield  {author} {\bibinfo {author} {\bibfnamefont {D.~K.}\ \bibnamefont
  {Mark}}\ and\ \bibinfo {author} {\bibfnamefont {O.~I.}\ \bibnamefont
  {Motrunich}},\ }\href {\doibase 10.1103/PhysRevB.102.075132} {\bibfield
  {journal} {\bibinfo  {journal} {Phys. Rev. B}\ }\textbf {\bibinfo {volume}
  {102}},\ \bibinfo {pages} {075132} (\bibinfo {year} {2020})}\BibitemShut
  {NoStop}%
\bibitem [{\citenamefont {Desaules}\ \emph {et~al.}(2021)\citenamefont
  {Desaules}, \citenamefont {Hudomal}, \citenamefont {Turner},\ and\
  \citenamefont {Papi\ifmmode~\acute{c}\else \'{c}\fi{}}}]{Desaules_2021}%
  \BibitemOpen
  \bibfield  {author} {\bibinfo {author} {\bibfnamefont {J.-Y.}\ \bibnamefont
  {Desaules}}, \bibinfo {author} {\bibfnamefont {A.}~\bibnamefont {Hudomal}},
  \bibinfo {author} {\bibfnamefont {C.~J.}\ \bibnamefont {Turner}}, \ and\
  \bibinfo {author} {\bibfnamefont {Z.}~\bibnamefont
  {Papi\ifmmode~\acute{c}\else \'{c}\fi{}}},\ }\href {\doibase
  10.1103/PhysRevLett.126.210601} {\bibfield  {journal} {\bibinfo  {journal}
  {Phys. Rev. Lett.}\ }\textbf {\bibinfo {volume} {126}},\ \bibinfo {pages}
  {210601} (\bibinfo {year} {2021})}\BibitemShut {NoStop}%
\bibitem [{\citenamefont {Moudgalya}\ \emph
  {et~al.}(2020{\natexlab{c}})\citenamefont {Moudgalya}, \citenamefont
  {Bernevig},\ and\ \citenamefont {Regnault}}]{Moudgalya_2020c}%
  \BibitemOpen
  \bibfield  {author} {\bibinfo {author} {\bibfnamefont {S.}~\bibnamefont
  {Moudgalya}}, \bibinfo {author} {\bibfnamefont {B.~A.}\ \bibnamefont
  {Bernevig}}, \ and\ \bibinfo {author} {\bibfnamefont {N.}~\bibnamefont
  {Regnault}},\ }\href {\doibase 10.1103/PhysRevB.102.195150} {\bibfield
  {journal} {\bibinfo  {journal} {Phys. Rev. B}\ }\textbf {\bibinfo {volume}
  {102}},\ \bibinfo {pages} {195150} (\bibinfo {year}
  {2020}{\natexlab{c}})}\BibitemShut {NoStop}%
\bibitem [{\citenamefont {Mukherjee}\ \emph {et~al.}(2020)\citenamefont
  {Mukherjee}, \citenamefont {Nandy}, \citenamefont {Sen}, \citenamefont
  {Sen},\ and\ \citenamefont {Sengupta}}]{Mukherjee_2020}%
  \BibitemOpen
  \bibfield  {author} {\bibinfo {author} {\bibfnamefont {B.}~\bibnamefont
  {Mukherjee}}, \bibinfo {author} {\bibfnamefont {S.}~\bibnamefont {Nandy}},
  \bibinfo {author} {\bibfnamefont {A.}~\bibnamefont {Sen}}, \bibinfo {author}
  {\bibfnamefont {D.}~\bibnamefont {Sen}}, \ and\ \bibinfo {author}
  {\bibfnamefont {K.}~\bibnamefont {Sengupta}},\ }\href {\doibase
  10.1103/PhysRevB.101.245107} {\bibfield  {journal} {\bibinfo  {journal}
  {Phys. Rev. B}\ }\textbf {\bibinfo {volume} {101}},\ \bibinfo {pages}
  {245107} (\bibinfo {year} {2020})}\BibitemShut {NoStop}%
\bibitem [{\citenamefont {Zhao}\ \emph {et~al.}(2020)\citenamefont {Zhao},
  \citenamefont {Vovrosh}, \citenamefont {Mintert},\ and\ \citenamefont
  {Knolle}}]{Zhao_2020}%
  \BibitemOpen
  \bibfield  {author} {\bibinfo {author} {\bibfnamefont {H.}~\bibnamefont
  {Zhao}}, \bibinfo {author} {\bibfnamefont {J.}~\bibnamefont {Vovrosh}},
  \bibinfo {author} {\bibfnamefont {F.}~\bibnamefont {Mintert}}, \ and\
  \bibinfo {author} {\bibfnamefont {J.}~\bibnamefont {Knolle}},\ }\href
  {\doibase 10.1103/PhysRevLett.124.160604} {\bibfield  {journal} {\bibinfo
  {journal} {Phys. Rev. Lett.}\ }\textbf {\bibinfo {volume} {124}},\ \bibinfo
  {pages} {160604} (\bibinfo {year} {2020})}\BibitemShut {NoStop}%
\bibitem [{\citenamefont {Mizuta}\ \emph {et~al.}(2020)\citenamefont {Mizuta},
  \citenamefont {Takasan},\ and\ \citenamefont {Kawakami}}]{Mizuta_2020}%
  \BibitemOpen
  \bibfield  {author} {\bibinfo {author} {\bibfnamefont {K.}~\bibnamefont
  {Mizuta}}, \bibinfo {author} {\bibfnamefont {K.}~\bibnamefont {Takasan}}, \
  and\ \bibinfo {author} {\bibfnamefont {N.}~\bibnamefont {Kawakami}},\ }\href
  {\doibase 10.1103/PhysRevResearch.2.033284} {\bibfield  {journal} {\bibinfo
  {journal} {Phys. Rev. Research}\ }\textbf {\bibinfo {volume} {2}},\ \bibinfo
  {pages} {033284} (\bibinfo {year} {2020})}\BibitemShut {NoStop}%
\bibitem [{\citenamefont {Sugiura}\ \emph {et~al.}(2021)\citenamefont
  {Sugiura}, \citenamefont {Kuwahara},\ and\ \citenamefont
  {Saito}}]{Sugiura_2021}%
  \BibitemOpen
  \bibfield  {author} {\bibinfo {author} {\bibfnamefont {S.}~\bibnamefont
  {Sugiura}}, \bibinfo {author} {\bibfnamefont {T.}~\bibnamefont {Kuwahara}}, \
  and\ \bibinfo {author} {\bibfnamefont {K.}~\bibnamefont {Saito}},\ }\href
  {\doibase 10.1103/PhysRevResearch.3.L012010} {\bibfield  {journal} {\bibinfo
  {journal} {Phys. Rev. Research}\ }\textbf {\bibinfo {volume} {3}},\ \bibinfo
  {pages} {L012010} (\bibinfo {year} {2021})}\BibitemShut {NoStop}%
\bibitem [{\citenamefont {Banerjee}\ and\ \citenamefont
  {Sen}(2021)}]{Banerjee_2021}%
  \BibitemOpen
  \bibfield  {author} {\bibinfo {author} {\bibfnamefont {D.}~\bibnamefont
  {Banerjee}}\ and\ \bibinfo {author} {\bibfnamefont {A.}~\bibnamefont {Sen}},\
  }\href {\doibase 10.1103/PhysRevLett.126.220601} {\bibfield  {journal}
  {\bibinfo  {journal} {Phys. Rev. Lett.}\ }\textbf {\bibinfo {volume} {126}},\
  \bibinfo {pages} {220601} (\bibinfo {year} {2021})}\BibitemShut {NoStop}%
\bibitem [{\citenamefont {Halimeh}\ \emph {et~al.}(2022)\citenamefont
  {Halimeh}, \citenamefont {Barbiero}, \citenamefont {Hauke}, \citenamefont
  {Grusdt},\ and\ \citenamefont {Bohrdt}}]{Halimeh_2022}%
  \BibitemOpen
  \bibfield  {author} {\bibinfo {author} {\bibfnamefont {J.~C.}\ \bibnamefont
  {Halimeh}}, \bibinfo {author} {\bibfnamefont {L.}~\bibnamefont {Barbiero}},
  \bibinfo {author} {\bibfnamefont {P.}~\bibnamefont {Hauke}}, \bibinfo
  {author} {\bibfnamefont {F.}~\bibnamefont {Grusdt}}, \ and\ \bibinfo {author}
  {\bibfnamefont {A.}~\bibnamefont {Bohrdt}},\ }\href {\doibase
  https://doi.org/10.48550/ARXIV.2203.08828} {\  (\bibinfo {year} {2022}),\
  https://doi.org/10.48550/ARXIV.2203.08828}\BibitemShut {NoStop}%
\bibitem [{\citenamefont {Aramthottil}\ \emph {et~al.}(2022)\citenamefont
  {Aramthottil}, \citenamefont {Bhattacharya}, \citenamefont
  {Gonz\'alez-Cuadra}, \citenamefont {Lewenstein}, \citenamefont {Barbiero},\
  and\ \citenamefont {Zakrzewski}}]{Aramthottil_2022}%
  \BibitemOpen
  \bibfield  {author} {\bibinfo {author} {\bibfnamefont {A.~S.}\ \bibnamefont
  {Aramthottil}}, \bibinfo {author} {\bibfnamefont {U.}~\bibnamefont
  {Bhattacharya}}, \bibinfo {author} {\bibfnamefont {D.}~\bibnamefont
  {Gonz\'alez-Cuadra}}, \bibinfo {author} {\bibfnamefont {M.}~\bibnamefont
  {Lewenstein}}, \bibinfo {author} {\bibfnamefont {L.}~\bibnamefont
  {Barbiero}}, \ and\ \bibinfo {author} {\bibfnamefont {J.}~\bibnamefont
  {Zakrzewski}},\ }\href {\doibase 10.1103/PhysRevB.106.L041101} {\bibfield
  {journal} {\bibinfo  {journal} {Phys. Rev. B}\ }\textbf {\bibinfo {volume}
  {106}},\ \bibinfo {pages} {L041101} (\bibinfo {year} {2022})}\BibitemShut
  {NoStop}%
\bibitem [{\citenamefont {McClarty}\ \emph {et~al.}(2020)\citenamefont
  {McClarty}, \citenamefont {Haque}, \citenamefont {Sen},\ and\ \citenamefont
  {Richter}}]{McClarty_2020}%
  \BibitemOpen
  \bibfield  {author} {\bibinfo {author} {\bibfnamefont {P.~A.}\ \bibnamefont
  {McClarty}}, \bibinfo {author} {\bibfnamefont {M.}~\bibnamefont {Haque}},
  \bibinfo {author} {\bibfnamefont {A.}~\bibnamefont {Sen}}, \ and\ \bibinfo
  {author} {\bibfnamefont {J.}~\bibnamefont {Richter}},\ }\href {\doibase
  10.1103/PhysRevB.102.224303} {\bibfield  {journal} {\bibinfo  {journal}
  {Phys. Rev. B}\ }\textbf {\bibinfo {volume} {102}},\ \bibinfo {pages}
  {224303} (\bibinfo {year} {2020})}\BibitemShut {NoStop}%
\bibitem [{\citenamefont {Kuno}\ \emph {et~al.}(2020)\citenamefont {Kuno},
  \citenamefont {Mizoguchi},\ and\ \citenamefont {Hatsugai}}]{Kuno_2020}%
  \BibitemOpen
  \bibfield  {author} {\bibinfo {author} {\bibfnamefont {Y.}~\bibnamefont
  {Kuno}}, \bibinfo {author} {\bibfnamefont {T.}~\bibnamefont {Mizoguchi}}, \
  and\ \bibinfo {author} {\bibfnamefont {Y.}~\bibnamefont {Hatsugai}},\ }\href
  {\doibase 10.1103/PhysRevB.102.241115} {\bibfield  {journal} {\bibinfo
  {journal} {Phys. Rev. B}\ }\textbf {\bibinfo {volume} {102}},\ \bibinfo
  {pages} {241115} (\bibinfo {year} {2020})}\BibitemShut {NoStop}%
\bibitem [{\citenamefont {Lee}\ \emph {et~al.}(2020)\citenamefont {Lee},
  \citenamefont {Melendrez}, \citenamefont {Pal},\ and\ \citenamefont
  {Changlani}}]{Lee_2020}%
  \BibitemOpen
  \bibfield  {author} {\bibinfo {author} {\bibfnamefont {K.}~\bibnamefont
  {Lee}}, \bibinfo {author} {\bibfnamefont {R.}~\bibnamefont {Melendrez}},
  \bibinfo {author} {\bibfnamefont {A.}~\bibnamefont {Pal}}, \ and\ \bibinfo
  {author} {\bibfnamefont {H.~J.}\ \bibnamefont {Changlani}},\ }\href {\doibase
  10.1103/PhysRevB.101.241111} {\bibfield  {journal} {\bibinfo  {journal}
  {Phys. Rev. B}\ }\textbf {\bibinfo {volume} {101}},\ \bibinfo {pages}
  {241111} (\bibinfo {year} {2020})}\BibitemShut {NoStop}%
\bibitem [{\citenamefont {Lee}\ \emph {et~al.}(2021)\citenamefont {Lee},
  \citenamefont {Pal},\ and\ \citenamefont {Changlani}}]{Lee_2021}%
  \BibitemOpen
  \bibfield  {author} {\bibinfo {author} {\bibfnamefont {K.}~\bibnamefont
  {Lee}}, \bibinfo {author} {\bibfnamefont {A.}~\bibnamefont {Pal}}, \ and\
  \bibinfo {author} {\bibfnamefont {H.~J.}\ \bibnamefont {Changlani}},\ }\href
  {\doibase 10.1103/PhysRevB.103.235133} {\bibfield  {journal} {\bibinfo
  {journal} {Phys. Rev. B}\ }\textbf {\bibinfo {volume} {103}},\ \bibinfo
  {pages} {235133} (\bibinfo {year} {2021})}\BibitemShut {NoStop}%
\bibitem [{\citenamefont {Yang}(1989)}]{Yang_1989}%
  \BibitemOpen
  \bibfield  {author} {\bibinfo {author} {\bibfnamefont {C.~N.}\ \bibnamefont
  {Yang}},\ }\href {\doibase 10.1103/PhysRevLett.63.2144} {\bibfield  {journal}
  {\bibinfo  {journal} {Phys. Rev. Lett.}\ }\textbf {\bibinfo {volume} {63}},\
  \bibinfo {pages} {2144} (\bibinfo {year} {1989})}\BibitemShut {NoStop}%
\bibitem [{\citenamefont {Vafek}\ \emph {et~al.}(2017)\citenamefont {Vafek},
  \citenamefont {Regnault},\ and\ \citenamefont {Bernevig}}]{Vafek_2017}%
  \BibitemOpen
  \bibfield  {author} {\bibinfo {author} {\bibfnamefont {O.}~\bibnamefont
  {Vafek}}, \bibinfo {author} {\bibfnamefont {N.}~\bibnamefont {Regnault}}, \
  and\ \bibinfo {author} {\bibfnamefont {B.~A.}\ \bibnamefont {Bernevig}},\
  }\href {\doibase 10.21468/SciPostPhys.3.6.043} {\bibfield  {journal}
  {\bibinfo  {journal} {SciPost Phys.}\ }\textbf {\bibinfo {volume} {3}},\
  \bibinfo {pages} {043} (\bibinfo {year} {2017})}\BibitemShut {NoStop}%
\bibitem [{\citenamefont {Schecter}\ and\ \citenamefont
  {Iadecola}(2019)}]{Schecter_2019}%
  \BibitemOpen
  \bibfield  {author} {\bibinfo {author} {\bibfnamefont {M.}~\bibnamefont
  {Schecter}}\ and\ \bibinfo {author} {\bibfnamefont {T.}~\bibnamefont
  {Iadecola}},\ }\href {\doibase 10.1103/PhysRevLett.123.147201} {\bibfield
  {journal} {\bibinfo  {journal} {Phys. Rev. Lett.}\ }\textbf {\bibinfo
  {volume} {123}},\ \bibinfo {pages} {147201} (\bibinfo {year}
  {2019})}\BibitemShut {NoStop}%
\bibitem [{\citenamefont {Pakrouski}\ \emph {et~al.}(2020)\citenamefont
  {Pakrouski}, \citenamefont {Pallegar}, \citenamefont {Popov},\ and\
  \citenamefont {Klebanov}}]{Pakrouski_2020}%
  \BibitemOpen
  \bibfield  {author} {\bibinfo {author} {\bibfnamefont {K.}~\bibnamefont
  {Pakrouski}}, \bibinfo {author} {\bibfnamefont {P.~N.}\ \bibnamefont
  {Pallegar}}, \bibinfo {author} {\bibfnamefont {F.~K.}\ \bibnamefont {Popov}},
  \ and\ \bibinfo {author} {\bibfnamefont {I.~R.}\ \bibnamefont {Klebanov}},\
  }\href {\doibase 10.1103/PhysRevLett.125.230602} {\bibfield  {journal}
  {\bibinfo  {journal} {Phys. Rev. Lett.}\ }\textbf {\bibinfo {volume} {125}},\
  \bibinfo {pages} {230602} (\bibinfo {year} {2020})}\BibitemShut {NoStop}%
\bibitem [{\citenamefont {Nakagawa}\ \emph {et~al.}(2022)\citenamefont
  {Nakagawa}, \citenamefont {Katsura},\ and\ \citenamefont
  {Ueda}}]{Nakagawa_2022}%
  \BibitemOpen
  \bibfield  {author} {\bibinfo {author} {\bibfnamefont {M.}~\bibnamefont
  {Nakagawa}}, \bibinfo {author} {\bibfnamefont {H.}~\bibnamefont {Katsura}}, \
  and\ \bibinfo {author} {\bibfnamefont {M.}~\bibnamefont {Ueda}},\ }\href
  {\doibase https://doi.org/10.48550/arXiv.2205.07235} {\bibfield  {journal}
  {\bibinfo  {journal} {arXiv}\ } (\bibinfo {year} {2022}),\
  https://doi.org/10.48550/arXiv.2205.07235}\BibitemShut {NoStop}%
\bibitem [{\citenamefont {Yoshida}\ and\ \citenamefont
  {Katsura}(2022)}]{Yoshida_2022}%
  \BibitemOpen
  \bibfield  {author} {\bibinfo {author} {\bibfnamefont {H.}~\bibnamefont
  {Yoshida}}\ and\ \bibinfo {author} {\bibfnamefont {H.}~\bibnamefont
  {Katsura}},\ }\href {\doibase 10.1103/PhysRevB.105.024520} {\bibfield
  {journal} {\bibinfo  {journal} {Phys. Rev. B}\ }\textbf {\bibinfo {volume}
  {105}},\ \bibinfo {pages} {024520} (\bibinfo {year} {2022})}\BibitemShut
  {NoStop}%
\bibitem [{\citenamefont {Alhambra}\ \emph {et~al.}(2020)\citenamefont
  {Alhambra}, \citenamefont {Anshu},\ and\ \citenamefont
  {Wilming}}]{Alhambra_2020}%
  \BibitemOpen
  \bibfield  {author} {\bibinfo {author} {\bibfnamefont {A.~M.}\ \bibnamefont
  {Alhambra}}, \bibinfo {author} {\bibfnamefont {A.}~\bibnamefont {Anshu}}, \
  and\ \bibinfo {author} {\bibfnamefont {H.}~\bibnamefont {Wilming}},\ }\href
  {\doibase 10.1103/PhysRevB.101.205107} {\bibfield  {journal} {\bibinfo
  {journal} {Phys. Rev. B}\ }\textbf {\bibinfo {volume} {101}},\ \bibinfo
  {pages} {205107} (\bibinfo {year} {2020})}\BibitemShut {NoStop}%
\bibitem [{\citenamefont {Ren}\ \emph {et~al.}(2021)\citenamefont {Ren},
  \citenamefont {Liang},\ and\ \citenamefont {Fang}}]{Ren_2021}%
  \BibitemOpen
  \bibfield  {author} {\bibinfo {author} {\bibfnamefont {J.}~\bibnamefont
  {Ren}}, \bibinfo {author} {\bibfnamefont {C.}~\bibnamefont {Liang}}, \ and\
  \bibinfo {author} {\bibfnamefont {C.}~\bibnamefont {Fang}},\ }\href {\doibase
  10.1103/PhysRevLett.126.120604} {\bibfield  {journal} {\bibinfo  {journal}
  {Phys. Rev. Lett.}\ }\textbf {\bibinfo {volume} {126}},\ \bibinfo {pages}
  {120604} (\bibinfo {year} {2021})}\BibitemShut {NoStop}%
\bibitem [{\citenamefont {O'Dea}\ \emph {et~al.}(2020)\citenamefont {O'Dea},
  \citenamefont {Burnell}, \citenamefont {Chandran},\ and\ \citenamefont
  {Khemani}}]{ODea_2020}%
  \BibitemOpen
  \bibfield  {author} {\bibinfo {author} {\bibfnamefont {N.}~\bibnamefont
  {O'Dea}}, \bibinfo {author} {\bibfnamefont {F.}~\bibnamefont {Burnell}},
  \bibinfo {author} {\bibfnamefont {A.}~\bibnamefont {Chandran}}, \ and\
  \bibinfo {author} {\bibfnamefont {V.}~\bibnamefont {Khemani}},\ }\href
  {\doibase 10.1103/PhysRevResearch.2.043305} {\bibfield  {journal} {\bibinfo
  {journal} {Phys. Rev. Research}\ }\textbf {\bibinfo {volume} {2}},\ \bibinfo
  {pages} {043305} (\bibinfo {year} {2020})}\BibitemShut {NoStop}%
\bibitem [{\citenamefont {Bariev}(1991)}]{Bariev_1991}%
  \BibitemOpen
  \bibfield  {author} {\bibinfo {author} {\bibfnamefont {R.~Z.}\ \bibnamefont
  {Bariev}},\ }\href {\doibase 10.1088/0305-4470/24/10/010} {\bibfield
  {journal} {\bibinfo  {journal} {Journal of Physics A: Mathematical and
  General}\ }\textbf {\bibinfo {volume} {24}},\ \bibinfo {pages} {L549}
  (\bibinfo {year} {1991})}\BibitemShut {NoStop}%
\bibitem [{\citenamefont {Chhajlany}\ \emph {et~al.}(2016)\citenamefont
  {Chhajlany}, \citenamefont {Grzybowski}, \citenamefont
  {Stasi\ifmmode~\acute{n}\else \'{n}\fi{}ska}, \citenamefont {Lewenstein},\
  and\ \citenamefont {Dutta}}]{Chhajlany_2016}%
  \BibitemOpen
  \bibfield  {author} {\bibinfo {author} {\bibfnamefont {R.~W.}\ \bibnamefont
  {Chhajlany}}, \bibinfo {author} {\bibfnamefont {P.~R.}\ \bibnamefont
  {Grzybowski}}, \bibinfo {author} {\bibfnamefont {J.}~\bibnamefont
  {Stasi\ifmmode~\acute{n}\else \'{n}\fi{}ska}}, \bibinfo {author}
  {\bibfnamefont {M.}~\bibnamefont {Lewenstein}}, \ and\ \bibinfo {author}
  {\bibfnamefont {O.}~\bibnamefont {Dutta}},\ }\href {\doibase
  10.1103/PhysRevLett.116.225303} {\bibfield  {journal} {\bibinfo  {journal}
  {Phys. Rev. Lett.}\ }\textbf {\bibinfo {volume} {116}},\ \bibinfo {pages}
  {225303} (\bibinfo {year} {2016})}\BibitemShut {NoStop}%
\bibitem [{\citenamefont {Ruhman}\ and\ \citenamefont
  {Altman}(2017)}]{Ruhman_2017}%
  \BibitemOpen
  \bibfield  {author} {\bibinfo {author} {\bibfnamefont {J.}~\bibnamefont
  {Ruhman}}\ and\ \bibinfo {author} {\bibfnamefont {E.}~\bibnamefont
  {Altman}},\ }\href {\doibase 10.1103/PhysRevB.96.085133} {\bibfield
  {journal} {\bibinfo  {journal} {Phys. Rev. B}\ }\textbf {\bibinfo {volume}
  {96}},\ \bibinfo {pages} {085133} (\bibinfo {year} {2017})}\BibitemShut
  {NoStop}%
\bibitem [{\citenamefont {Gotta}\ \emph
  {et~al.}(2021{\natexlab{a}})\citenamefont {Gotta}, \citenamefont {Mazza},
  \citenamefont {Simon},\ and\ \citenamefont {Roux}}]{Gotta_2021}%
  \BibitemOpen
  \bibfield  {author} {\bibinfo {author} {\bibfnamefont {L.}~\bibnamefont
  {Gotta}}, \bibinfo {author} {\bibfnamefont {L.}~\bibnamefont {Mazza}},
  \bibinfo {author} {\bibfnamefont {P.}~\bibnamefont {Simon}}, \ and\ \bibinfo
  {author} {\bibfnamefont {G.}~\bibnamefont {Roux}},\ }\href {\doibase
  10.1103/PhysRevLett.126.206805} {\bibfield  {journal} {\bibinfo  {journal}
  {Phys. Rev. Lett.}\ }\textbf {\bibinfo {volume} {126}},\ \bibinfo {pages}
  {206805} (\bibinfo {year} {2021}{\natexlab{a}})}\BibitemShut {NoStop}%
\bibitem [{\citenamefont {Gotta}\ \emph
  {et~al.}(2021{\natexlab{b}})\citenamefont {Gotta}, \citenamefont {Mazza},
  \citenamefont {Simon},\ and\ \citenamefont {Roux}}]{Gotta_2021bis}%
  \BibitemOpen
  \bibfield  {author} {\bibinfo {author} {\bibfnamefont {L.}~\bibnamefont
  {Gotta}}, \bibinfo {author} {\bibfnamefont {L.}~\bibnamefont {Mazza}},
  \bibinfo {author} {\bibfnamefont {P.}~\bibnamefont {Simon}}, \ and\ \bibinfo
  {author} {\bibfnamefont {G.}~\bibnamefont {Roux}},\ }\href {\doibase
  10.1103/PhysRevB.104.094521} {\bibfield  {journal} {\bibinfo  {journal}
  {Phys. Rev. B}\ }\textbf {\bibinfo {volume} {104}},\ \bibinfo {pages}
  {094521} (\bibinfo {year} {2021}{\natexlab{b}})}\BibitemShut {NoStop}%
\bibitem [{\citenamefont {Gotta}\ \emph {et~al.}(2022)\citenamefont {Gotta},
  \citenamefont {Mazza}, \citenamefont {Simon},\ and\ \citenamefont
  {Roux}}]{Gotta_2022}%
  \BibitemOpen
  \bibfield  {author} {\bibinfo {author} {\bibfnamefont {L.}~\bibnamefont
  {Gotta}}, \bibinfo {author} {\bibfnamefont {L.}~\bibnamefont {Mazza}},
  \bibinfo {author} {\bibfnamefont {P.}~\bibnamefont {Simon}}, \ and\ \bibinfo
  {author} {\bibfnamefont {G.}~\bibnamefont {Roux}},\ }\href {\doibase
  10.1103/PhysRevB.105.134512} {\bibfield  {journal} {\bibinfo  {journal}
  {Phys. Rev. B}\ }\textbf {\bibinfo {volume} {105}},\ \bibinfo {pages}
  {134512} (\bibinfo {year} {2022})}\BibitemShut {NoStop}%
\bibitem [{\citenamefont {Pitaevskii}\ and\ \citenamefont
  {Stringari}(2003)}]{Stringari_2003}%
  \BibitemOpen
  \bibfield  {author} {\bibinfo {author} {\bibfnamefont {L.~P.}\ \bibnamefont
  {Pitaevskii}}\ and\ \bibinfo {author} {\bibfnamefont {S.}~\bibnamefont
  {Stringari}},\ }\href@noop {} {\emph {\bibinfo {title} {Bose-Einstein
  Condensation}}}\ (\bibinfo  {publisher} {Oxford University Press},\ \bibinfo
  {year} {2003})\BibitemShut {NoStop}%
\bibitem [{\citenamefont {Weinberg}\ and\ \citenamefont
  {Bukov}(2017)}]{Quspin_1}%
  \BibitemOpen
  \bibfield  {author} {\bibinfo {author} {\bibfnamefont {P.}~\bibnamefont
  {Weinberg}}\ and\ \bibinfo {author} {\bibfnamefont {M.}~\bibnamefont
  {Bukov}},\ }\href {\doibase 10.21468/SciPostPhys.2.1.003} {\bibfield
  {journal} {\bibinfo  {journal} {SciPost Phys.}\ }\textbf {\bibinfo {volume}
  {2}},\ \bibinfo {pages} {003} (\bibinfo {year} {2017})}\BibitemShut {NoStop}%
\bibitem [{\citenamefont {Weinberg}\ and\ \citenamefont
  {Bukov}(2019)}]{Quspin_2}%
  \BibitemOpen
  \bibfield  {author} {\bibinfo {author} {\bibfnamefont {P.}~\bibnamefont
  {Weinberg}}\ and\ \bibinfo {author} {\bibfnamefont {M.}~\bibnamefont
  {Bukov}},\ }\href {\doibase 10.21468/SciPostPhys.7.2.020} {\bibfield
  {journal} {\bibinfo  {journal} {SciPost Phys.}\ }\textbf {\bibinfo {volume}
  {7}},\ \bibinfo {pages} {20} (\bibinfo {year} {2019})}\BibitemShut {NoStop}%
\bibitem [{\citenamefont {Oganesyan}\ and\ \citenamefont
  {Huse}(2007)}]{Oganesyan_2007}%
  \BibitemOpen
  \bibfield  {author} {\bibinfo {author} {\bibfnamefont {V.}~\bibnamefont
  {Oganesyan}}\ and\ \bibinfo {author} {\bibfnamefont {D.~A.}\ \bibnamefont
  {Huse}},\ }\href {\doibase 10.1103/PhysRevB.75.155111} {\bibfield  {journal}
  {\bibinfo  {journal} {Phys. Rev. B}\ }\textbf {\bibinfo {volume} {75}},\
  \bibinfo {pages} {155111} (\bibinfo {year} {2007})}\BibitemShut {NoStop}%
\bibitem [{\citenamefont {Atas}\ \emph {et~al.}(2013)\citenamefont {Atas},
  \citenamefont {Bogomolny}, \citenamefont {Giraud},\ and\ \citenamefont
  {Roux}}]{Atas_2013}%
  \BibitemOpen
  \bibfield  {author} {\bibinfo {author} {\bibfnamefont {Y.~Y.}\ \bibnamefont
  {Atas}}, \bibinfo {author} {\bibfnamefont {E.}~\bibnamefont {Bogomolny}},
  \bibinfo {author} {\bibfnamefont {O.}~\bibnamefont {Giraud}}, \ and\ \bibinfo
  {author} {\bibfnamefont {G.}~\bibnamefont {Roux}},\ }\href {\doibase
  10.1103/PhysRevLett.110.084101} {\bibfield  {journal} {\bibinfo  {journal}
  {Phys. Rev. Lett.}\ }\textbf {\bibinfo {volume} {110}},\ \bibinfo {pages}
  {084101} (\bibinfo {year} {2013})}\BibitemShut {NoStop}%
\bibitem [{\citenamefont {Mazza}\ \emph {et~al.}(2018)\citenamefont {Mazza},
  \citenamefont {Iemini}, \citenamefont {Dalmonte},\ and\ \citenamefont
  {Mora}}]{Mazza_2018}%
  \BibitemOpen
  \bibfield  {author} {\bibinfo {author} {\bibfnamefont {L.}~\bibnamefont
  {Mazza}}, \bibinfo {author} {\bibfnamefont {F.}~\bibnamefont {Iemini}},
  \bibinfo {author} {\bibfnamefont {M.}~\bibnamefont {Dalmonte}}, \ and\
  \bibinfo {author} {\bibfnamefont {C.}~\bibnamefont {Mora}},\ }\href {\doibase
  10.1103/PhysRevB.98.201109} {\bibfield  {journal} {\bibinfo  {journal} {Phys.
  Rev. B}\ }\textbf {\bibinfo {volume} {98}},\ \bibinfo {pages} {201109}
  (\bibinfo {year} {2018})}\BibitemShut {NoStop}%
\bibitem [{\citenamefont {Tamura}\ and\ \citenamefont
  {Katsura}(2022)}]{Tamura_2022}%
  \BibitemOpen
  \bibfield  {author} {\bibinfo {author} {\bibfnamefont {K.}~\bibnamefont
  {Tamura}}\ and\ \bibinfo {author} {\bibfnamefont {H.}~\bibnamefont
  {Katsura}},\ }\href {\doibase https://doi.org/10.48550/arXiv.2207.06040}
  {\bibfield  {journal} {\bibinfo  {journal} {arXiv}\ } (\bibinfo {year}
  {2022}),\ https://doi.org/10.48550/arXiv.2207.06040}\BibitemShut {NoStop}%
\end{thebibliography}
\end{document}